\documentclass[prb,preprint,aps,superscriptaddress,longbibliography]{revtex4-1}

\usepackage{graphicx, epstopdf}
\usepackage{subfigure}
\usepackage{float}
\usepackage{amsmath}
\usepackage{amsfonts}
\usepackage{amssymb}
\usepackage{bm}
\usepackage{subfigure}
\usepackage{caption}
\usepackage{hyperref}

\usepackage{color,soul}
\usepackage{xcolor}
\usepackage{hyperref}
\hypersetup{
     colorlinks = true,
     citecolor  = red,  
     linkcolor  = blue  
}

\begin{document}


\title{Quantum Anomalous Hall Effect and Related Topological Electronic
  States}

\author{Hongming Weng}
\affiliation{Beijing National Laboratory for Condensed Matter Physics,
      and Institute of Physics, Chinese Academy of Sciences, Beijing
      100190, China}
      
\affiliation{Collaborative Innovation Center of Quantum Matter,
  Beijing, China}

\author{Rui Yu}
\affiliation{International Center for
      Materials Nanoarchitectonics (WPI-MANA), 
      National Institute for Materials Science, Tsukuba 305-0044, Japan}
\author{Xiao Hu}
\email{HU.Xiao@nims.go.jp}
\affiliation{International Center for
      Materials Nanoarchitectonics (WPI-MANA), 
      National Institute for Materials Science, Tsukuba 305-0044, Japan}
\author{Xi Dai}
\email{daix@iphy.ac.cn}
\affiliation{Beijing National Laboratory for Condensed Matter Physics,
      and Institute of Physics, Chinese Academy of Sciences, Beijing
      100190, China}
      
\affiliation{Collaborative Innovation Center of Quantum Matter,
  Beijing, China}

\author{Zhong  Fang}
\email{zfang@iphy.ac.cn}
\affiliation{Beijing National Laboratory for Condensed Matter Physics,
      and Institute of Physics, Chinese Academy of Sciences, Beijing
      100190, China}
      
\affiliation{Collaborative Innovation Center of Quantum Matter,
  Beijing, China}


\begin{abstract}
  Over a long period of exploration, the successful observation of
  quantized version of anomalous Hall effect (AHE) in thin film of
  magnetically-doped topological insulator completed a quantum Hall
  trio---quantum Hall effect (QHE), quantum spin Hall effect (QSHE),
  and quantum anomalous Hall effect (QAHE).  On the theoretical front,
  it was understood that intrinsic AHE is related to Berry curvature
  and U(1) gauge field in momentum space. This understanding
  established connection between the QAHE and the topological
  properties of electronic structures characterized by the Chern
  number.  With the time reversal symmetry broken by magnetization, a
  QAHE system carries dissipationless charge current at edges, similar
  to the QHE where an external magnetic field is necessary. The QAHE
  and corresponding Chern insulators are also closely related to other
  topological electronic states, such as topological insulators and
  topological semimetals, which have been extensively studied recently
  and have been known to exist in various compounds.  First-principles
  electronic structure calculations play important roles not only for
  the understanding of fundamental physics in this field, but also
  towards the prediction and realization of realistic compounds.  In
  this article, a theoretical review on the Berry phase mechanism and
  related topological electronic states in terms of various
  topological invariants will be given with focus on the QAHE and
  Chern insulators. We will introduce the {Wilson} loop method and the
  band inversion mechanism for the selection and design of topological
  materials, and discuss the predictive power of first-principles
  calculations. Finally, remaining issues, challenges and possible
  applications for future investigations in the field will be
  addressed.
\end{abstract}

\maketitle

\centerline{\bfseries List of Contents}\medskip


\noindent
{I.} Introduction\\
\hspace*{10pt}{I.A.}  {Hall Effect and Anomalous Hall Effect (AHE)}\\
\hspace*{10pt}{I.B.}  Berry Phase Mechanism for Intrinsic AHE\\

\noindent
{II.}    Quantum Hall Family and Related Topological Electronic States\\
\hspace*{10pt}{II.A.}  Chern Number, Chern Insulator, and Quantum AHE (QAHE)\\
\hspace*{10pt}{II.B.}  $Z_2$ Invariant, Topological Insulator, and
Quantum Spin Hall Effect (QSHE) \\
\hspace*{10pt}{III.C.}  Magnetic Monopole, Weyl Node and Topological Semimetal\\

\noindent
{III.}    First-Principles Calculations for Topological Electronic States\\
\hspace*{10pt}{III.A.}  Calculations of Berry Connection and Berry
Curvature \\
\hspace*{10pt}{III.B.}  Wilson Loop Method for Evaluation of
Topological Invariants \\
\hspace*{10pt}{III.C.}  Boundary States Calculations\\

\noindent
{IV.}    Material Predictions and Realizations of QAHE\\
\hspace*{10pt}{IV.A.}   Band Inversion Mechanism\\
\hspace*{10pt}{IV.B.}   QAHE in Magnetic Topological Insulators\\
\hspace*{10pt}{IV.C.}   QAHE in Thin Film of Weyl semimetals\\
\hspace*{10pt}{IV.D.}   QAHE on Honeycomb Lattice\\

\noindent
{V.}    Discussion and Future Prospects\\

\noindent
{VI.}    Acknowledgements \\

\noindent
{VII.}    References \\


\section{Introduction}
{\bf I.A. Hall Effect and Anomalous Hall Effect (AHE)}
\\


Edwin H. Hall discovered, in 1879, that when a conductor carrying
longitudinal current was placed in a vertical magnetic field, the
carriers would be pressed towards the transverse side of the
conductor, which led to observed transverse voltage.  This is called
Hall effect (HE)~\cite{halleffect}, and it was a remarkable discovery,
although it was difficult to understand at that time since the
electron was not to be discovered until 18 years later.  We now know
that the HE is due to the Lorentz force experienced by the moving
electrons in the magnetic field, {which is balanced by a transverse
  voltage for a steady current in the longitudinal direction.}  The
Hall resistivity $\rho_{xy}$ under perpendicular external magnetic
field ${\bf H}$ (along the $z$ direction) can be written as
$\rho_{xy}=R_0 H$, where $R_0$ is the Hall coefficient, which can be
related to the carrier density $n$ as $R_0=-\frac{1}{ne}$ (in the free
electron gas approximation).  The HE is a fundamental phenomenon in
condensed matter physics, and it has been widely used as as
experimental tool to identify the type of carrier and to measure the
carrier density or the strength of magnetic fields.

In 1880, Edwin H. Hall further found that this ``pressing electricity
effect'' in ferromagnetic (FM) conductors was larger than in
non-magnetic (NM) conductors.  This enhanced Hall effect was then
called as the anomalous Hall effect (AHE)~\cite{ahe}, in order to
distinguish it from the ordinary HE.  Later experiments on Fe, Co and
Ni~\cite{kundt_hall_1893} suggested that the AHE was related to the
sample magnetization ${\bf M}$ (along $z$), and an empirical relation
for the total Hall effect in FM conductors was established
as~\cite{pugh_hall_1930,pugh_hall_1932}
\begin{equation}
\rho_{xy}=R_0H+R_sM,
\label{eq:AHE_rxy}
\end{equation}
where the second term is the anomalous Hall resistivity, and its
coefficient $R_s$ is material-dependent (in contrast to $R_0$, which
depends only on carrier density $n$).

\begin{figure}[tbp]
\begin{centering}
 \includegraphics[clip,width=0.95\textwidth]{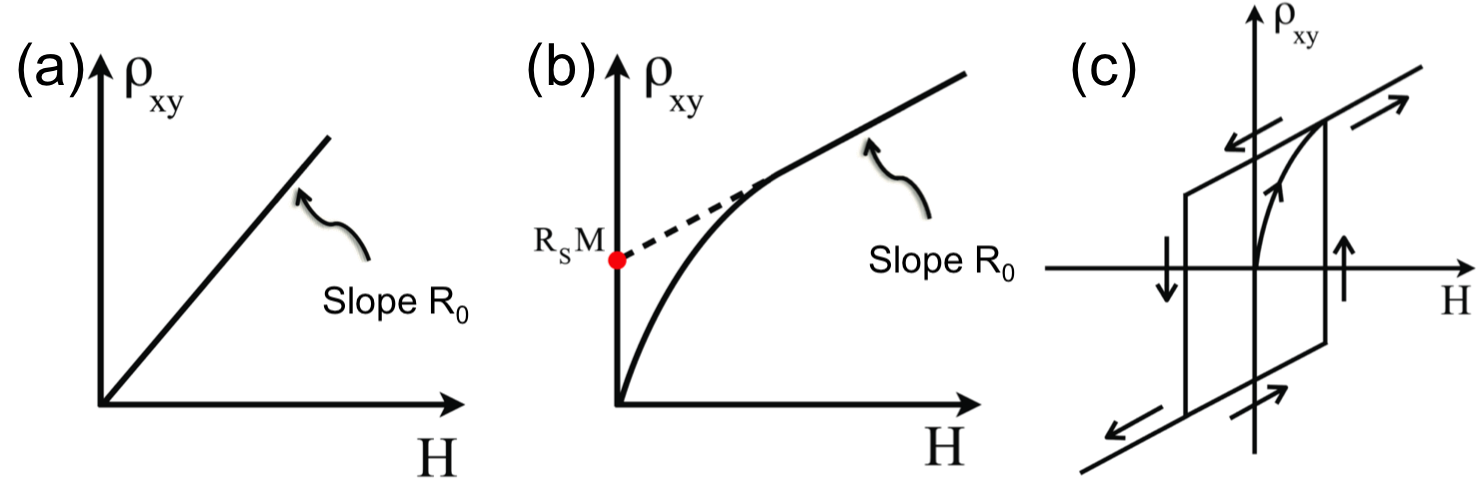} \par
\end{centering}
\caption{(Color online) Typical behaviors of the Hall and anomalous
  Hall effects. The Hall resistivity $\rho_{xy}$ is plotted versus
  external magnetic field ${\bf H}$.  (a) The Hall effect; (b) The
  anomalous Hall effect; and (c) The hysteresis loop measured from the
  anomalous Hall effect.}
  \label{fig:HE_AHE}
\end{figure}

As schematically shown in Fig. \ref{fig:HE_AHE}, if we plot the Hall
resistivity $\rho_{xy}$ versus external magnetic field ${\bf H}$, we
will generally expect a straight line (which crosses the origin of the
coordinates) for NM conductors (see Fig. \ref{fig:HE_AHE}(a));
however, non-linear behavior will be expected for FM conductors,
wherein $\rho_{xy}$ increases sharply at low field and crosses over
into a linear region under high field (see Fig. \ref{fig:HE_AHE}(b)).
The initial sharp increase of $\rho_{xy}$ is due to the saturation of
magnetization of the sample under external field. After the
saturation of magnetization, $\rho_{xy}$ will depend on ${\bf H}$
linearly (for high field region), which is dominated by the ordinary
Hall contribution.  Therefore, the slope of the linear part under high
field gives us $R_0$.  If we extrapolate this linear part to the zero
field limit (${\bf H}$=0), it will not go through the origin of
coordinates, and its intercept on $y$-axis gives us $R_sM$ as can be
learned from Eq. \eqref{eq:AHE_rxy}.  This is a surprising result,
which immediately implies two important facts: (1) A kind of Hall
effect can be observed even in the absence of external magnetic field
(i.e., no Lorentz force); (2) The anomalous Hall resistivity is
sensitive to magnetic moment ${\bf M}$, and it has been suggested that
this property would be useful for the detection of magnetization of
conducting carriers, particularly for the cases of weak itinerant
magnetism, such as surface/interface
magnetization~\cite{bergmann_transition_1978}, dilute magnetic
semiconductors~\cite{ohno_magnetotransport_1992,ohno_gamnas:_1996},
etc.  Experimentally, measurements are usually done in magnetization
loops by scanning the magnetic field from positive to negative, and a
hysteresis loop in the $\rho_{xy}$ vs. ${\bf H}$ relationship is to be
expected.  This is very similar to the familiar hysteresis loop
observed in the ${\bf M}$ vs. ${\bf H}$ curve, as usually found in FM
conductors.  The empirical relation Eq. \eqref{eq:AHE_rxy} is very
simple and widely used. Unfortunately, its correctness is not well
justified.  As will be pointed out in a latter part of this article,
the anomalous Hall resistivity may have a very complicated form, which
is usually non-linear in ${\bf M}$.

Although the HE and AHE look quite similar phenomenologically, their
underlying physics are completely different.  The HE is due to the
Lorentz force's influence on the moving electrons under magnetic
field; however the AHE exists in the absence of external magnetic
field, where there is no orbital effect of moving electrons.  In other
words, why do the electrons move towards the transverse direction in
the absence of Lorentz force?  The mechanism of AHE has been an
enigmatic problem since its discovery, a problem that lasted almost a
century.  This problem involves concepts deeply related to topology
and geometry that have been formulated only in recent years
\cite{jungwirth, onoda-2, fang_anomalous_2003, yao_first_2004,
  sinova_magneto_transport_2004,
  nagaosa_anomalous_2010,XiaoDi_RMP_2010} after the Berry phase was
recognized in 1984~\cite{berry}.

Karplus and Luttinger provided a crucial step in unraveling this
problem as early as 1954~\cite{kl}. They showed that moving electrons
under external electric field can acquire an ``anomalous velocity'',
which is perpendicular to the electric {field} and contributes to the
transverse motion of electrons and therefore to the AHE.  This
``anomalous velocity'' comes from the occupied electronic states in FM
conductors with spin-orbit coupling (SOC). They suggested that the
mechanism leads to an anomalous Hall resistivity proportional to the
square of the longitudinal resistivity:
$\rho_{xy}\propto\rho_{xx}^{2}$.  Since this contribution depends only
on the electronic band structures of perfect periodic crystal and is
completely independent of scattering from impurities or defects, it is
called {\it intrinsic} AHE, and it was not widely accepted until the
concept of Berry phase was well established.

For a long time, two other {\it extrinsic} contributions had been
considered to be the dominant mechanisms that give rise to the AHE.
Smit~\cite{skew, smit_spontaneous_1958} suggested that there always
exist defects or impurities in real materials, which will scatter the
moving electrons. In the presence of SOC and ferromagnetism, this
scattering is asymmetric and should lead to unbalanced transverse
motion of electrons.  This is called {\it skew scattering}, and Smit
argued that it is the main source of the AHE. This mechanism predicted
$\rho_{xy}\sim \rho_{xx}$, in contrast with the intrinsic
contribution.  Berger~\cite{side}, on the other hand, argued that
electrons should experience a difference in electric field when
approaching and leaving an impurity, and this leads to another
asymmetric scattering process {called} {\it side jump}, which also
{contributes} to the AHE. This mechanism marvelously predicts
$\rho_{xy} \sim \rho_{xx}^2$, the same as the intrinsic contribution.

The debates about the origin of AHE lasted for a long time, and no
conclusion could be drawn unambiguously. From the experimental point
of view, defects or impurities in samples are unavoidable and are
usually complicated with rich varieties.  The contributions from both
{\it intrinsic} and {\it extrinsic} mechanisms should generally
coexist~\cite{Onoda_PRL_2006}.  Experimentalists have tried hard to
distinguish them~\cite{Miyasato_PRL_2007, tian_proper_2009, Ye_2012,
  zeng_linear_2006}. The controversy also arises because of a lack of
quantitative calculations that could be compared with experiments.

Still, the discovery of the quantum Hall effect in 1980s~\cite{qhe}
and the later studies on the geometric
phase~\cite{wilczek1989geometric} and the topological properties of
quantum Hall states~\cite{tknn, prange_quantum_1987,
  stone_quantum_1992}, promoted the fruitful study of AHE
significantly.  Around the early years of this century, the {\it
  intrinsic} AHE mechanism proposed by Karplus and Luttinger was
completely reformulated in the language of Berry phase and topology in
electronic structures~\cite{jungwirth, onoda-2, fang_anomalous_2003,
  yao_first_2004, sinova_magneto_transport_2004}.  It was recognized
for the first time that the so-called ``anomalous velocity''
originates from the Berry curvature of occupied eigen wave functions,
which can be understood as effective magnetic field in momentum space.
This effective magnetic field modifies the equation of motion of
electrons and leads to the {\it intrinsic} AHE~\cite{jungwirth, onoda-2,
  fang_anomalous_2003, yao_first_2004, sinova_magneto_transport_2004,
  XiaoDi_RMP_2010,nagaosa_anomalous_2010}.  From this understanding,
we now know the {\it intrinsic} AHE can be evaluated quantitatively
from the band structure calculations, thanks to the rapid development
in the field of first-principles calculations. Detailed calculations
for various compounds portray the contributions from {\it intrinsic}
AHE in a way that convincingly confirms the existing experimental
results~\cite{fang_anomalous_2003, yao_first_2004,
  yao_theoretical_2007,zeng_linear_2006, tian_proper_2009,
  lee_dissipationless_2004}, establishing the dominant role of the
{\it intrinsic} contribution to the AHE.  In the meantime, the {\it
  extrinsic} AHE was also analyzed
carefully~\cite{nagaosa_anomalous_2010}, and it was understood that
its contribution dominates for the clean limit case (rather than the
dirty limit), where the deviation of distribution of electronic states
from the equilibrium distribution is significant.  In this article, we
will not discuss the {\it extrinsic} AHE.  For readers who want to
learn more details, please refer to the review article by Nagaosa et
al.~\cite{nagaosa_anomalous_2010} Introduction of the Berry phase
mechanism for the understanding of {\it intrinsic} AHE was a big step
forward in the field, and it is also the fundamental base for
understanding the quantum anomalous Hall effect (QAHE), which is the
main subject of this article.

~\\~
\hspace*{15pt}{\bf I.B. Berry Phase Mechanism for Intrinsic AHE}
\\

In quantum {mechanics}, the Berry phase is the quantal phase acquired by
the adiabatic evolution of wave function associated with the adiabatic
change of the Hamiltonian in a parameter space $\bm{\mathcal{R}}$,
{with $\bm{\mathcal{R}}=(\mathcal{R}_1,\mathcal{R}_2,......,\mathcal{R}_m)$
being the set of parameters (a vector).
Let $|n(\bm{\mathcal{R}})\rangle$ be the $n$-th eigenstate of the
Hamiltonian $H(\bm{\mathcal{R}})$.}   The overlap of two wave
functions infinitesimally separated by $\Delta\bm{\mathcal{R}}$ in the
$\bm{\mathcal{R}}$-space can be evaluated as
\begin{equation}
  \langle n(\bm{\mathcal{R}}) | n(\bm{\mathcal{R}}+\Delta\bm{\mathcal{R}})\rangle=
  1+\Delta\bm{\mathcal{R}}\langle
  n(\bm{\mathcal{R}})|\nabla_{\bm{\mathcal{R}}}|n(\bm{\mathcal{R}})\rangle
  =\rm{exp}[-i\Delta\bm{\mathcal{R}} \cdot {\bf A}_n(\bm{\mathcal{R}})],
\end{equation}
where ${\bf A}_n(\bm{\mathcal{R}})=i \langle
n(\bm{\mathcal{R}})|\nabla_{\bm{\mathcal{R}}}|n(\bm{\mathcal{R}})\rangle$
is called the Berry connection.  {Here} ${\bf A}_n(\bm{\mathcal{R}})$
is an important quantity, because it can be viewed as a vector
potential and its curl
$\bm{\Omega}_n(\bm{\mathcal{R}})=\nabla_{\bm{\mathcal{R}}}\times {\bf
  A}_n(\bm{\mathcal{R}})$, called the Berry curvature, gives an
effective magnetic field in the parameter space $\bm{\mathcal{R}}$ (as
will be addressed below).  The Berry phase $\gamma_n$ can then be
defined as the integral of the Berry connection along a closed loop
$\mathcal{C}$ in the parameter space, or according to Stoke's theorem,
equivalently as the integral of Berry curvature on the surface
$\mathcal{S}$ enclosed by the adiabatic loop
$\mathcal{C}\equiv \partial \mathcal{S}$ (i.e., the surface with the
loop $\mathcal{C}$ as boundary):
\begin{equation}
  \gamma_n=\oint_C {\bf A}_n(\bm{\mathcal{R}}) \cdot d\bm{\mathcal{R}}
  =\int_{\mathcal{S}} {\bm{\Omega}}_n(\bm{\mathcal{R}}) \cdot d{\mathcal{S}},
\label{eq:berry-phase}
\end{equation}
where the second equality suggests that the Berry phase can also be
regarded as the effective magnetic flux passing through a surface
$\mathcal{S}$.

Although the concept of the Berry phase has broad applications in
physics, its relevance to the band structure in solids has been
recognized only in limited situations, such as the quantum Hall effect
under a strong magnetic field~\cite{tknn} and the calculation of
electronic polarization in
ferroelectrics~\cite{King_Smith_PRB_Berry_Polar_1993, Resta_RMP_1994}.
Here we will show that the Berry phase concept is also important for
understanding intrinsic AHE. In this case, we treat the crystal
momentum ($\bf k$) space as the parameter space.  We consider a
crystalline solid with discrete translational symmetry. The eigen equation of
the system is given as $H({\bf r})\psi_{n\bf k}({\bf
  r})=\varepsilon_{n\bf k}\psi_{n\bf k}({\bf r})$, where $H({\bf r})$
is the Hamiltonian, $\varepsilon_{n\bf k}$ and $\psi_{n\bf k}({\bf
  r})$ are the eigen energy and the eigen wave function, and $n$ is
the band index.  Because of the translational symmetry, the eigen wave
function $\psi_{n\bf k}({\bf r})$ of the system is $\bf k$-dependent,
where $\bf k$ is the momentum defined in the first Brillouin zone (BZ)
of momentum space.  According to {Bloch's theorem}, the eigen wave
function should take the form of Bloch state $\psi_{n\bf k}({\bf
  r})=e^{i{\bf k \cdot r}}u_{n\bf k}({\bf r})$, with $u_{n\bf k}({\bf
  r})$ being the periodic part of the wave function. Then the eigen
equation of the system can be recast as $H_{\bf k}({\bf r})u_{n\bf
  k}({\bf r})=\varepsilon_{n\bf k}u_{n\bf k}({\bf r})$, where $H_{\bf
  k}=e^{-i{\bf k \cdot r}} H e^{i{\bf k \cdot r}} $ is the ${\bf
  k}$-dependent Hamiltonian. Following the above discussion, we can
now define the Berry connection and Berry curvature in the parameter
(momentum {\bf k}) space as
\begin{eqnarray}
  {\bf A}_n({\bf k}) &=& i\langle u_{n\bf k}|\nabla_{\bf
    k}|u_{n\bf k}\rangle  \\
\label{eq:berryConnection}
  \bm{\Omega}_n({\bf k}) &=& \nabla_{\bf k}\times {\bf
    A}_n({\bf k})=i\langle \nabla_{\bf k}u_{n\bf k}|\times|\nabla_{\bf
    k}u_{n\bf k}\rangle.
\label{eq:berryC}
\end{eqnarray}

These two quantities are crucial in understanding the Berry phase
mechanism for intrinsic AHE.  Before we can go further, we have to
clarify several important properties of the Berry connection and Berry
curvature.
\begin{itemize}
\item {\it Berry connection is gauge dependent}: As we have learned
  from the textbooks of solid state physics, there exists an arbitrary
  phase factor for the eigen wave function $|u_{n\bf k}\rangle$ that
  is not uniquely determined by the eigen equation of the
  system. Under a U(1) gauge transformation, the eigen wave function
  $|u_{n\bf k}\rangle$ is transformed into $|u^\prime_{n\bf k}\rangle$
  as
\begin{equation}
  |u^\prime_{n\bf  k}\rangle=e^{i\phi_n({\bf k})}|u_{n\bf k}\rangle,
\label{eq:gaugeT}
\end{equation}
where $\phi_n({\bf k})$ is a real and smooth scalar function of $\bf
k$. It is easy to see that $|u^\prime_{n\bf k}\rangle$ is still the
eigen wave function of the system for the same eigen state, i.e.,
$H_{\bf k}|u^\prime_{n\bf k}\rangle=\varepsilon_{n\bf
  k}|u^\prime_{n\bf k}\rangle$ is satisfied. However, the
corresponding Berry connection will be changed by such a gauge
transformation, becoming
\begin{equation} {\bf A}_{n}^\prime({\bf k})= i\langle
  u^\prime_{n\bf k}|\nabla_{\bf k}|u^\prime_{n\bf k}\rangle = {\bf
    A}_{n}({\bf k})-\nabla_{\bf k}\phi_n({\bf k}).
\label{eq:gauge-for-A}
\end{equation}
If we regard the scalar function $\phi_n({\bf k})$ as a kind of scalar
potential in the ${\bf k}$-space, this form of transformation is the
same as that of a vector potential of magnetic field in {real space,}
and therefore the Berry connection ${\bf A}_{n}({\bf k})$ can be
viewed as an effective vector potential in momentum space. The gauge
dependence of the Berry connection suggests that it is not physically
observable. However, it becomes physical after integrating around a
closed path (i.e., the Berry phase $\gamma_n$ defined in
Eq.~\eqref{eq:berry-phase}). This is because the integration of the
second term of Eq.~\eqref{eq:gauge-for-A} around a closed path will
only contribute an integer {multiple} of $2\pi$, and the Berry phase
$\gamma_n$ is therefore invariant modulo $2\pi$.

\medskip

\item {\it Berry curvature is gauge invariant}: This conclusion can be
  drawn directly from the factor that $\nabla_{\bf k}\times\nabla_{\bf
    k}\phi_n({\bf k}) = 0$; therefore $\bm{\Omega}_n({\bf k}) =
  \nabla_{\bf k}\times {\bf A}^\prime_n({\bf k})=\nabla_{\bf k}\times
  {\bf A}_n({\bf k})$ is unchanged under the U(1) gauge transformation
  of Eq. \eqref{eq:gaugeT}. This vector form of Berry curvature
  suggests that it can be viewed as an effective magnetic field in
  momentum space. It is a gauge invariant local manifestation of the
  geometric properties of the wave function in the parameter ($\bf k$)
  space, and has {been} proven to be an important physical ingredient
  for the understanding of a variety of electronic
  properties~\cite{wilczek1989geometric, XiaoDi_RMP_2010,
    King_Smith_PRB_Berry_Polar_1993, Resta_RMP_1994,
    fang_anomalous_2003}.

\medskip

\item {\it Symmetry consideration}: Here we consider two important
  symmetries in solid state physics, namely time reversal symmetry
  (TRS) and inversion symmetry (IS).  Following the above discussion,
  it is easy to learn that the Berry curvature has the following
  symmetry properties:
\begin{eqnarray}
  \bm{\Omega}_n({\bf k}) &=& \bm{\Omega}_n(-{\bf k}) \qquad \; \enskip  {\rm for \
    inversion \ symmetry},        \\
  \bm{\Omega}_n({\bf k}) &=& -\bm{\Omega}_n(-{\bf k}) \qquad  {\rm for \
    time \ reversal \ symmetry},
\label{eq:TRS}
\end{eqnarray}
which suggests that $\bm{\Omega}_n({\bf k})\equiv 0$ for a system with
both TRS and IS. They are strong symmetry {constraints,} which means
that in order to study the possible physical effects related to the
Berry curvature, a system with either broken TRS or broken IS is
generally required. For example, the intrinsic AHE is observed in a
ferromagnetic (FM) system with SOC, where the TRS is broken. For a
system with {TRS} only, in general $\bm{\Omega}_n({\bf k})\ne 0$,
however its integration over the whole Brillouin zone should be zero
as implied by {Eq.~(\ref{eq:TRS}).} In such a case, the Berry
curvature may take effect only through ${\bf k}$-selective or band $n$
sensitive probes.

\medskip

\item {\it The choice of periodic gauge}: For the convenience of
  studies on solid states with discrete translational symmetry, we usually use
  the periodic gauge in practice. That is, we require the Bloch wave
  function to be periodic in momentum space, and use the condition
  $\psi_{\bf k}({\bf r})=\psi_{\bf k+G}({\bf r})$ (where ${\bf G}$ is
  a translational vector for the reciprocal lattice) to partly fix the
  phases of wave functions. This periodic gauge is used in most of the
  existing first-principles electronic structure calculation codes.
  However, we have to note that it is a very weak gauge-fixing
  condition, and abundant {gauge degrees of freedom} still remain.

\medskip

\item {\it Gauge field:} Spreading out each component of the vector
  form discussed above (Eq. \eqref{eq:berryC}), the Berry curvature
  can be written explicitly as an anti-symmetric second-rank tensor
  $\mathcal{F}_{\mu\nu}$,
\begin{eqnarray}
  \epsilon_{\mu\nu\xi}\Omega_{n,\xi}=\mathcal{F}_{n,\mu\nu} & = &
  \frac{\partial}{\partial k_\mu}A_{n,\nu}({\bf k}) -
  \frac{\partial}{\partial  k_\nu}A_{n,\mu}({\bf k})     \nonumber  \\
  & =& i \left[ \langle \frac{\partial u_{n\bf k}}{\partial k_\mu}|
    \frac{\partial u_{n\bf k}}{\partial k_\nu} \rangle -
  \langle \frac{\partial u_{n\bf k}}{\partial k_\nu}|
    \frac{\partial u_{n\bf k}}{\partial k_\mu} \rangle \right],
\end{eqnarray}
where $\epsilon_{\mu\nu\xi}$ is the Levi-Civita antisymmetric tensor,
and $\mu, \nu, \xi$ can be simply treated as the direction indices
(i.e., $x, y, z$) for our purposes. For example, for many of our
following discussions on the transport properties in a two dimensional
{(2D)} system, we will use the equality,
\begin{equation}
  \Omega_{n,z}=\mathcal{F}_{n,xy}.
\end{equation}
Writing it in such a way also helps us understand that Berry curvature
is simply the U(1) gauge field ($\mathcal{F}_{\mu\nu}$) in the
language of gauge theory. In the presence of gauge freedom, in order
to preserve the gauge invariance of Lagrangian under the gauge
transformation, we need to use the gauge covariant derivative instead
of the original derivative. For example, for the position operator
$\hat{x}_{\mu}=i \partial_{k_\mu}$, its gauge covariant form is
written as $\hat{x}_{\mu}=i \partial_{k_\mu} - A_{\mu}(k)$, where
$A_{\mu}(k)$ is the vector potential (i.e., the Berry connection). In
such a case, the gauge covariant position operators $\hat{x}_\mu$ and
$\hat{x}_\nu$ are no longer commutable, and their commutation relation
is given as $[\hat{x}_\mu, \hat{x}_\nu]=-i \mathcal{F}_{\mu\nu}$,
which gives the gauge field
$\mathcal{F}_{\mu\nu}=\partial_{k_\mu}A_\nu-\partial_{k_\nu}A_\mu$.

\end{itemize}

\medskip

Keeping in mind all these properties discussed above for the Berry
connection and Berry curvature, we are now ready to understand the
intrinsic AHE from the viewpoint of gauge field and geometric Berry
phase.  {For simplicity, let us} first consider a {2D} system with
Hamiltonian $H_0(x,y)$ defined in the $xy$-plane. Under an external
electric field $E_y$, the whole Hamiltonian of the system becomes
$H(x,y)=H_0(x,y)+eE_yy$. By evaluating the velocity operator as
$\dot{x}=-\frac{i}{\hbar}[x, H]=-\frac{i}{\hbar}[x,
H_0]-\frac{i}{\hbar}[x, eE_yy]$, we find that the presence of electric
field in the system leads to an additional velocity term,
$v_x=-\frac{i}{\hbar} [x,
eE_yy]=-\frac{e}{\hbar}\mathcal{F}_{xy}E_y=-\frac{e}{\hbar}\Omega_zE_y$,
which is along the $x$ direction (being transverse to the external
electric field $E_y$). This result comes from the non-vanishing
commutation relation between the gauge covariant position
operators. This additional velocity term is nothing but the anomalous
velocity term proposed by Karplus and Luttinger~\cite{kl}, and is the
origin of the intrinsic anomalous Hall effect.

Now let us generalize the above discussion to three dimensional {(3D)}
space and considering both the electric and magnetic external fields,
we find that the equation of motion of electrons, in the presence of
gauge field, should be recast as \cite{XiaoDi_RMP_2010}
\begin{eqnarray}
  \dot{\mathbf{r}} & = &\frac{1}{\hbar}\frac{\partial\epsilon_{n}(\mathbf{k})}{\partial\mathbf{k}}
  -\dot{\mathbf{k}}\times\mathbf{\Omega}_{n}({\bf  k}), \label{eq:eom-for-r} \\
  \dot{\mathbf{k}} & = & -\frac{e}{\hbar}(\mathbf{E}
  +\dot{\mathbf{r}}\times\mathbf{B}),  \label{eq:eom_of_Bloch_ele}
\end{eqnarray}
where ${\bf E}$ and ${\bf B}$ are the external electric and magnetic
fields, respectively. It is apparent that the group velocity
Eq. \eqref{eq:eom-for-r} is different from the textbook one by the
inclusion of an additional contribution (i.e., the second term on the
right-hand side). This term, $\dot{\bf k}\times{\bm\Omega}_n$, is now
known as the anomalous velocity term, and it is proportional to the
${\bf k}$-space Berry curvature ${\bm\Omega}_n({\bf k})$.  This
equation of motion can also be obtained from the semiclassical theory
of Bloch electron dynamics. To do this, we have to consider a narrow
wave-packet made out of the superposition of the Bloch state of a
band. Here we will not discuss the detailed derivation of this
approach; readers please refer to the review article by Xiao et
al. for more details \cite{XiaoDi_RMP_2010}.

For a ferromagnetic system in the presence of SOC, the time reversal
symmetry is broken; therefore the non-vanishing Berry curvature
contributes to the anomalous velocity term. Consider a system under
external electric field $E_y\ne 0$ and without magnetic field (i.e.,
$B=0$). { The current density} $j_x$ can be then obtained from the
equation of motion shown above as
\begin{eqnarray}
  j_{x} = -e v_x & = & -e
  \sum_{n}\int_{BZ}\frac{d^3\mathbf{k}}{(2\pi)^{3}}
  f_{n}(\mathbf{k})(-\dot{\mathbf{k}}\times{\bm\Omega}_{n}(\mathbf{k}))_x\nonumber \\
  & = &
  -\frac{e^{2}}{\hbar}\sum_{n}\int_{BZ}\frac{d^3\mathbf{k}}{(2\pi)^{3}}
  f_{n}(\mathbf{k})(\mathbf{E}\times{\bm\Omega}_{n}(\mathbf{k}))_{x}\nonumber \\
  & = &
  -\frac{e^{2}}{\hbar}\sum_{n}\int_{BZ}\frac{d^3\mathbf{k}}{(2\pi)^{3}}
  f_{n}(\mathbf{k})E_{y}\Omega_{n,z}(\mathbf{k}), \label{eq:tran_current}
\end{eqnarray}
where $f_{n}(k)$ is the Fermi distribution function. The dc Hall
conductivity is then given as
\begin{eqnarray}
  \sigma_{xy}=\frac{j_{x}}{E_{y}} & =&
  -\frac{e^{2}}{\hbar}\sum_{n}\int_{BZ}\frac{d^3\mathbf{k}}{(2\pi)^{3}}
  f_{n}(\mathbf{k})\Omega_{n,z}(\mathbf{k})\label{eq:hall_cond_Berry}  \\
  & = &
  -\int_{-\pi}^{\pi} \left [\frac{e^{2}}{2\pi h}\sum_{n}\int_{-\pi}^{\pi} \int_{-\pi}^{\pi}
    \left (f_{n}(\mathbf{k})\Omega_{n,z}(\mathbf{k}) \right ) dk_xdk_y
  \right ]
  \frac{dk_z}{2\pi},
  \label{eq:hall_cond}
\end{eqnarray}
which is simply the Brillouin zone (BZ) integral of the Berry
curvature weighted by the occupation factor $f_{n}(\mathbf{k})$ of
each state. For the convenience of our later discussions, in the last
part of the equation, we have explicitly written the {3D} integral as
a {2D} integral in the $k_xk_y$-plane, supplemented with a line
integral along $k_z$.

On the other hand, we know from textbooks that the DC Hall
conductivity can be also derived from the Kubo formula as
\begin{equation}
  \sigma_{xy}=\frac{e^{2}}{\hbar}\sum_{m\ne
    n}\int_{BZ}\frac{d^3\mathbf{k}}{(2\pi)^{3}}
  (f_{n}(\mathbf{k})-f_m(\mathbf{k})) \times { Im}
  \frac{\langle \psi_{n\mathbf{k}}|v_x|\psi_{m\mathbf{k}}\rangle
    \langle \psi_{mk}|v_y|\psi_{n\mathbf{k}}\rangle}
  {(\varepsilon_{n\mathbf{k}}-\varepsilon_{m\mathbf{k}})^{2}}
\label{eq:hall_cond_Kubo} {,}
\end{equation}
which is basically equivalent to Eq. \eqref{eq:hall_cond_Berry}, by
expanding each term of the vector form of the Berry curvature
explicitly with some algebra. However, we have to emphasize that the
form of Eq. \eqref{eq:hall_cond_Berry} for Hall conductivity is not
only compact, but also fundamentally important for the underlying
physics. It demonstrates the Berry phase mechanism of the intrinsic
AHE, and it is the key equation for our following discussions on the
geometric meaning and topological nature of {QAHE}.  For example, for
a two dimensional system, the BZ integral of Berry curvature for a
fully occupied band must give rise to an integer multiple of
$2\pi$. Therefore, the Hall conductivity $\sigma_{xy}$ for a 2D
insulator can be finite and be an integer multiple of $e^{2}/h$.  This
is the underlying physics for the quantization of the AHE.

In addition to the understanding of the Berry phase mechanism for the
intrinsic AHE discussed above, another important step forward in this
field is the quantitative and accurate evaluation of the anomalous
Hall conductivity (AHC) for realistic materials, yielding results that
can be compared with experimental observations.  Due to the rapid
increase of computational power and the development of
first-principles calculation methods, such quantitative comparisons
(though still difficult) become possible and contribute greatly to the
development of the field.  In a series of papers reporting
quantitative first-principles calculations for
SrRuO$_{3}$~\cite{fang_anomalous_2003}, Fe~\cite{yao_first_2004} and
CuCr$_2$Se$_{4-x}$Br$_x$~\cite{yao_theoretical_2007}, it was
demonstrated that the calculated AHC can be reasonably compared with
experiments, suggesting the importance of the intrinsic AHE. In more
recent years, accurate calculations of AHC have been achieved by using
the Wannier function interpolation
scheme~\cite{WangXJ_ab_AHE_Wannier_2006,
  Wang_ahe_Fermi_surf_2007}. The presence of SOC and the breaking of
TRS are crucially important for the intrinsic AHE in those systems.

In summary of this part, we have discussed the Berry phase mechanism
for intrinsic AHE. Conceptually, this is an important increment in our
understanding because the intrinsic AHE is now directly linked to the
geometric and topological properties of the Bloch states in momentum
space. This understanding has also led to a great number of research
projects on the exotic properties of electronic systems with SOC, such
as QAHE and the topological insulators.




\section{Quantum Hall Family and Related Topological Electronic
  States}

One of the main subjects of condensed matter physics is the study and
classification of different phases and various phase transitions of
materials that possess rich varieties of physical properties. Landau
developed a general symmetry-breaking theory to understand the phases
and phase transitions of materials. He pointed out that different
phases really correspond to different symmetries of compounds. An
ordered phase, in general, can be described by a local order
parameter, and the symmetry of the system changes as a material
changes from one phase to another.  Landau's symmetry-breaking theory
is very successful in explaining a great many kinds of phases (or
states) in materials, but it does not work for the topological quantum
states. A topological quantum state cannot be described by a local
order parameter, and it can change from one state to another without
any symmetry-breaking. Topology, a word mostly used in mathematics, is
now used to describe and classify the electronic structures of
materials. ``Topological electronic state'' means an electronic state
which carries certain topological properties (in momentum space
usually), such as the states in topological insulators~\cite{
  haldane_model_1988, Kane_Z2_2005, Kane_Mele_Graphene_PRL_2005,
  bernevig_quantum_2006, Bernevig_HgTe_QW_science_2006,
  Molenkamp_HgTe_science_2007, ZhangHJ_Bi2Se3_2009NP,
  TI_exp_Yazdani_2009, TI_exp_XueQK_2009, TI_exp_Hasan_2009,
  chen_experimental_2009,TI_exp_SZX_2010, Hasan_Kane_RMP_2010,
  QiXL_RMP_2011, DuRuiRui_InAs_GaSb_2011, TI_exp_Yazdani_2_2011,
  Ando_TI_materials_2013, schnyder_classification_2008}, topological
semimetals~\cite{ WanXG_WeylTI_2011, XuGang_HgCrSe_2011_PRL,
  Burkov_Weyl_2011, Burkov_Weyl_semimetal_2012PRB, WangZJ_Na3Bi_2012,
  young_dirac_2012, turner_quantized_2012, WangZJ_Cd3As2_2013,
  ChenYL_Na3Bi_2014Science, ChenYL_Cd3As2_2014NatMa,
  ZhouXJ_Cd3As2_2014, neupane_observation_2013,
  yang_classification_2014, jeon_landau_2014,
  xu_observation_2014,Burkov_Topological_nodal_semimetals_2011PRB,lufu,allcarbon_nodeLine2014},
topological superconductors~\cite{ LiangFu_MF_TI_2008PRL,
  Hasan_Kane_RMP_2010, QiXL_RMP_2011, QiXL_TRInvariant_2008,
  QiXL_MF_from_QHE_2010PRB, FuLiang_Odd_Parity_TSC_2010,
  Hor_CuBiSe_TSC_2010, zhang_pressure-induced_2011, NaCoO2_Weng}, etc.  One of the
most important characteristics of topology is the robustness against
local deformations, or in the language of physics, the insensitivity
to environmental perturbations, which makes topological electronic
states promising for future applications. To characterize the order of
topological states, new parameters, namely the topological invariants,
are needed. In this section, we will review some of the topological
electronic states and their quantum physics, including the integer
quantum Hall (IQH) state, the quantum anomalous Hall (QAH) state, the
quantum spin Hall (QSH) state, and the topological semimetal (TSM)
state. We will use the concept of Berry phase in momentum space to
discuss the topological nature of topological states and the related
topological invariants.

~\\~
\hspace*{15pt}{\bf II.A. Chern Number, Chern Insulator, and Quantum AHE}
\\

In 1980, Klaus von Klitzing discovered~\cite{qhe} that the Hall
conductance, viewed as a function of strength of the magnetic field
applied normal to the two-dimensional electron gas plane, at very low
temperature, was quantized and exhibited a staircase sequence of wide
plateaus. The values of Hall conductance were integer multiples of a
fundamental constant of nature: $e^{2}/h=1/(25812.807572\;\Omega)$,
with totally unanticipated precision, and independent of the geometry
and microscopic details of the experiment. This is an important and
fundamental effect in condensed matter physics, and it is called the
integer quantum Hall effect (IQHE) --- the quantum version of the Hall
effect.

The IQHE is now well understood in terms of single particle orbitals
of an electron in a magnetic filed (i.e., the Landau levels), and the
phenomenon of ``exact quantization'' has been shown to be a
manifestation of gauge invariance.  The robustness and the remarkable
precision of Hall quantization can also be understood from the
topological nature of the electronic states of 2D electron gas under
magnetic field. The integer number, originally known as the TKNN
number~\cite{tknn} in the Hall conductance derived from the Kubo
formula, is now characterized as a topological invariant called
``Chern number''.  This topological understanding of the IQHE is a
remarkable leap of progress, opening up the field of topological
electronic states in condensed matter physics. IQHE is therefore
regarded as the first example of topologically non-trivial electronic
states to be identified and understood.

This conceptual breakthrough with regard to the topological nature of
IQHE, though important, was not widely generalized, and the
relationship of IQHE to the rich variety of condensed materials was
not revealed for a long time. This is because the IQHE is observed
only in a very particular system, i.e. 2D electron gas under strong
external magnetic field, and the formation of Landau levels (usually
at very low temperature) is required. Under such extreme conditions,
the material's details and its electronic band structure become
irrelevant to the physics. In this sense, the lattice model proposed
by Haldane in 1988~\cite{haldane_model_1988} is very stimulating for
the study of topological electronic states. He proposed that a
spinless fermion model on a periodic 2D honeycomb lattice without net
magnetic flux can in principle support a similar IQHE. Although
Haldane's model is very abstract and unrealistic (at least at that
time), his result suggested that a kind of quantized Hall effect can
exist even in the absence of magnetic field and the corresponding
Landau levels. In other words, certain materials, other than the 2D
electron gas under magnetic field, can have topologically non-trivial
electronic band structures of their own, which can be characterized by
a non-zero Chern number. Such materials were called Chern
insulators later. In this way, the concept of topological electronic
state was generalized and was connected to the electronic band
structure of materials.

The progress in the study of IQHE was very inspiring in the 1980s; on
the other hand, not much of it was related to the field of AHE. Around
that time, the development of the AHE field was rather independent,
with almost no intersection with the study of IQHE. But the
understanding of Berry phase mechanism for intrinsic
AHE~\cite{jungwirth, onoda-2, fang_anomalous_2003, yao_first_2004,
  sinova_magneto_transport_2004}, reformulated around the early years
of the 21st century, changed the situation significantly. This
understanding established the connection between the AHE and the
topological electronic states. By this connection, it is now
understood~\cite{onoda-2,fang_anomalous_2003,onoda_quantized_2003}
that intrinsic AHE can have a quantum version --- the quantum
anomalous Hall effect (QAHE), which is in principle similar to IQHE
but without external magnetic field and the corresponding Landau
levels. It turns out that this is exactly the same effect discussed by
Haldane for a Chern insulator with a non-zero Chern number. Currently,
the fields of IQHE and AHE are closely combined, and the realization
of a Chern insulator becomes possible by finding the QAHE in suitable
magnetic compounds with strong SOC.

The IQHE and the QAHE have their own characteristics; however, their
underlying physics, in terms of the topological properties of their
electronic structures, are basically the same. They are all related to
the Berry connection, Berry curvature and Berry phase in momentum
space, as discussed in the previous section. In this part, we will not
discuss the particular details associated with each effect, but rather
we will concentrate on their common features, namely the topological
property and topological invariant (Chern number). To begin with, we
will discuss several important concepts:

\begin{figure}
\begin{centering}
\includegraphics[width=0.9\textwidth]{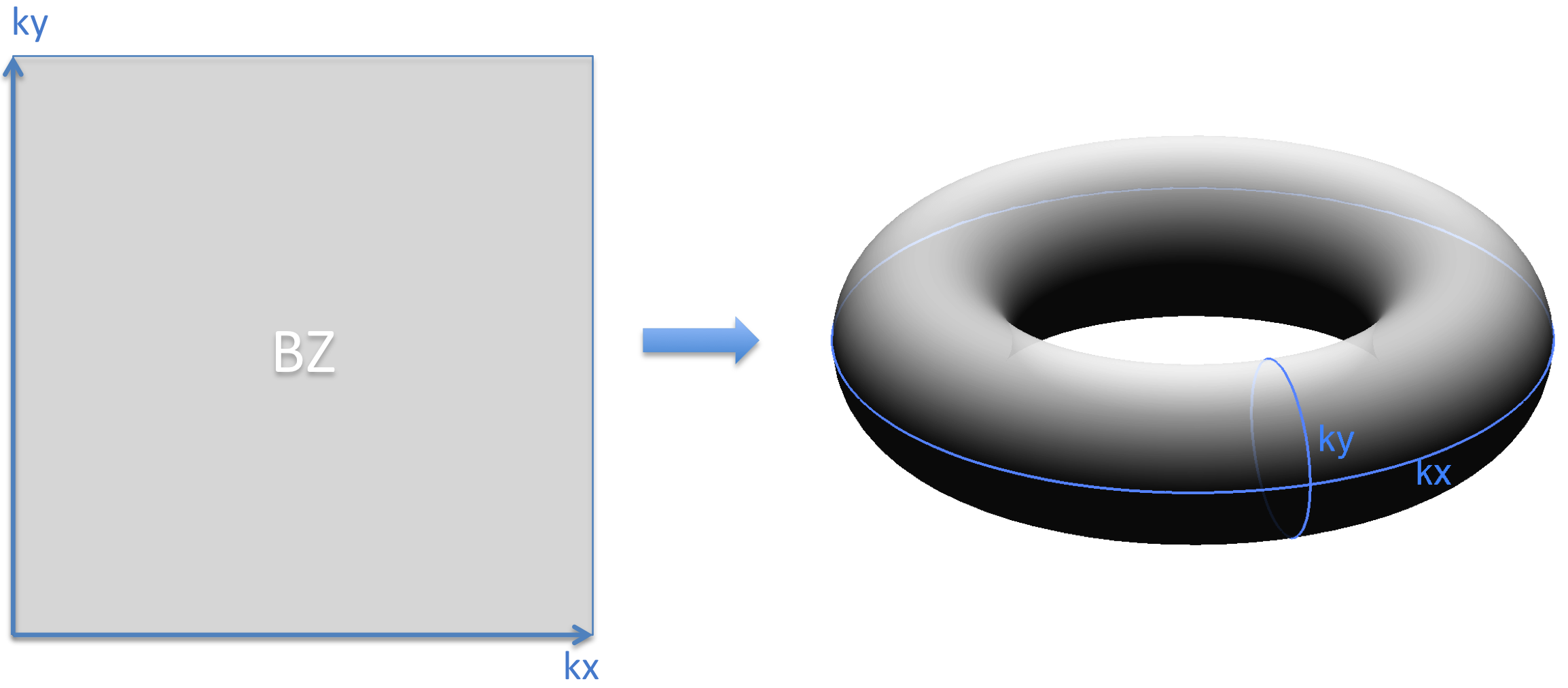}
\par\end{centering}
\caption{Under the periodic boundary condition for the Bloch states,
  the 2D Brillouin zone is equivalent to a torus.  }
\label{fig:Brillouin_Torus}
\end{figure}

\begin{itemize}
\item{\it Hall Conductivity as Integral of Berry Curvature:}
  Considering a 2D insulating system with broken TRS, the Hall
  conductivity of the system at low enough temperature can be written
  as
\begin{eqnarray}
  \sigma_{xy} &=& -\frac{e^2}{\hbar}\sum_n \int_{BZ} \frac{d^2{\bf
      k}}{(2\pi)^2}f_n({\bf k})\Omega_{n,z}({\bf k})  \nonumber  \\
  &=& -\frac{e^2}{2\pi h} \int_{BZ} d^2{\bf k} \sum_{n (occ)}
  \Omega_{n,z}({\bf k}) \nonumber \\
  &=& -\frac{e^2}{2\pi h} \int_{BZ} d^2{\bf k}
  \Omega_{z}({\bf k}),
\label{eq:sigxy}
\end{eqnarray}
where the second equality follows because of the existence of an
energy gap and $f_n({\bf k})=1$ (=0) for the occupied (unoccupied)
state {(see Eqs.~(\ref{eq:tran_current})-(\ref{eq:hall_cond}) for a 3D
  version)}.  This form of Hall conductivity can be derived either
from the Berry phase formalism or from the Kubo formula, as discussed
in the previous section. It is also important to note that this
formula can be equally applied to the anomalous Hall conductivity of a
2D FM insulator without external field and to the Hall conductivity of
a 2D electron gas under magnetic field.  For the former it is
straightforward; for the latter case, however, special considerations
have to be taken. First, we assume that the external magnetic field is
strong enough for the system to form the Landau levels. Second, the
system must be insulating with the chemical potential located within
the gap (i.e., between two neighboring Landau levels). Third, since
the presence of external magnetic field will break the translational
symmetry, we have to use the concept of magnetic translational
symmetry to recover the periodicity of the system. As a result,
magnetic Bloch states must be used for the evaluation of Berry
curvature, and the magnetic Brillouin zone is used for the
integration, correspondingly.

\medskip

\item{\it 2D Brillouin Zone as a Torus:} By adopting the periodic
  gauge, the wave function $\psi_{n{\bf k}}({\bf r})$ is periodic in
  momentum space. Under such a periodic boundary condition, the 2D BZ
  becomes a torus as schematically shown in
  Fig. \ref{fig:Brillouin_Torus}.  Therefore, the integral of the
  Berry curvature $\Omega_z$ over the 2D BZ can be recast as a vector
  integral over the surface of the torus as
\begin{equation}
  \int_{BZ}d^2{\bf k}\Omega_{z}({\bf
    k})=\int_{\mathcal{S}}\bm\Omega({\bf
    k})\cdot d\mathcal{S}.
\label{eq:p-2-t}
\end{equation}
Here we have used an effective Berry curvature ${\bm\Omega}({\bf k})$
defined over the torus surface, which {satisfies} the condition
$\bm\Omega({\bf k})\cdot {\bf n}=\Omega_{z}({\bf k})$, where ${\bf n}$
is the unit vector along the normal of the surface $\mathcal{S}$.

\medskip

\item{\it Quantization and Chern Number:} The torus is a closed
  manifold without boundary, so according to Chern's
  theorem~\cite{chern_curvatura_1945} in differential geometry, the
  vector integral of Berry curvature over a torus surface must be an
  integer multiple of $2\pi$,
\begin{equation}
  \int_{\mathcal{S}}\bm\Omega({\bf
    k})\cdot d\mathcal{S}  = 2\pi Z,
\label{eq:Chern}
\end{equation}
where $Z$ is an integer number called Chern number. Having this
property, it is now easy to see that the Hall conductivity for a 2D
insulator must be quantized as
\begin{equation}
  \sigma_{xy}=-\frac{e^2}{2\pi h}\times 2\pi Z = -\frac{e^2}{h} Z.
\end{equation}

\medskip

\item{\it Absence of Smooth Gauge:} Based on Stokes\textquoteright{}
  theorem, the surface integral of Berry curvature
  (Eq. \eqref{eq:Chern}) can also be evaluated as an loop integral of
  the Berry connection ${\bf A}({\bf k})$ along the boundary of the
  BZ, i.e., $\int_{\mathcal{S}}{\bm\Omega}({\bf k})\cdot
  d\mathcal{S}=\oint_{\mathcal{C}}{\bf A}({\bf k})\cdot d{\bf k}$.
  Since the BZ is a torus, which has no boundary, the integral must be
  vanishing if ${\bf A}({\bf k})$ is smoothly defined in the whole
  BZ.   Therefore, a non-zero Chern number indicates that we cannot
  choose a smooth gauge transformation such that ${\bf A}({\bf k})$ is
  continuous and single valued over the whole BZ.
  Thus, non-zero Chern number can be viewed as obstruction to continously 
  define the phase of the occupied wave functions on a 2D BZ, which is a torus.~\cite{thouless_wannier_1984,
    thonhauser_insulator/chern-insulator_2006,brouder_exponential_2007}.

\end{itemize}

For a system with multiple bands, the Berry curvature should be
understood as the summation of contributions coming from all occupied
bands. Having the {properties} discussed above, we can now define a
Chern insulator as a 2D insulator whose electronic structure gives a
non-zero Chern number. The above discussions about the Chern number
and the topology of electronic states in momentum space may be still
too abstract, in the following we will give a more explicit
explanation, in which the none zero Chern number manifests itself as the winding number of the Wannier center evolution for the effective 1D systems with constant $k_y$.
Here we consider the simplest case: a 2D
lattice with only one single occupied band (the band index $n$ can
therefore be neglected). {Although it is impossible to choose a single
  smooth gauge over the whole BZ for a Chern insulator according to
  our above discussions,} it is possible to choose a special gauge
that is smooth and periodic along one direction (say $k_x$), but not
necessarily along the other (say $k_y$)~\cite{thouless_wannier_1984,
  thonhauser_insulator/chern-insulator_2006, brouder_exponential_2007,
  Soluyanov_wannier_Z2_2011}. Thus, we can do the integration in
Eq. \eqref{eq:p-2-t} explicitly for the $k_x$ direction with fixed
$k_y$ and obtain (we set the lattice parameter $a$=1)
\begin{eqnarray}
  2\pi Z &=& -\int_{-\pi}^{\pi} \int_{-\pi}^{\pi} dk_x dk_y (\partial_{k_x} A_y
  - \partial_{k_y} A_x)   \nonumber \\
  &=& \int_{-\pi}^{\pi} dk_y \partial_{k_y} \left(\int_{-\pi}^{\pi} dk_x A_x(k_x, k_y )
  \right) \nonumber \\
  &=& \int_{-\pi}^{\pi} d \theta(k_y).
\end{eqnarray}

\begin{figure}[th]
\begin{centering}
\includegraphics[angle=-0,width=0.59\textwidth]{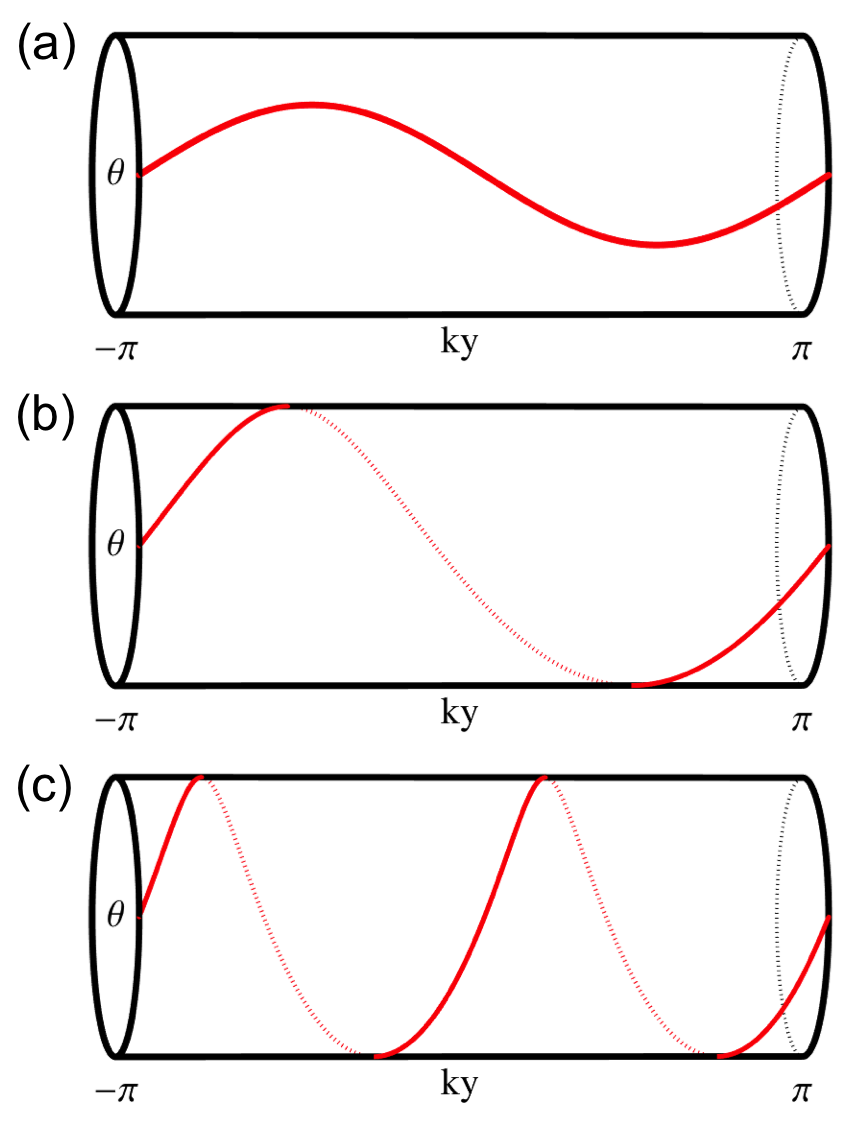}
\par\end{centering}

\caption{The evolution of $\theta(k_y)$ (or 1D Wannier center) on
  cylinder surface for system with (a) Z=0, (b) Z=1 and (c) Z=2.}
\label{fig:IQHE_Wannier_center_winding}
\end{figure}

Here, $\theta(k_y)=\int_{-\pi}^{\pi}dk_x A_x$ is an angle (i.e.,
Berry phase) calculated from the 1D integration of $A_x(k_x, k_y)$
along the $k_x$ axis (a closed loop) for each fixed $k_y$.  We can
then plot $\theta(k_y)$ as a function of $k_y$. As shown in
Fig. \ref{fig:IQHE_Wannier_center_winding}, over a cylinder surface
(in the cylinder coordinates), $k_y$ is along the longitudinal
direction, and the azimuth is the angle $\theta (k_y)$ for each fixed
$k_y$. Moving from $k_y$=$-\pi$ to $k_y$=$\pi$, we can see the
difference between the trivial insulator ($Z$=0) and the Chern
insulator ($Z\ne 0$). The winding number of $\theta(k_y)$ over the
cylinder surface is zero for the former
(Fig. \ref{fig:IQHE_Wannier_center_winding}(a)), and non-zero for the
{latter} (Fig. \ref{fig:IQHE_Wannier_center_winding}(b), (c)). In this way, we
relate the Chern number
to some kind of winding number defined by the eigen functions,
and see that the Chern insulator has a ``twisted'' energy band.

It is also interesting to note that the $k_y$-dependent Berry phase
calculated as
\begin{equation}
  P_n(k_{y})=\frac{\theta_n(k_y)}{2\pi}=
  \frac{1}{2\pi}\int_{-\pi}^{\pi}dk_{x}A_{n,x}(k_{x},k_{y})
\label{eq:1D_wannier_center}
\end{equation}
can be related to the Wannier center of a 1D system (where we have
recovered the band index $n$ dependence). To see this, we consider a
generic 2D system and denote the Wannier function in cell ${\bf R}$
associated with band $n$, given in terms of the Bloch states, as
\begin{equation}
  |n{\bf R}\rangle = \frac{1}{(2\pi)^2}\int_{\rm BZ} d^2{\bf k} e^{-i{\bf k}\cdot {\bf
      R}}|\psi_{n\bf k}\rangle,
\end{equation}
with
\begin{equation}
  |\psi_{n\bf k}\rangle = \sum_{\bf R} e^{i{\bf k}\cdot {\bf R}} |n{\bf R}\rangle,
\end{equation}
where ${\bf k}=(k_x,k_y)$ is considered, and the periodic part of the
Bloch function is defined as
\begin{equation}
u_{n\bf k}({\bf r})= e^{-i{\bf k}\cdot {\bf r}}\psi_{n\bf k}({\bf r}).
\end{equation}
Then the matrix elements of the position operator ${\bf r}$ between
Wannier functions take the form
\begin{equation}
  \langle n{\bf R} | {\bf \hat r} | m{\bf 0} \rangle = i \frac{1}{(2\pi)^2}\int_{\rm BZ} d^2{\bf
    k} e^{i{\bf k}\cdot {\bf R}}\langle u_{n\bf k} | {\bm\nabla}_{\bf k}
  | u_{m\bf k}\rangle,
\end{equation}
from which we get the center of the Wannier function in the {\bf 0}-th
cell as
\begin{equation}
  \overline{\bf r}=\langle n{\bf 0}|{\bf \hat r}|n{\bf 0}\rangle = i
  \frac{1}{(2\pi)^2}\int_{\rm BZ} d^2{\bf k}\langle u_{n\bf k} | {\bm\nabla}_{\bf
    k}| u_{n\bf k}\rangle.
\end{equation}

Now {let us} consider the $k_x$ and $k_y$ directions separately, and
define the 1D Wannier function for each fixed-$k_y$ as
\begin{equation}
  |n,R_x, k_y\rangle = \frac{1}{(2\pi)}\int_{-\pi}^{\pi} dk_x e^{-ik_xR_x}|\psi_{n\bf k}\rangle.
\end{equation}
Then the 1D Wannier center (along $x$ direction) as function of $k_y$
can be defined as the average value of position operator $\hat{x}$ as
\begin{equation}
  \overline{x}(k_y)=\langle n, R_{x}=0,k_y|\hat{x}|n,R_{x}=0,k_y\rangle =
  \frac{i}{(2\pi)}\int_{-\pi}^{\pi} dk_x\langle u_{n\bf k} | {\nabla}_{k_x}|
  u_{n\bf k}\rangle
  =P_n(k_y),
\end{equation}
from which we see that $P_n(k_y)$ is nothing but the 1D Wannier center
of band $n$. The Chern number can therefore be understood as the
winding number of 1D Wannier center when {it evolves} as a function of
$k_y$ (see Fig. \ref{fig:IQHE_Wannier_center_winding}). We can also
see that when the Chern number is not zero, we cannot choose a smooth
gauge transformation such that ${\bf A}({\bf k})$ is continuous and
single valued over the whole BZ.
Although the above discussion is only for the simplest systems with single occupied band, generalization to the multi-band situation is quite straightforward.
For a general band insulator with multiple occupied bands, the Berry connection
${\bf A}_{nm}({\bf k})=i\langle u_{n\bf k}|\nabla_{\bf k}|u_{m\bf k}\rangle$ contains band index $n$, $m$ and becomes nonabelian
Then  the Berry curvature that defines the Chern number is obtained from the
trace of the Berry connection ${\bf A}({\bf k})$.

The Hall conductivity of a Chern insulator with a non-zero Chern
number must be quantized as an integer multiple of
$\frac{e^2}{h}$. Different Chern numbers give different states whose
Hall conductivities are different, but their local symmetry can be the
same. Therefore, to distinguish the states, we cannot use local order
parameters (in the language of Landau's symmetry breaking theory);
instead we need a topological invariant, the Chern number, as a global
order parameter of the system. By rewriting Hall conductivity in terms
of the Berry curvature and Berry phase, we can now unify QHE and
QAHE. Readers will recall that the QAHE is nothing but the quantum
version of the AHE realized in a Chern insulator without the presence
of external magnetic field.

\begin{figure}
  \begin{centering}
\includegraphics[width=0.99\textwidth]{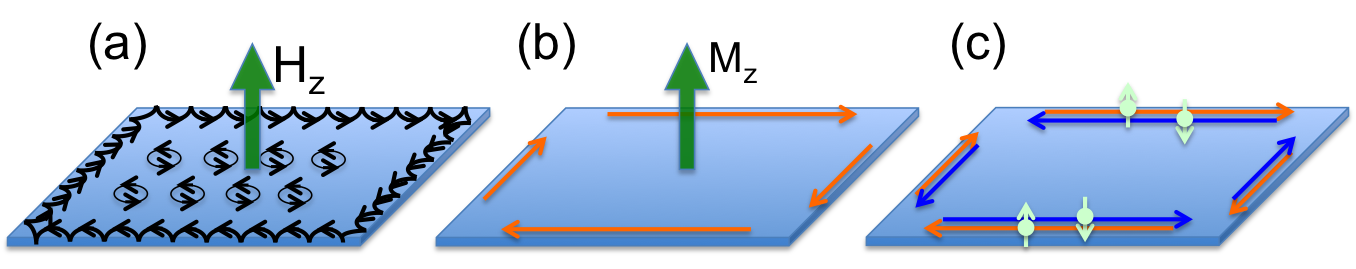}
\par\end{centering}
\caption{\label{fig:IQHE_edge_state} Schematic illustrations of the
  edge states in (a) QHE; (b) QAHE, and (c) QSHE. }
\label{fig:IQHE_edgestates}
\end{figure}

Similar to QHE, where the TKNN number is related to the number of edge
states in a real 2D sample with boundary~\cite{tknn}, the Chern number
can also be physically related to the number of edge states for a 2D
Chern insulator~\cite{haldane_model_1988}. The existence of edge
states is a direct result of the topological property of the bulk
electronic structure, and is due to the phenomenon discussed in the
literature as the bulk-boundary-correspondence
\cite{Hatsugai_1993,Bulk-boundary-correspondence_PRB_2011}.  In this
case, due to the broken TRS, the edge state must be chiral (i.e, the
electrons of the edge state can move only in one direction surrounding
the sample boundary, either left- or right-handed, as shown in
Fig.~\ref{fig:IQHE_edge_state}(b)). As discussed with regard to IQHE,
the charge transport of the edge state is in principle
dissipationless, and back-scattering is absent due to the lack of an
edge state with opposite
velocity~\cite{MacDonald_1984PhRvB,Halperin_1982PhRvB}.

Although the topological properties of Chern insulators and the
related QAHE are fundamentally the same as that discussed with regard
to IQHE, they are conceptually broadly generalized to a wide field and
to a rich variety of materials. This generalization is an important
step forward, providing the building blocks for subsequent discussions
of many possible topological electronic states. From the {application}
point of view, the Chern insulator {(or the related QAHE)} is also
important because the quantized Hall conductivity can be realized in
the absence of magnetic field, greatly simplifying measurement
conditions.

~\\~
\hspace*{15pt}{\bf II.B. $Z_2$ Invariant, Topological Insulator, and Quantum Spin Hall Effect (QSHE)}
\\

Although the Chern insulator {(or the related QAHE)} is the simplest
topological electronic state, its realization occurred rather later
and was much stimulated by the rapid development in the field of
topological insulators---another interesting topological electronic
state protected by TRS.  Considering symmetry, it is easy to prove
that the Chern number in Eq. (\ref{eq:Chern}) must be vanishing for an
insulator with TRS. However, this does not mean that the electronic
state carries no topological property in this case. Kane and
Mele~\cite{Kane_Z2_2005} introduced a new topological invariant, the
$Z_{2}$ number, to classify an insulating system with TRS. They
proposed that a time reversal invariant insulator can be further
classified as a trivial insulator with $Z_{2}=0$ or a non-trivial
topological insulator with $Z_2=1$, which is a typical example of a
symmetry protected topological
state~\cite{gu_tensor-entanglement-filtering_2009}. A $Z_2$ topological
insulator in 2D is also called a quantum spin Hall insulator (QSHI)
because it can support the quantum spin Hall effect
(QSHE)~\cite{Kane_Mele_Graphene_PRL_2005,Bernevig_HgTe_QW_science_2006,bernevig_quantum_2006,Molenkamp_HgTe_science_2007},
which shares certain features with the IQHE and QAHE and can be
understood from the viewpoint of ``band-twisting'' or winding number
of Berry phase in momentum space. A QSHI is different from a trivial
insulator in the sense that it has gapped insulating states in the
bulk but gapless states on the edge due to its topological
property. It is also distinguished from the Chern insulator in the
sense that QSHI has even number of edge states, composed of pairs
of counter-propagating chiral edge states, as shown in
Fig.~\ref{fig:IQHE_edge_state}(c)). The QSHE has been explicitly
discussed for graphene lattice~\cite{Kane_Mele_Graphene_PRL_2005} and
HgTe quantum well
structures~\cite{Bernevig_HgTe_QW_science_2006,bernevig_quantum_2006,Molenkamp_HgTe_science_2007}. In
this part, however, we will take a simple task and discuss the $Z_2$
invariant and the topological state with TRS from the viewpoint used
above for the Chern insulator.

For a 2D insulator with TRS, the total number of occupied electronic states
must be even. Suppose we can divide them into two subspaces, I and II , which are
related by TRS and smoothly defined on the whole BZ. Evaluating the Chern number 
for each subspace independently, if one has Chern number Z, the other one must have Chern
number -Z (due to the TRS) and the total Chern number of the whole system is always zero.
It seems that we can use Z as the topological index to classify the band insulators
with TRS. While unfortunately, the smooth partition of the occupied states into
two subspaces with one being the time reversal of another is only possible
with extra good quantum numbers, such
as $S_z$, where the Chern number obtained within the spin up subspace can be used to
describe the topology of the system and is called ``spin Chern number". But generically 
such a smooth partition of the occupied states can only be made for half of the BZ (not
the whole BZ) and we need at least two patches (A and B) to fully cover the whole BZ. 
The winding number of the U(2N) gauge transformation matrix $\hat t_{AB}(k)$ at half of the 
boundary between two patches can be used to define a new classification of the band insulators with TRS. 
As further proved by Fu and
Kane, only the even and odd feature of the above winding number is unchanged under the U(2N) gauge transformation and all the band 
insulators with TRS can be classified into topological trivial and non-trivial classes, which is called $Z_2$ invariance accordingly. Fu and Kane further
derived the expression for $Z_2$ invariance in terms of both Berry connection and Berry curvature as~\cite{FuLiang_Z2Pump_2006}
\begin{equation}
  Z_{2}=\frac{1}{2\pi}\bigg(\oint_{\partial(BZ/2)}{\bf A}({\bf
    k})\cdot d{\bf k}
  -\int_{BZ/2}d^{2}{\bf k}\Omega_{z}({\bf k})\bigg)\; mod\;2,
\end{equation}
where integral of Berry curvature is performed over the half BZ
(i.e., $BZ$/2), and the integral of Berry connection is performed
along the boundary of the half BZ (i.e., $\partial(BZ/2)$).
To evaluate $Z_2$ invariance using the above equation, one needs to find a smooth gauge for the wave functions on half of the BZ, which is very
difficult for the band structure calculations for realistic materials, and as a result, this formula is rarely used in practice. 
On the other hand, the $Z_2$ invariance is a gauge invariant quantity, one should 
be able to compute it without any gauge fixing condition. For that purpose, some of the authors of the present paper developed an alternative
expression for the $Z_2$ invariance called Wilson loop method, which will be introduced in detail in section 3.2. Here we just sketch its main idea 
briefly. Similar with the Wilson loop method for Chern number calculation, the purpose of the Wilson loop method is to calculate the
``Wannier center" of each band for the effective 1D insulators with fixed $k_y$ and determine the $Z_2$ invariance by looking at their evolution
 with $k_y$. With the presence of TRS, a generic system contains 2N occupied bands. As mentioned previously, now the Berry connection
 becomes 2N*2N matrix. The loop integral of such U(2N) Berry connection gives a 2N*2N unitary matrix $\hat D(k_y)$. The U(1) part of
 this matrix contributes to the Chern number as discussed before and the $Z_2$ invariance can be obtained from the remaining SU(2) part by
 taking the phases of its eigenvalues $\theta_n(k_y)$, where $n$ denotes the band index. As proved in Ref.~\onlinecite{YuRui_Z2_2011PRB}, the 
 TRS only guarantees the double degeneracy of $\theta_n(k_y)$ at two time reversal invariant loops $k_y=0$ and $k_y=\pi$. The $Z_2$ invariance of 
 the system is determined by whether the $\theta_n(k_y)$s switch parters or not as they evolves from $k_y=0$ to $\pi$. 

\begin{figure}
\begin{centering}
\includegraphics[angle=-0,width=0.59\textwidth]{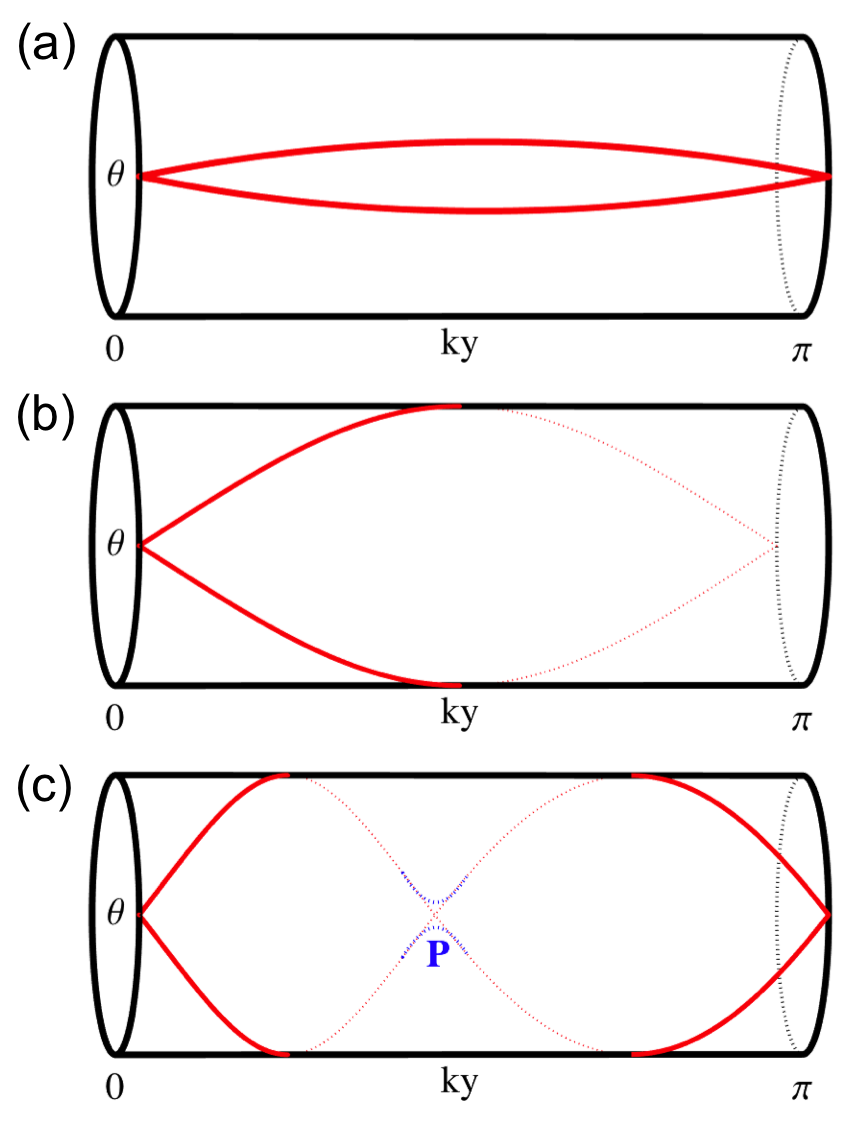}
\par\end{centering}
\caption{\label{fig:winding_number_Z2} Evolution of Wannier centers
  ($P_n(k_y)$) for systems with (a) $Z_2$=0, (b) $Z_2$=1, (c) $|Z|$=2
  but $Z_2$=0.  The Wannier centers wind the cylinder surface even (a,
  c) and odd (b) times respectively. Figure from {Ref.}~\onlinecite{YuRui_Z2_2011PRB}}
\end{figure}

To be explicit, let us again consider the simplest example of 2D
insulator, with only two occupied bands, $\psi_{I,\bf k}$ and
$\psi_{II,\bf k}$, which are related by the TRS (say, for example,
$\psi_{I,-\bf k}$=$\hat{T} \psi_{II,\bf k}$, where $\hat{T}$ is the
time reversal operator). The two states form a Kramers pair and must
be degenerate at the time-reversal-invariant momentum (TRIM) of the BZ
(i.e, with $k_{x}$ or $k_{y}=0, \pi$). We can evaluate $\theta(k_y)$ for
all occupied bands (as we have done above for the Chern insulator). As mentioned previously, the two Wannier centers must be degenerate at $k_y=0$ and $\pi$, which leads to  three situations in general (as shown in Fig. \ref{fig:winding_number_Z2}(a)-(c)).
First, if the evolution path of the $\theta(k_y)$ doesn't enclose the half BZ,
this is the trivial situation with $Z_2$ invariance $\nu$=0
(Fig. \ref{fig:winding_number_Z2}(a)).  Second, if the evolution path of the $\theta(k_y)$ enclose the half BZ once, 
this is the non-trivial topological 
insulator with $Z_2$ index $\nu$=1, where the
crossing of two $\theta(k_y)$ curves at $k_y=0$ and $\pi$ is protected
by the TRS (Fig. \ref{fig:winding_number_Z2}(b)).  Third, if the evolution path of the $\theta(k_y)$ enclose the half BZ twice,
the two $\theta(k_y)$ curves must cross at some $k_y$ other than the
TRIM. Such crossings are not protected by the TRS and are removed by
small perturbations, which drive the crossings into anti-crossings. As
the result, the system becomes equivalent to the trivial case with $Z_2$ index
$\nu$ =0 (Fig. \ref{fig:winding_number_Z2}(c)).

The edge states of 2D $Z_2$ topological insulators must appear in
pairs due to the existence of TRS, and each pair of edge states is
composed of two counter-propagating chiral edge states, which are
related to each other by the TRS. In this case, if one state at the
edge has velocity $v$, we can always find another state with opposite
velocity $-v$. However, back-scattering between the two states is
again forbidden because the two states must have opposite spins. 
Existence of such pair of edge states leads to the observable QSHE~\cite{Kane_Mele_Graphene_PRL_2005, Kane_Z2_2005,
  Bernevig_HgTe_QW_science_2006}, where the spin Hall conductivity is
quantized in units of $\frac{e}{2\pi}$ (if $S_z$ is conserved). According to our above
discussions of the $Z_2$ topology, if there exist multiple pairs of
edge states, two pairs can in principle couple together and open up a
gap (because scattering between the pairs is not forbidden). As a
result, the $Z_2$ number is physically related to the number of pairs
of edge states mod 2, in contrast to the case of Chern insulators.

Using the 2D topological invariants (such as $Z$ or $Z_2$) as building
blocks, a 3D crystal can in principle be characterized by a triplet of
2D topological numbers for the three crystal orientations,
respectively. In such a way, we can extend the topological classes
from 2D to 3D. There are in general two situations by doing this: (1)
one is the trivial extension leading the ``weak'' 3D topological
class, which can be viewed as a simple stacking of 2D topological
insulating layers along a certain crystal orientation; (2) another is
a non-trivial extension leading to a ``strong'' topological class,
which is a new and stable topological state in 3D. For example, by a
simple stacking of 2D Chern insulating layers along the $z$ direction,
if the band dispersion along $k_z$ is weak enough, each 2D layer with
fixed $k_z$ will keep its original topological property (characterized
by a non-zero Chern number). As a result, a 3D ``weak'' Chern
insulator can be obtained. In a similar way, a 3D ``weak'' topological
insulator can be obtained too. However, it is important to note that,
if we concentrate on insulators only, there will be no ``strong''
Chern insulator in 3D; on the other hand, a new 3D state with TRS,
called ``strong'' topological insulator, can be
obtained~\cite{FuLiang_3dTI_2007PRL, Moore_TI_TRS_2007,
  roy_topological_2009} by extending the $Z_2$ number from 2D to
3D. (As will be discussed in the next section, the missing ``strong''
Chern class in 3D has to be a metal---the topological semimetal).

For a 3D insulator with TRS, there will be 8 TRIM points in the BZ
(different from 2D BZ which has only 4 TRIM points). For any 4 TRIM
points that lie in the same plane, we can define a 2D $Z_2$ number as
discussed above. In total, there will be 16  invariant configurations and can be
distinguished by four independent $Z_2$ indices, $\nu_0$($\nu_x\nu_y\nu_z$), as discussed in
Ref.~\onlinecite{FuLiang_3dTI_2007PRL}. Here $\nu_0$ is the total $Z_2$
number, and the $\nu_{x,y,z}$ are the 2D $Z_2$ numbers along the $x,
y, z$ directions, respectively. The three weak $Z_2$ indices
$\nu_{x,y,z}$ can be evaluated for the $k_x=\pi$, $k_y=\pi$, and $k_z=\pi$ planes, as a convention, respectively. The total
$Z_2$ index $\nu_0$, however, is a strong index, and it has to be
evaluated globally by considering the 3D structure. That is, we have
to consider the change of 2D $Z_2$ number between two parallel planes
(perpendicular to certain reciprocal lattice vector). For example, we
can choose the $k_x$=0 and $\pi$ planes, or $k_y$=0 and $\pi$ planes,
or $k_z$=0 and $\pi$ planes. If the $Z_2$ numbers of two planes are
different, we have $\nu_0=1$, otherwise, if they are the same, we have
$\nu_0=0$. A realistic example will be given in Section 3. For a 3D
strong topological insulator with strong topological index $\nu_0\ne
0$, on its boundary (i.e, surface), we will expect the Dirac cone type
surface states, which are protected by the
TRS~\cite{Hasan_Kane_RMP_2010, QiXL_RMP_2011}.


~\\~
\hspace*{15pt}{\bf II.C. Magnetic Monopole, Weyl node and Topological Semimetal}
\\

As mentioned above, if we attempt to consider the possible ``strong''
topological state characterized by Chern numbers in 3D, we will find
that it has to be a metal, and this leads to a very interesting new
topological state of quantum matters --- topological metals or
semimetals. We will see that this is a much more general state, and
the 2D Chern insulating state can be regarded as a special cut of the
3D topological semimetal state along a certain plane in momentum
space. From another point of view, we can also raise a question like
the following. Since we know from the above discussions that
insulators can be further classified as topologically trivial and
non-trivial insulators, can we do the same classification for metals?
If possible, what will be the topological invariant for the proper
description of the topological metallic state? This is a very
interesting question, and we will show in this part that it is related
to the magnetic monopoles~\cite{fang_anomalous_2003,
  volovik_universe_2009} and Weyl nodes in momentum space.

Berry phase is called quantum geometric phase since it has a very
intuitive geometric picture. It is proportional to the solid angle
subtended at the magnetic monopole by the adiabatic loop of
Hamiltonian in momentum space.  The magnetic monopole is the
source or drain of the gauge field ${\bf \Omega}_n({\bf k})$. Since
\begin{equation}
\label{berrycurvature}
{\bf \Omega}_{n}({\bf k})=-{Im} \sum_{m\neq n}\frac{\langle u_{m}|\nabla_{{\bf k}}{H}({\bf k})|u_{n}\rangle\times \langle u_{n}|\nabla_{{\bf k}}{H}({\bf k})|u_{m}\rangle}{(E_m({\bf k})-E_n({\bf k}))^2},
\end{equation}
it diverges at the point ${\bf k=k}_0$ where $E_m({\bf
  k}_0)$=$E_n({\bf k}_0)$ {(see Eqs.~(\ref{eq:hall_cond}) and
  ~(\ref{eq:hall_cond_Kubo}))}.  This means that the magnetic monopole
is formed by energy level crossing, and a two-energy-level system is
the simplest case.  Consider the generic form of the Hamiltonian for
any two-level system,
\begin{equation}
\label{2x2H}
H=d_0({\bf k})I_{2\times2}+\mathbf{d}({\bf k})\cdot \mathbf{\sigma},
\end{equation}
where $I_{2\times2}$ is {the} identity matrix, $\mathbf{d}$ is a 3D
vector depending on momentum ${\bf k}$, and $\mathbf{\sigma}$ are
Pauli matrices. There are two eigen states $\psi_\pm$ with eigen
energies $E_{\pm}=d_0(\mathbf{k})\pm\sqrt{\mathbf{d}\cdot\mathbf{d}}$.
The term $d_0(\mathbf{k})$ is just a shift of zero energy level, and
can be neglected. At the energy degeneracy or level crossing point
($\bf k=k_0$), $E_+$=$E_-$ (i.e, $\mathbf{d}(\mathbf{k}_0)$=0) is
required. Obviously, $\mathbf{k}_0$ is not necessarily on the path of
a adiabatic loop change of the Hamiltonian, which means that the band
gap between $E_+$ and $E_-$ can be well preserved if the Fermi level
is away from the band-crossing point. Around the neighborhood of
the $\mathbf{k}_0$ point, $\mathbf{d}(\mathbf{k})$ can be expanded as
$\mathbf{d}(\mathbf{k}) \approx
\mathbf{d}(\mathbf{k}_0)+(\mathbf{k}-\mathbf{k}_0)\cdot \nabla
\mathbf{d}(\mathbf{k})$.  Taking the zero point of the parameter space
as $\mathbf{k}_0=0$, we have $\mathbf{d}(\mathbf{k})$=$(\mathbf{k}
\cdot \nabla) \mathbf{d}(\mathbf{k})$. Now for the simplest case,
where $\mathbf{d}(\mathbf{k})=\pm \mathbf{k}$, we can have
$H(\mathbf{k})=\pm \mathbf{k}\cdot \mathbf{\sigma}$, and the
corresponding Berry curvature can be obtained as
\begin{equation}
\label{magmono}
\mathbf{\Omega}_{\pm}(\mathbf{k})=\mp\frac{\mathbf{k}}{2|k|^3} {.}
\end{equation}
Obviously, such a magnetic field distribution in momentum space is
similar to the electric field distribution of a point charge in real
space, and can be understood as magnetic field around a ``magnetic
charge''--magnetic monopole~\cite{fang_anomalous_2003}. In other
words, the divergence of magnetic field ${\bm\Omega}({\bf k})$ is no
longer zero (i.e., $\nabla_{\bf k}\cdot {\bm\Omega}({\bf k})\ne 0$)
but is related to the magnetic charge $\pm\frac{1}{2}$ at the source or
drain. The degeneracy or level crossing point $\mathbf{k}_0$ is the
place where the magnetic monopole with strength $\frac{1}{2}$ is
located.

Joshua Zak~\cite{Zak_Berry_1989} pointed out that such Berry phase and
Berry curvature can also exist in periodic systems where the
eigenstates are Bloch wave functions and the parameter space is the
crystal momenta $\mathbf{k}$ which can vary in closed loops, such as
Brillouin zone or Fermi surface. Then, the adiabatic evolution {loop
  forms} a compact manifold that has no boundary. For example, the
Brillouin zone of a 2D system is a torus, and the Fermi surface of a
3D metal (in the simplest case) is an enclosed sphere. Obviously,
Gauss\textquoteright{s} law ensures that the total flux penetrating
the closed surface must be quantized and is equal to the magnetic
charge of monopoles enclosed by the surface (either torus or Fermi
sphere).

Now we can see that the magnetic monopole in momentum space (or
parameter space in general) plays a crucial role in determining the
topology of electronic band structures. Besides the insulating case,
in the following we concentrate on a 3D metal with a well-defined
Fermi surface in momentum space. Taking the above two-level system as
an example and assuming that $d_1(\mathbf{k})=k_x$,
$d_2(\mathbf{k})=k_y$ and $d_3(\mathbf{k})=k_z$, the magnetic monopole
is located at the origin $k_x=k_y=k_z=0$ and the {low-energy}
Hamiltonian can be written as
 \begin{eqnarray}
\label{eq:weyl}
H(\mathbf{k}) & = & \pm\mathbf{k}\cdot \mathbf{\sigma}  \nonumber  \\
& = &  \pm\left[\begin{array}{cc}k_z & k_x-i k_y  \\ k_x+i k_y & -k_z
  \end{array}\right].
\end{eqnarray}
This Hamiltonian was first proposed by Weyl~\cite{weyl_elektron_1929},
who found that, for massless fermions, the $4\times4$ Dirac
representation is reducible, and is composed of two $2\times2$
(irreducible) Weyl fermions with positive (+) and negative (-)
chirality (and opposite magnetic charge). Similar to the definition of
Chern number for a torus (in a 2D insulator), the total flux of the
gauge field passing through a Fermi surface of 3D metal must be
quantized as a multiple of $2\pi$. We can then define the Fermi
surface Chern number $C_{FS}$ as
\begin{equation}
\label{eq:cfs}
C_{FS}=\frac{1}{2\pi}\int_{FS}{\bm\Omega}({\bf k})\cdot d{\mathcal{S}},
\end{equation}
and use it as a new topological invariant to {describe} topological
metallic states. The Fermi surface Chern number $C_{FS}$ is non-zero
if a Weyl node (or a magnetic monopole) is enclosed by a Fermi
surface, and this leads to a non-trivial topological metallic state
(called Weyl metal). If the Fermi level happens to be exactly at the
Weyl node, we will get a topological semimetal state (i.e, Weyl
semimetal).  Unfortunately, according to the ``no-go
theorem''~\cite{nielsen_absence_1981-1,nielsen_absence_1981-2}, for
any lattice model, the Weyl nodes with opposite chirality have to
appear in pairs (although they might be separated in momentum space),
and the summation of $C_{FS}$ for all pieces of Fermi surfaces must be
vanishing~\cite{haldane_attachment_2014}. This makes it difficult to
give a proper definition for Weyl metal. Nevertheless, we should note
that Weyl nodes (and magnetic monopoles) are stable topological
objects, which can be well defined in 3D momentum
space~\cite{Balents_Physics_Weyl_2011,murakami_phase_2007}. Two Weyl
nodes with opposite signs may be separated in momentum space and lead
to two pieces of Fermi surfaces, each of which has a non-zero Fermi
surface Chern number $C_{FS}\ne 0$. The two Weyl nodes may annihilate
each other if and only if they overlap in ${\bf
  k}$-space~\cite{WanXG_WeylTI_2011, Burkov_Weyl_2011,
  Balents_Physics_Weyl_2011, Burkov_Weyl_semimetal_2012PRB,
  WangZJ_Na3Bi_2012}.

Special attentions must be paid to the systems with both TRS and IS,
where Kramers degeneracy exists for every momentum ${\bf k}$. In such
cases, each pair of Weyl nodes, if any, must overlap exactly in the
${\bf k}$-space. In other words, the minimum effective Hamiltonian to
describe such system must be at least 4$\times$4 and contain two Weyl
nodes (with opposite signs) simultaneously, as shown below,
 \begin{eqnarray}
   H(\mathbf{k}) = \left[
   \begin{array}{cc}
   {\bf k}\cdot {\bm\sigma} &   0  \\
    0  & -{\bf k}\cdot {\bm\sigma}
  \end{array}
  \right].
\end{eqnarray}

This is called a 3D Dirac node, and is a {straightforward} extension
of the 2D graphene to 3D space~\cite{ZhangWei_NJP_BiSe_2010}.  As we
have discussed above, in such a case, a perturbative mass term can be
introduced in principle, opening up a gap and leading to an insulating
state. However, if we consider additional symmetries in the system,
such as the crystalline symmetry, the mass term may again be
forbidden, and this would stabilize the 3D Dirac cone and lead to a 3D
Dirac metal or semimetal
state~\cite{young_dirac_2012,WangZJ_Na3Bi_2012}. Therefore, the 3D
Dirac node is not as stable as the Weyl node, however, it is a good
starting point for us to reach the true Weyl semimetal state by
breaking either TRS or IS in the 3D Dirac
semimetals~\cite{Burkov_Weyl_semimetal_2012PRB}.

In general, the energy level crossing may happen in many materials and
at any energy and momentum; however, it is quite rare to have
band-crossing exactly at Fermi level, particularly for the cases with
broken TRS or IS. Up to now, only a few materials have been
theoretically proposed to host such Weyl
semimetal~\cite{WanXG_WeylTI_2011,XuGang_HgCrSe_2011_PRL,Burkov_Weyl_2011,WS_Vanderbilt_2014}
or Dirac semimetal
states~\cite{young_dirac_2012,WangZJ_Na3Bi_2012,WangZJ_Cd3As2_2013}. Among
them, only Na$_3$Bi~\cite{WangZJ_Na3Bi_2012,ChenYL_Na3Bi_2014Science}
and Cd$_3$As$_2$~\cite{WangZJ_Cd3As2_2013,ChenYL_Cd3As2_2014NatMa}
have been confirmed experimentally to be Dirac semimetals. The
proposed Weyl semimetals breaking TRS, such as pyrochlore
Iridate~\cite{WanXG_WeylTI_2011} and HgCr$_2$Se$_4$~\cite{XuGang_HgCrSe_2011_PRL}, 
have not been confirmed yet due to experimental difficulties.
The proposals for Weyl semimetals keeping
TRS but breaking IS are thought to be a way to overcome those
difficulties. Presently the following are typical proposals: One is a
super-lattice system formed by alternately stacking normal and
topological insulators~\cite{multilayerTRI}. The second involves
Tellurium, Selenium crystals or BiTeI under
pressure~\cite{SeTe,WS_Vanderbilt_2014}. The third one is the solid
solutions LaBi$_{1-x}$Sb$_x$Te$_3$ and
LuBi$_{1-x}$Sb$_x$Te$_3$~\cite{WS_Vanderbilt_2014} tuned around the
topological transition points~\cite{murakami_phase_2007}. The fourth
is TaAs-family compounds, including TaAs, TaP, NbAs and NbP, which are
natural Weyl semimetals and each of them possesses a total of 12 pairs
of Weyl points~\cite{TaAs_Weng, TaAs_Hassan}.

\begin{figure}
\begin{centering}
\includegraphics[width=0.8\textwidth]{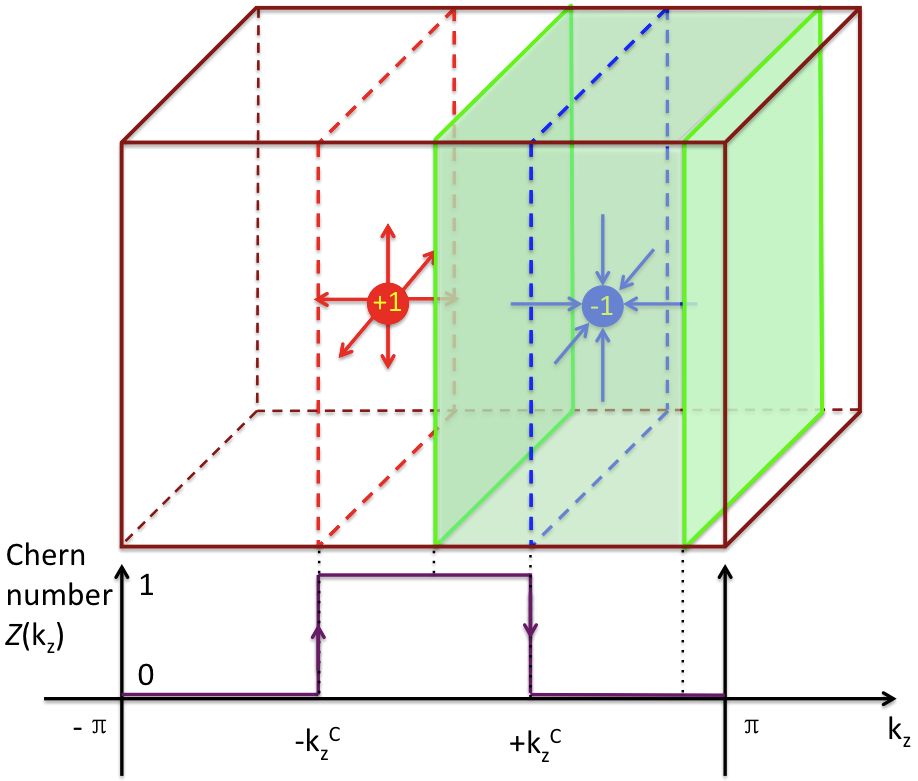}
\par\end{centering}
\caption{Two Weyl nodes with opposite chirality ($\pm 1$) distribut along $k_z$ in 3D momentum space. 
The Chern number Z($k_z$), evaluated for each 2D plane with fixed $k_z$, is shown at the bottom as a function of $k_z$. Jumps of Chern number are seen when the plane moves across a Weyl point. A cube containing one Weyl node is shadowed.}
\label{fig:weylnogo}
\end{figure}

The Weyl semimetal is a new state of quantum matters. It is of particular interest
here that the Weyl semimetal with broken TRS is closely related to the Chern
insulators and provides a unique way to realize QAHE. 
Let us  first consider a
Weyl semimetal with a single Weyl node in a continuous model defined as $H({\bf k}) = v_F {\bf k}\cdot{\bm \sigma}$ , 
 which as mentioned previously generates a monopole in Berry curvature right at the 
 origin. The Gauss's theorem then requires that the total flux flowing through out any 
 closed surface enclosing the Weyl point must be $\pm 1$, i.e, equal to the chirality of the Weyl point (Eq.~\eqref{eq:cfs}). 
 For a lattice system, there is an important ``no-go theorem" indicating that Weyl points
 must appear in pairs with opposite chirality. Originally this theorem has been proved by field theory.~\cite{nielsen_absence_1981-1, nielsen_absence_1981-2} Here we provide a much
 intuitive way to understand it in terms of the concept of Chern number.
 Given a Weyl semimetal in 3D lattice (assuming all trivial states are far away from the Fermi level), first let's consider a 2D plane with fixed $k_z$, where 
 band structures within the plane should be fully gapped unless the plane cut through a Weyl point exactly. The integral of Berry curvature over such a 2D 
 insulating plane must be quantized and gives rise to a well-defined integer Chern number. Moving the plane (i.e, $k_z$) from -$\pi$ to $\pi$, we will then 
 get the Chern number as a function of $k_z$, as shown in Fig.~\ref{fig:weylnogo}. Now, it is important to note that the Chern number as a function of $k_z$ 
 should jump by +1 or -1 (depending on the chirality of the Weyl point), whenever the moving-plane goes across a Weyl point. This is because there exists a 
 topological phase transition at the Weyl point, and the band gap of 2D plane is closed and re-opened when the plane goes across the Weyl point. This jump 
 can also be understood by the following consideration. Selecting two planes at the opposite sides of a Weyl point, we can construct a closed manifold surrounding 
 the Weyl point, i.e, the cube formed by the two parallel planes and the four side-surfaces, shown in Fig.~\ref{fig:weylnogo} as the shaded area. The flux 
 flowing through the four side-surfaces should exactly cancel each other, being net zero. Then, the total flux flowing through the closed manifold (i.e, the cube), 
 which is now equivalent to the difference of Chern numbers for the two parallel planes, must be equal to the chirality of the Weyl point enclosed by the 
 manifold (namely $\pm 1$) as discussed above. Having understood the jump, let's look at the Chern number evolution as the function of $k_z$ from -$\pi$ to $\pi$.  
 In order to satisfy the periodic boundary condition in the BZ of lattice, the positive jump (+1) and the negative jump (-1) must appear in pairs, if there is any. 
 This leads to the important conclusion that Weyl points with opposite chirality must appear in pairs.

Finally, the total Hall conductivity of the system is given by the integral of $k_z$ as
\begin{eqnarray}
\label{eq:3dahc}
\sigma_{xy}^{total}=\frac{1}{2\pi}\int_{-\pi}^{\pi}dk_z\sigma_{xy}(k_z) {.}
\end{eqnarray}
In such a way, we see the relationship between 2D Chern insulators and
3D Weyl semimetals. Given a 3D Weyl semimetal as defined above, we can
form a 2D thin film (or quantum well structure) such that $k_z$ is
quantized by the sample thickness, and only particular {values of}
$k_z$ are allowed. If it happens that, for some particular
thicknesses, the 2D Chern number are non-zero, we will expect
quantized total Hall conductivity, and this leads to a 2D Chern
insulator and the QAHE.

\begin{figure}
\begin{centering}
\includegraphics[width=0.8\textwidth]{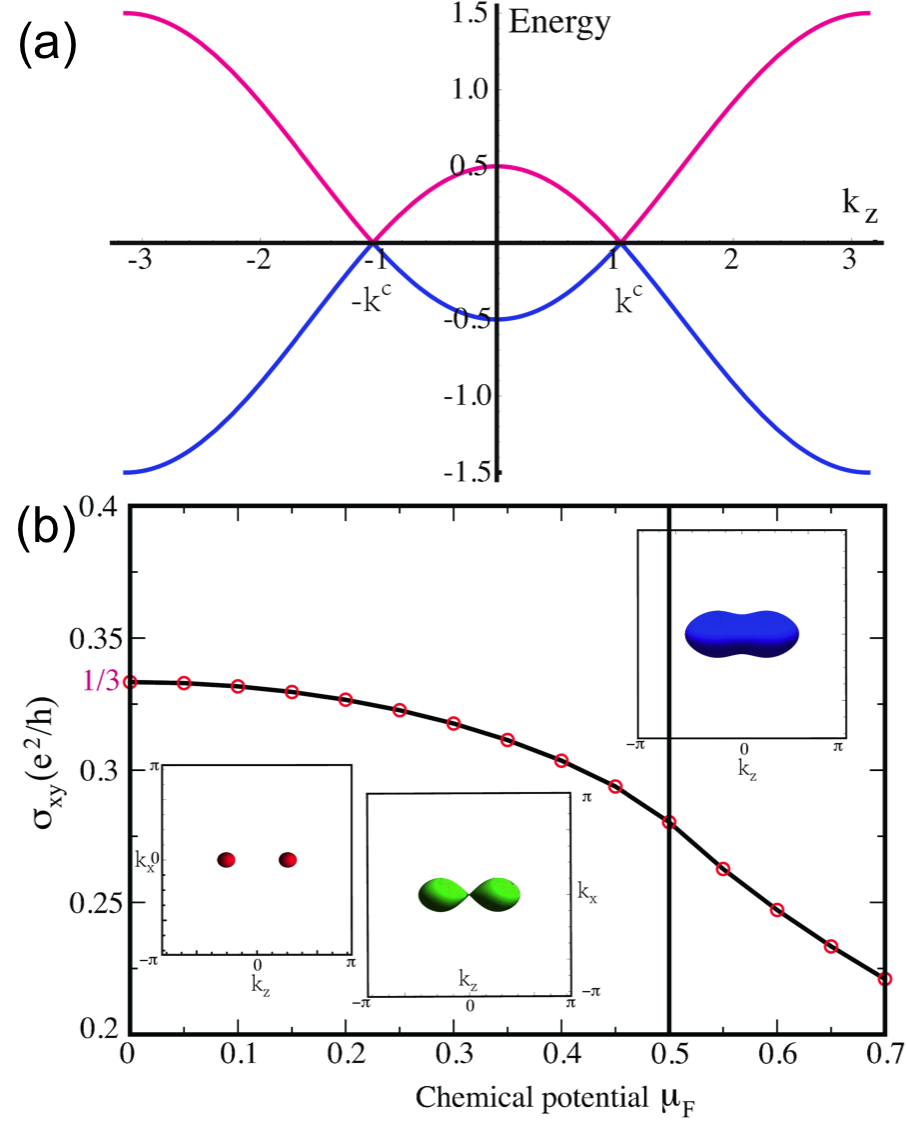}
\par\end{centering}
\caption{(a) The $k_z$ band dispersion of FM Weyl semimetal. (b) The AHC dependence on the position of chemical potential $\mu_F$.
Inset: Fermi surface for different $\mu_F$.}
\label{fig:WeylAHC}
\end{figure}


To explicitly show the Hall conductivity of a FM Weyl semimetal, we
introduce a simple Hamiltonian based on a cubic lattice BHZ
model~\cite{qi_topological_2006},
\begin{eqnarray}
\label{eq:weylbhz}
H^{WS}={\rm sin}(k_x)\sigma_x+{\rm sin}(k_y)\sigma_y
+[2+e_s-{\rm cos}(k_x)-{\rm cos}(k_y)-{\rm cos}(k_z)]\sigma_z.
\end{eqnarray}
It has two Weyl nodes at (0,0, $\pm k_z^c$) with $\cos(k_z^c)$=$e_s$
when $0< e_s <1$. The energy dispersion along $k_z$ axis is shown in
Fig.~\ref{fig:WeylAHC} for $e_s$=$\frac{1}{2}$. The Hall conductivity
$\sigma_{xy}(k_z)$ is calculated by using Eqs. ~\eqref{eq:sigxy} and
~\eqref{berrycurvature} for each plane with fixed $k_z$.  When
chemical potential $\mu_{F}$ is 0 and passes exactly through the Weyl
nodes, $\sigma_{xy}(k_z)$ is quantized to be 1 for $|k_z|<k_z^c$ and 0
for $|k_z|>k_z^c$, consistent with the above discussions based on the
continuum model. Integrating $\sigma_{xy}(k_z)$ over $k_z$, the total
Hall conductivity $\sigma_{xy}$ is given as
$\frac{e^2}{h}\frac{k_z^c}{\pi}$~\cite{Burkov_ahe_weyl_prb_2013,
  Burkov_ahe_FM_2014, Burkov_ahe_Weyl_2014}, which depends only on the
separation of Weyl nodes in momentum space (i.e, $k_z^c$).  Shifting
the chemical potential $\mu_F$ away from 0 leads to partially occupied
bands, and the corresponding Fermi surfaces change from two isolated
spheres ($0<\mu_F <0.5$) to two touched ones ($\mu_F=0.5$) and finally
they merge into a peanut shape ($0.5<\mu_F<1.5$), as shown in
Fig.~\ref{fig:WeylAHC}. The total Hall conductivity decreases with
rising chemical potential $\mu_F$, because the Berry curvature from
the upper band tends to cancel out the contribution from the lower
band. Finally, when both bands are fully occupied the sum of Berry
curvature over all bands must vanish~\cite{TI&TSC_book_Bernevig}. It
is important to note that a fully occupied band (without a
corresponding Fermi surface) may still contribute to the Hall
conductivity, and this should be carefully considered in practical
calculations. In such a case, the Hall conductivity is usually
evaluated by the volume integrals of the Berry curvature of occupied
bands over the whole BZ, as is done for the calculation of intrinsic
AHE~\cite{fang_anomalous_2003, yao_first_2004,
  WangXJ_ab_AHE_Wannier_2006}; this is also discussed by Chen et
al.~\cite{Burkov_ahe_weyl_prb_2013} and Vanderbilt et
al.~\cite{Vanderbilt_comment_ahe_Weyl_2014,
  Burkov_Reply_Vanderbilt_2014,
  Haldane_ahe_Fermi_surf_2004,Wang_ahe_Fermi_surf_2007}.


\section{First-Principles Calculations for Topological Electronic
  States}


Theoretical predictions, particularly first-principles electronic
structure calculations based on density functional
theory~\cite{martin_electronic_2004}, have played important roles in
the exploration of topological electronic states and materials. This
is not an accidental success, but rather due to several deep reasons:
(1) To describe complicated electronic band structures of real
materials, particularly with the spin-orbit coupling (SOC) included,
first-principles calculations are necessary; (2) First-principles
calculations nowadays can reach the accuracy even up to 90\% for most
of the physical properties of ``simple'' materials (i.e., weakly
correlated electronic materials), which makes prediction possible; (3)
The topological electronic properties of materials are robust,
non-perturbative, and not sensitive to small error bars, if any. Given
those advantages of first-principles calculations and its great
success in this field, however, we have to be aware that the numerical
determination of topological invariants (such as integer $Z$ or Z$_2$
numbers) is still technically demanding, because those numbers are
related to the phases of eigen wave functions, which are gauge
dependent and randomized in most of the calculations.  In addition to
that, we also have to note that: (1) We still have a band-gap problem
in either the local density approximation (LDA) or the generalized
gradient approximation (GGA) for the exchange-correlation potential;
(2) The present first-principles calculations based on LDA or GGA can
not treat strongly correlated systems properly; (3) The evaluation of
Berry phase~\cite{wilczek1989geometric} and topological numbers may
require some additional complicities, such as fine ${\bf k}$-points
meshes, gauge-fixing condition, etc. In this section, we will discuss
some important issues and techniques related to the first principles
studies of topological electronic states.

~\\~
\hspace*{15pt}{\bf III.A. Calculations of Berry Connection and Berry Curvature}
\\

From the computational point of view, a self-consistent
first-principles electronic structure calculation for a real material
is typically performed based on the LDA or GGA for the
exchange-correlation potential. Such a calculation will generate a set
of single particle eigen states with wave functions $u_{n\bf k}({\bf
  r})$, eigen values $\varepsilon_{n}({\bf k})$, and occupation
$f_{n}({\bf k}$) for the ground state of the system. Having obtained
those quantities, the remaining task is to evaluate the Berry
connection ${\bf A}_n({\bf k})=i\langle u_{n\bf k}|\nabla_{\bf
  k}|u_{n\bf k}\rangle$ and the Berry curvature ${\bf \Omega}_n({\bf
  k})=\nabla_{\bf k}\times {\bf A}_n({\bf k})$. The Hall conductivity
can be obtained either from the summation of Berry curvature over the
BZ (Eq.~\eqref{eq:hall_cond}) or from the Kubo formula directly
(Eq.~\eqref{eq:hall_cond_Kubo}). In the latter case, the matrix
element of velocity operator $\hat{{\bf v}}$ is given as
\begin{eqnarray}
 \langle \psi_{m{\bf k}}|\hat{\bf v}|\psi_{n{\bf
      k}}\rangle=\frac{1}{\hbar}\langle u_{m{\bf
      k}}|[\frac{\partial}{\partial {\bf k}}, H_{\bf k}]|u_{n{\bf
      k}}\rangle
=\frac{1}{\hbar}(E_n({\bf k})-E_m({\bf k}))\langle u_{m{\bf
      k}}| \frac{\partial}{\partial {\bf k}} |u_{n{\bf
      k}}\rangle .
\label{eq:velocity}
\end{eqnarray}

The strategy is simple, but the computational task is hard, because
although the Berry curvature does not depend on the gauge choice, the
Berry connection does. It is also true, according to our above
discussions, that choosing a smooth gauge for a topologically
non-trivial state is generally difficult or even impossible. In most
calculations, the Berry connection is a rapidly varying function of
momentum ${\bf k}$, which makes the convergence of the calculation
difficult. To overcome this problem, a straightforward way is to
increase the number of ${\bf k}$-points in momentum space. Another
possible way is to use the Wannier representation, such as the
maximally localized Wannier functions
\cite{Vanderbilt_Wannier_RMP_2012}.

The first-principles calculations of Berry curvature and its
integration over the BZ to get the intrinsic anomalous Hall
conductivity (AHC) have been performed for SrRuO$_3$ and body-centered
cubic Fe by Fang et al.~\cite{fang_anomalous_2003} and Yao et
al.~\cite{yao_first_2004}, respectively. These were the first
quantitative demonstrations of strong and rapid variation of Berry
curvature over the BZ. The sharp peaks and valleys make the
calculation of AHC challenging. Millions of $\bf k$ sampling points
are required to ensure convergence. Wang et
al.~\cite{WangXJ_ab_AHE_Wannier_2006} proposed an efficient approach
employing the Wannier interpolation technique. They first perform the
normal ${\it ab~initio}$ self-consistent calculation by using
relatively coarse $k$-grid --- good enough to ensure the total energy
convergence. Then, Maximally Localized Wannier Functions (MLWF) are
constructed for the states around the Fermi level. The number of
Wannier functions should be carefully chosen so that the isolated
group of bands around the Fermi level can be well reproduced,
especially those bands passing through the Fermi level (because the
small energy splitting due to spin-orbit coupling or band
anti-crossing can produce sharp peaks or valleys of Berry
curvature). Once MLWFs are obtained, the interpolation technique is
applied to obtain the necessary quantities, such as eigen energies,
eigen states, and velocity operator, on a much finer $k$-grid. Since
the number of MLWFs is typically much smaller than the number of basis
functions used in ${\it ab~initio}$ calculations, this approach saves
a lot of computation cost, both in time and in storage.

In additional to the direct methods discussed above, some advanced
techniques with well-controlled convergency with respect to the random
phases can be developed. This is because Berry connection is not the
physical quantity that we can measure, and our ultimate goal is to
evaluate the Berry curvature which is gauge-invariant. For example, it
was proposed that for calculation of electrical polarization within
the modern theory of
polarization~\cite{King_Smith_PRB_Berry_Polar_1993,Resta_RMP_1994} the
change of polarization is directly proportional to the sum of Berry
phase terms of occupied states. Since Berry phase is the integral of
the Berry connection along a closed loop in the parameter space (see
Eq.~\eqref{eq:berry-phase}), it is well defined and gauge-independent.
To calculate the change of polarization along the direction of $z$, a
series of loops (called Wilson loops
\cite{Wilson:1974ji,Giles:1981hj}) in momentum space are
constructed. Each loop has fixed $\bf k_{\bot}$=($k_x, k_y$), but is
periodic along $k_z$. We can then discretize $k_z$ as
$k_z^j$=$\frac{j}{J} G_z$ ($j$=0, 1, ..., J-1), where $G_z$ is the
reciprocal lattice vector along $k_z$. Taking the periodic gauge
$u_{n, ({\bf k}_{\bot}, k_z^J)}=e^{-i G_{z}\cdot {\bf r}} u_{n, ({\bf
    k}_{\bot}, k_z^0)}$, the Berry phase along each loop can be
calculated by
\begin{eqnarray}
  \gamma({\bf k}_{\bot})={ Im}\left[ \ln  \det  \prod_{j=0}^{J-1} \langle u_{n, ({\bf k}_{\bot}, k_z^j)}|u_{m, ({\bf k}_{\bot}, k_z^{j+1})}\rangle \right]
\label{eq:berryWilsonloop}
\end{eqnarray}
with band indices $n$ and $m$ for occupied states. By doing this, the
random phase problem associated with eigen wave functions can be
largely eliminated, and the main computational task is to evaluate the
inner product of two neighboring wave functions along the loop.

For a metallic system, the intrinsic AHC can also be calculated
through this method, as has been implemented by Wang et
al.~\cite{Wang_ahe_Fermi_surf_2007}. It was justified by Haldane
~\cite{Haldane_ahe_Fermi_surf_2004} that the non-quantized part of
intrinsic AHC in a FM metal can be recast as a Fermi surface
integration of the Berry curvature, so we can construct the Wilson
loops over the Fermi surface rather than the Brillouin
zone. Considering a FM metal with magnetization $M$ along the $z$
direction, a 2D plane (in momentum space) perpendicular to $k_z$ will
cut through the Fermi surface. The intersection between the plane and
the Fermi surface defines loops, which are the Wilson loops used for
calculating the AHC. Following our above discussions, we can
discretized each loop and evaluate the Berry phase along the loop
efficiently by using Eq.~\eqref{eq:berryWilsonloop}. To do this,
several things should be borne in mind: (1) The periodic boundary
condition has to be adapted as $u_{n, ({\bf k}_{\bot}, k_z^J)}=u_{n,
  ({\bf k}_{\bot}, k_z^0)}$ if the loop does not cross the Brillouin
zone boundary; (2) The direction of Wilson loops should be defined
consistently; (3) For the case with multiple branches, a continuity
condition must be used to choose each branch carefully. Finally, in
additional to the non-quantized part of AHC, the quantized
contribution to the AHC (coming from the deep occupied states) can be
determined by the Fermi-sea integration of the Berry curvature, which
is gauge invariant and has no ambiguity.

This Wilson loop method is a convenient tool, and it can also be
applied to investigate the topological invariant for systems with or
without TRS as will be discussed below.

~\\~
\hspace*{15pt}{\bf III.B. Wilson Loop Method for Evaluation of Topological Invariants}
\\

In real materials, the band structures are usually very complicated,
and the determination of topological invariants becomes essential and
computationally demanding. The band degeneracy, either accidentally or
due to symmetry, makes the numerical determination of phases of wave
functions a tough task. Here we will present the Wilson loop method
for determining topological indices efficiently~
\cite{YuRui_Z2_2011PRB, Vanderbilt_wannierCenter_TI_2014PRB,
  Soluyanov_wannier_Z2_2011, ringel_determining_2011}.  This method is
computationally easy and is equally applicable for Chern insulators,
$Z_2$ topological insulators (2D and 3D), and the topological
crystalline insulators~\cite{FuLiang_Topological_Crystalline_2011PRL,
  Hsieh_EXP_TCI_2012}, which have attracted lots of interest recently.

For band insulators with inversion symmetry, the $Z_2$ indices can
easily be computed as the product of parity eigen values for half of
the occupied states (Kramers pairs have identical parities) at the
time-reversal-invariant-momentum (TRIM)
points~\cite{FuLiang_TI_with_inversionSymmetry_2007}. In such a case,
parity is a good quantum number for the TRIM points and can be
computed from the eigen wave functions obtained from the
first-principles calculations. This method has been used frequently
and is very efficient.  Unfortunately, for general cases where
inversion symmetry is absent, the evaluation of the $Z_2$ number
becomes difficult. Three different methods have been considered for
such a case: (i) Directly compute the $Z_2$ numbers from the
integration of the Berry curvature $\bf \Omega (k)$ over half of the
Brillouin zone~\cite{Fukui_Hatsugai_Z2_2007}. In order to do so, one
has to set up a fine mesh in the $\bf k$-space and calculate the
corresponding quantity for each $\bf k$ point. Since the calculation
involves the Berry connection $\bf A(k)$, one has to numerically fix
the gauge on the half BZ, which is not easy for the realistic wave
functions obtained by first-principles calculations.  (ii) Start from
an artificial system with inversion symmetry, and then adiabatically
deform the Hamiltonian towards the realistic one.  If the energy gap
never closes at any point in the BZ during the deformation process,
the realistic system must share the same topological nature with the
initial reference system, whose $Z_2$ number can easily be counted by
the parity eigenvalue
formula~\cite{FuLiang_TI_with_inversionSymmetry_2007}.  Unfortunately,
making sure that the energy gap remains open on the whole BZ is very
difficult numerically, especially in 3D. (iii) Calculate the boundary
(edge or surface) states. Due to the open boundary condition, the
first-principles calculation for the boundary states is numerically
heavy. The Wilson loop method, which we will present here, has the
following advantages: firstly, it uses only the periodic bulk system;
second, it does not require a gauge-fixing condition -- thereby
greatly simplifying the calculation; third, it can easily be applied
to a general system with or without inversion symmetry.

\begin{figure}[tbp]
\centering
 \includegraphics[width=0.8\textwidth, angle=-90]{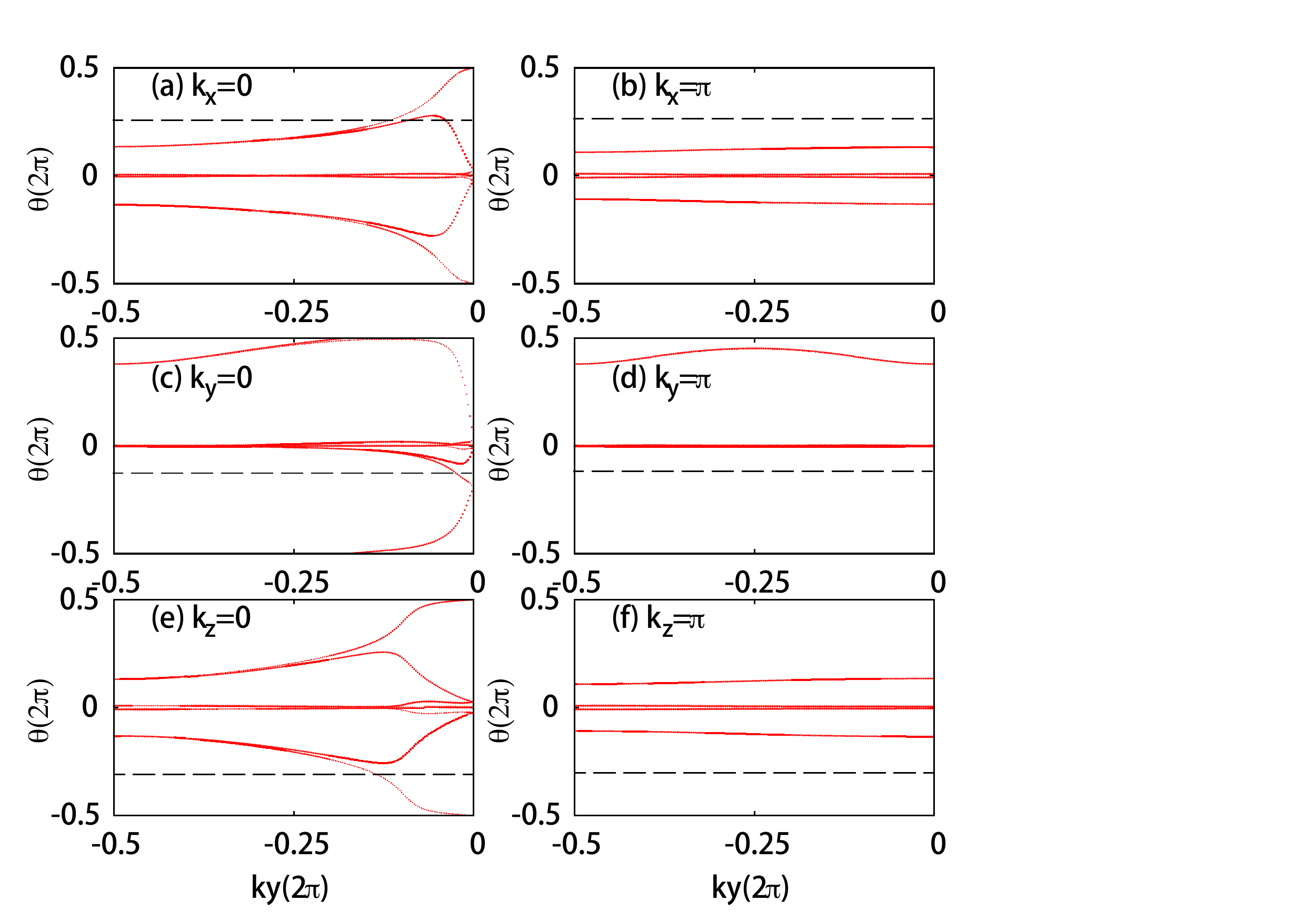}
 \caption{(Color online) Example of Wilson loop method for the 3D
   strong topological insulator TlN with $Z_2$ index (1;000). The
   calculated $\theta_m(k_y)$ for all occupied bands ($m$ is the band
   index) are plotted as functions of $k_y$. The dashed lines are
   arbitrary reference lines. The $\theta_m(k_y)$ crosses the
   reference line an odd number of times in half of $k_y$ direction
   ($k_y$ $\in$ [0,$\pi$] or [$\pi$, 2$\pi$]) for (a) $k_x=0$, (c)
   $k_y=0$ and (e) $k_z=0$ plane (left panel), but an even number of
   times for (b) $k_x=\pi$, (d) $k_y=\pi$ and (f) $k_z=\pi$ plane
   (right panel). Figures from Ref.~\onlinecite{XL_Sheng:2014vq}. }
\label{fig:wilson_loop}
\end{figure}

As discussed in the previous section for 2D insulators, to determine
the topological number (either $Z$ or $Z_2$), the key is the $\theta
(k_y)$ angle and its 1D evolution. In the single band model, the
$\theta (k_y)$ angle is given simply as the 1D integration of
$A_x(k_x, k_y)$ along the $k_x$ axis. However, for real compounds with
multiple bands, its computation requires the Wilson loop
method~\cite{YuRui_Z2_2011PRB}. Once the $\theta$ angle is computed,
all the topological numbers can be determined easily, following our
previous discussions.  For the mathematic details of this method,
please refer to our paper~\cite{YuRui_Z2_2011PRB}. Here we will focus
on the practical steps for real calculations.

Considering an insulator with $N$ occupied states, after the
self-consistent electronic structure calculations, we get the
converged charge density, the total energy, the band structure, etc.
Then, without loss of generality, we consider a periodic plane of
($k_x$, $k_y$) with $-\pi<k_x<\pi$ and $-\pi<k_y<\pi$, and consider
the $k_x$ axis as a closed loop. For each fixed $k_y$, we discretize
the $k_x$ line into an $N_k$ mesh with mesh-index $i$ ranging from 0
to $N_k-1$. Due to the periodic condition, the $i$=$N_k$ point is
equivalent to the $i$=0 point. The Wilson loop method consists of the
following steps:

(1) We need to calculate the eigen wave functions of occupied states
for all mesh points. This can be done by continuing from the converged
charge density and with no need for charge self-consistency
furthermore.

(2) Using the obtained wave functions, we can calculate the inner
products among them and evaluate the $N\times N$ matrix
$F_{i,i+1}^{m,n}=\langle m,k_{x,i},k_y | n, k_{x,i+1}, k_y \rangle$,
where $m$ and $n$ are band indices of occupied states ranging from 1 to $N$. The $F$
matrix~\cite{YuRui_Z2_2011PRB} is nothing but the discretized
version of Berry connection $A_x$.

(3) Now we can construct the $N\times N$ matrix $D$ as the product of
$F$, $D(k_y)=F_{0,1}F_{1,2}...F_{N_k-2,N_k-1}F_{N_k-1,0}$.

(4) Diagonalization of the $D(k_y)$ matrix gives us $N$ eigenvalues
$\lambda_n(k_y)$, whose phase angle can be obtained as
$\theta_n(k_y)$=Im[log$\lambda_n(k_y)$].

This phase angle $\theta_n(k_y)$, which is now band dependent, is
exactly what we need to determine the winding number and the
corresponding topological numbers, as discussed in section II. In
practice, for a system with many bands, we can draw an arbitrary
reference line on the cylinder surface, parallel to the $k_y$ axis,
and then count how many times the evolution lines of $\theta_n(k_y)$
cross the reference line~\cite{YuRui_Z2_2011PRB}. For the calculation
of $Z_2$ in time-reversal invariant systems, we can further reduce the
computational task by considering only the $0<k_y<\pi$ part. As an
example, we show in Fig.~\ref{fig:wilson_loop} the calculated
$\theta_n$ and their evolutions for one 3D topological insulators TlN
proposed recently~\cite{XL_Sheng:2014vq}. All the six planes, each of
them passing through four TRIM points, are shown. We see that the
$\theta_n$ lines cross the reference line an odd number of times for
all three planes containing $\Gamma$, but an even number of times for
those planes containing ($\pi$,$\pi$,$\pi$). Therefore, we conclude
that it is topologically non-trivial with $Z_2$ index (1;000).

Finally, through the whole calculation process, we can always take
advantage of Wannier functions, particularly the maximally localized
Wannier functions (MLWF) as proposed by
Vanderbilt~\cite{Vanderbilt_Wannier_RMP_2012,
  Souza_Vanderbilt_MLWF_2001}.  For example, after the self-consistent
electronic structure calculations, we can construct the MLWF for the
energy bands close to the Fermi level, which can usually be reproduced
up to a very high accuracy. In this process, however, some of the
unoccupied bands should be
included~\cite{Souza_Vanderbilt_MLWF_2001}. Once the MLWF are
obtained, we can easily calculate the eigen wave functions and their
inner products for any ${\bf k}$-points using the interpolation
technique~\cite{WangXJ_ab_AHE_Wannier_2006, yates_spectral_2007}. This
will greatly simplify our task for the computation of topological
numbers. In addition, the MLWF are also useful for the calculations of
boundary states, as discussed below.


~\\~
\hspace*{15pt}{\bf III.C. Boundary States Calculations}
\\

After the determination of bulk topological numbers by either of the
methods presented above, the calculation of boundary states of real
compounds becomes necessary. This is not just for the purpose of
confirming topological properties -- it is also very useful for
experiments, such as angle-resolved photo-emission spectroscopy
(ARPES) and scanning tunneling microscope (STM), because it provides
necessary information for direct comparison.

In practice, two methods are usually used for the calculation of
boundary states. First, the boundary states can be directly calculated
from first-principles calculations by using a supercell of slab
geometry, as is usually done for other surface or edge
calculations~\cite{yip_handbook_2007}. In this way, the atomic
details, such as the surface or edge terminations, the atomic
relaxations or reconstructions, the possible defects or absorptions,
can be carefully treated and studied. This method is straightforward,
although it is computationally demanding. Due to the presence of
boundary states, we have to use a very large supercell with a thick
vacuum region to avoid possible mutual coupling between the
boundaries. The convergence test as a function of cell size is usually
necessary. This method can generate detailed surface eigen states,
on the basis of which the spin texture of topological surface states
(showing characteristic spin-momentum locking effect) can be obtained.
With spin-orbit coupling included, the eigen states are expressed as
two-component spinors, $\psi_{n{\bf k}}=\left(\begin{array}{c}
    \phi_{n{\bf k}\uparrow} \\ \phi_{n{\bf k}\downarrow}
  \end{array}\right)$, and the expectation value of the spin operator ${\bf
  S}=\frac{\hbar}{2}{\bf \sigma}$ can be calculated easily by
$\psi_{n{\bf k}}^\dagger {\bf S} \psi_{n{\bf
    k}}$~\cite{ZhangWei_NJP_BiSe_2010, XL_Sheng:2014vq}. In addition,
the layer-dependent distribution of surface states (i.e, the
penetration depth) can also be studied carefully through these eigen
states~\cite{ZhangWei_NJP_BiSe_2010}.

Second, we can take advantage of MLWF and calculate the boundary
states from the Green's functions of a semi-infinite
system~\cite{harrison_electronic_1989}. We can use the MLWF as a basis
to construct an effective low-energy Hamiltonian $H_{WF}$, which can
be regarded as a tight-binding model with its hopping parameters
determined from the first-principles electronic structure
calculations.  The maximally localized property of MLWF guarantees the
short range hopping. Using this $H_{WF}$ as building blocks, we can
construct a large Hamiltonian $H_{semi}$ for the semi-infinite system
with only one boundary. This $H_{semi}$ is a block tridiagonal matrix
if we consider only the hopping terms between the nearest-neighbor
building blocks. In practice, we find that the thickness of the
building block should also be tested to ensure convergence since in
some cases the hopping among MLWFs might extend to more than the
nearest-neighbor. Then we can use the iterative
method~\cite{godfrin_method_1991,sancho_quick_1984,sancho_highly_1985}
to solve the problem and get the projected Green's functions onto the
boundary. From these boundary Green's functions, we can get necessary
information such as the charge density of states and the spin density
of states at the
boundary~\cite{ZhangHJ_BiSb_2009PRB,Dai_HgTe_2008PRB}. This procedure
is an approximate way and it cannot treat the atomic details at the
boundary precisely, however, it provides the most important
information for the gapless nature of the boundary states of
topological insulators.

\section{Material Predictions and Realizations of QAHE}

The topological electronic states discussed in previous sections are
important and interesting, not only because they are conceptually new
but also because they are realizable and testable in condensed
materials. In recent years, this field has been developing very fast.
Many candidate compounds with non-trivial topological electronic
states are proposed, and some of them have been successfully
synthesized and confirmed experimentally. The realizations of various
topological electronic states have strongly promoted this field, and a
lot of interesting experiments on their non-trivial topological
properties are now becoming possible. In this section we will focus on
the topic of how to realize the QAHE in real materials.

In spite of great success in recent studies of TIs and QSHE, the
realization of QAHE has taken a long time. The key to achieving this
goal is to find a proper Chern insulator. Any realistic material
system in which QAHE can be realized must satisfy the following
conditions:
\begin{enumerate}
\item   It must be two dimensional;
\item   It must be insulating in the bulk;
\item It must break the TRS with a certain magnetic ordering (the
  simplest case is a FM long-range order);
\item The occupied bands must carry a non-zero Chern number (therefore
  strong SOC is usually required).
\end{enumerate}
It is not difficult for a material to satisfy one or two conditions in
the above list. But finding a realistic material that satisfies all of
these conditions simultaneously turns out to be very challenging. This
is the reason the Chern insulator is theoretically proposed long
before the discovery of TIs, but its realization came much
later. Nevertheless, the discovery of QSHE and TIs
\cite{Kane_Z2_2005,Kane_Mele_Graphene_PRL_2005,Bernevig_HgTe_QW_science_2006,FuLiang_3dTI_2007PRL,ZhangHJ_Bi2Se3_2009NP,QiXL_RMP_2011,Hasan_Kane_RMP_2010}
have greatly stimulated the field of QAHE. It is natural to expect
that starting from known 2D TIs and then breaking the TRS will be one
of the simple ways to achieve the QAHE. The 2D TIs already satisfy
three out of those four conditions, and these TIs can be effectively
viewed as two Chern insulator layers related by TRS. Breaking of TRS
can, in principle, destroy one layer and keep the other layer active,
which will lead to the QAHE. Based on this idea, we can go further: if
the exchange-splitting is strong enough, we can even realize the QAHE
starting from a 2D narrow gap semiconductor, where the strong
exchange-splitting may cause band inversion for half of the states
(i.e, for only one of the spin subspaces, as will be addressed
below). Unfortunately, it is still very hard to induce
exchange-splitting in 2D TIs or narrow gap semiconductors; two known
procedures, the magnetic proximity effect and magnetic doping, are
most commonly employed. Despite the issue of magnetic ordering
temperature $T_c$ (which can be very low), magnetic doping will
obviously introduce impurities that may destroy the insulating state
of parent compounds. On the other hand, using the magnetic proximity
effect may have the advantage of avoiding the problem of impurity
doping, while the possible exchange-splitting induced by this
procedure is typically very weak. Therefore, in practice, we have to
choose the candidate compounds and procedures carefully in order to
balance various conditions and requirements. In this sense, the first-principles calculations play important, predictive roles for the final
realization of the QAHE. As listed below, up to now, there have been
several different types of proposals for the realization of the QAHE,
and most of them start from TIs as the parent compounds.
\begin{itemize}
\item To sandwich a TI thin film with FM insulators. The FM proximity
  effect will open up a gap for the Dirac surface states on both
  surfaces of the TI, which gives rise to the
  QAHE~\cite{QiXL_TRInvariant_2008}.
\item To dope the magnetic ions into the TI thin film and make it a FM
  semiconductor~\cite{CXLiu_HgMnTe_2008, yu_quantized_2010,
    LiuCX_QAHE_InAs-GaSb_2013}. The QAHE can be obtained by proper
  control of the doping. This is the only one of these proposals that
  has succeeded so far~\cite{chang_experimental_2013,
    Tokura_2014NatPhys}.
\item To make a thin film (or quantum well structure) of a FM Weyl
  semimetal (such as HgCr$_2$Se$_4$). The quantum size effect gives
  rise to the quantization of the crystal momentum along the growing
  axis, which at the same time quantizes the anomalous Hall
  coefficient at a certain film
  thickness~\cite{XuGang_HgCrSe_2011_PRL, yang_quantum_2011}.
\item To grow graphene, silicene, or other honeycomb lattice on top of
  magnetic insulators or with magnetic adsorbates. The low energy
  bands (which are tunable by external electric field) acquire both
  the Zeeman exchange-splitting and the SOC, which drive the system
  towards QAHE~\cite{qiao_quantum_2010, nandkishore_quantum_2010,
    Tse_QAHE_graphene_2011PhRvB, MacDonald_2011_PRL,
    ding_engineering_2011, Qiao:2011fe, zhang_electrically_2012,
    QiaoZhenhua_PRB_2012, ZHQiao_QAHE_AFM_2014_PRL,
    ezawa_valley-polarized_2012, EzawaNJP12, EzawaPRL13}.
\item To mimic the honeycomb lattice by growing double-layer thin
  films of a transition metal oxide with perovskite structure along
  the (111) direction, or by forming a sheet of organic moleculars
  that contain transition-metal ions. The magnetic long-range order
  might be stabilized among the $d$ electrons of the transition metal.
  The QAHE can be realized in such systems if a suitable exchange
  field is established~\cite{DiXiao_NatComm_2011_LaAlO3_LaAuO3,
    KYYangPRB11, GAFietePRB11, GAFietePRB12, Fiete_PRB_2012,
    XiaoHu_NJP_2013, wang_interaction-induced_2014,
    wang_quantum_2013}.
\item Other proposals, such as heavy metal on magnetic insulator
  substrate~\cite{Vanderbilt_QAHE_2013}, strained epitaxial film with
  interface band inversion in EuO/GdN~\cite{Vanderbilt_QAHE_2014} or
  CdO/EuO~\cite{Zhang_QAHE_2014}, etc.
\end{itemize}

Most of the proposals mentioned above are supported by
first-principles calculations, and we will start this section with a
brief introduction about how to explore new topological materials from
the viewpoint of a band inversion mechanism. Then we will discuss
details of some of the proposals for the prediction and realization of
the QAHE, with no attempting to cover them all.
%
%
%
%
%
%
%
%

~\\~
\hspace*{15pt}{\bf IV.A. Band Inversion Mechanism}
\\

Though we have found, in the previous sections, that the Wilson loop
method can reduce computation effort in calculating various
topological invariants, it is still quite tedious and inefficient for
wide-range searching or design of topological materials. From this
point of view, band inversion is the most intuitive picture and most
practically useful guideline for the initial screening of possible
topological materials. Band inversion involves the energy order
switching of low energy electronic bands around a certain TRIM point
in the Brillouin Zone when compared with the energy ordering in the
atomic limit (which is topologically trivial). Band inversion may
happen when atoms form crystalline solids and acquire strong enough
band dispersions. The mechanism was first demonstrated for a HgTe/CdTe
quantum well structure~\cite{Bernevig_HgTe_QW_science_2006} and later
was found to be applicable to many other materials, including 2D and
3D
TIs~\cite{ZhangHJ_Bi2Se3_2009NP,ZhangHJ_2013_RRL,Ando_TI_materials_2013,
  ZrTe5_PRX_2014,WengHM_YbB6_2014PRL}, Chern
insulators~\cite{yu_quantized_2010}, topological Weyl
semimetals~\cite{WanXG_WeylTI_2011, XuGang_HgCrSe_2011_PRL} and
topological Dirac
semimetals~\cite{WangZJ_Na3Bi_2012,WangZJ_Cd3As2_2013}.

\begin{figure}[tbp]
\centering
  \includegraphics[width=0.95\textwidth]{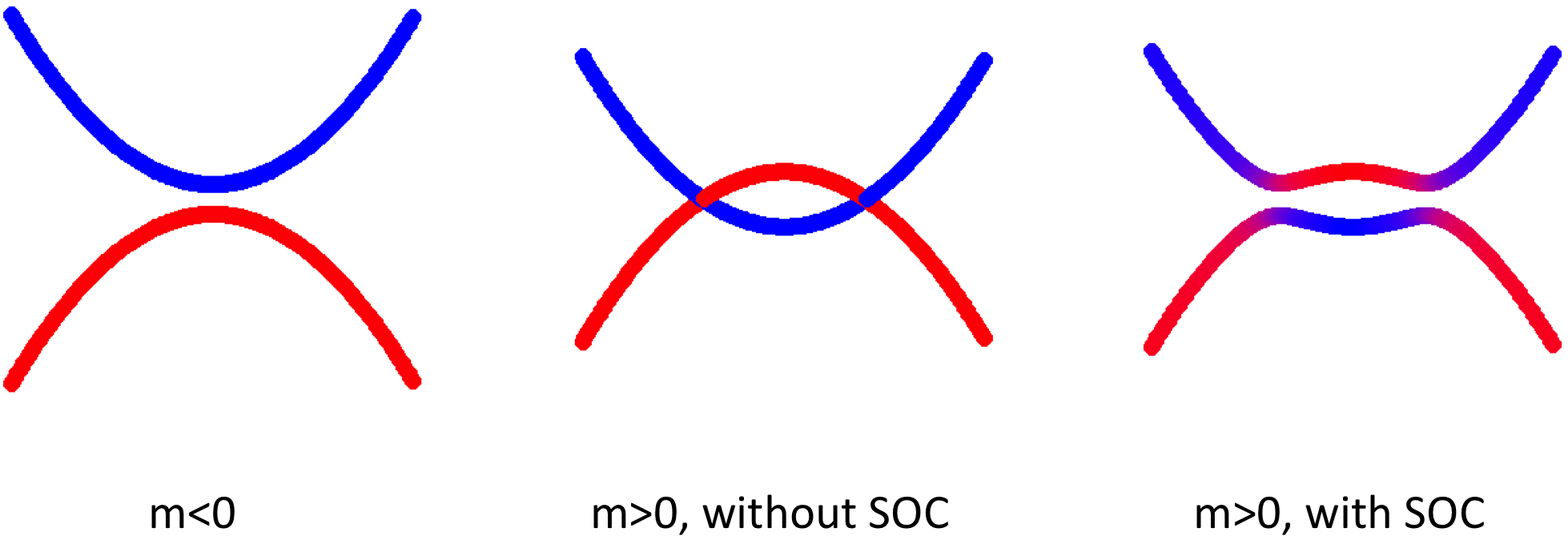}
  \caption{(Color online) Illustration of the band inversion
    mechanism. Schematic plot for the band structure with normal order
    (m$<$0), inverted order (m$>$0) without SOC, and inverted order
    with SOC, respectively. (See main text for detailed explanations).}
    \label{bandinversion}
\end{figure}

As schematically illustrated in Fig. \ref{bandinversion}, we consider
two bands close to the Fermi level around the $\Gamma$ point.  Far
away from the $\Gamma$ point, the red state is occupied and
energetically lower than the blue state. However, close to $\Gamma$
point, they have an inverted energy ordering with the blue state lower
and occupied at $\Gamma$. Let us assume that two bands do not couple
when the SOC is absent, and they must cross at certain $k$-point away
from the $\Gamma$ point. For time-reversal-invariant systems with SOC,
the spin degree of freedom has to be taken into account, and at least
four bands must be included in this simple model. The band crossing
points are, in general, not stable and will open up a gap (in the
presence of SOC). As a result, the system becomes an insulator if the
chemical potential is located within the gap, and most importantly,
this gapped state may be topologically non-trivial depending on the
coupling terms. It should be noted that if additional crystal
symmetries are considered, the four-fold degenerated band crossings
are possibly protected. This situation is demonstrated in topological
Dirac semimetals~\cite{WangZJ_Na3Bi_2012,WangZJ_Cd3As2_2013}.

To be more explicit, we consider two parabolic bands with energy
dispersion $E_{\pm}({\bf k})=\pm (-m+{\bf k}^2)$. The $+$ and $-$
signs are for the blue and red bands, respectively. The band inversion
(around ${\bf k}$=0) happens for the parameter region of $m>0$. We
first focus on a 2D system with broken TRS. Using two bands as a
basis, we can construct an effective 2$\times$2 Hamiltonian,
\begin{equation}
H_{eff}({\bf k})=\left[
\begin{array}{cc}
E_+({\bf k}) &  M({\bf k})^*  \\
 M({\bf k}) & E_-({\bf k})
\end{array}
\right],
\end{equation}
where $M({\bf k})$ is the coupling between the two bands that
determines the topological properties of the resulting electronic
structure. If $m<0$, i.e., without band inversion, we will in general
expect nothing but a normal insulator. However, if $m>0$ for the
band-inverted system, we may find many interesting topological phases,
depending on the choice of $M({\bf k})$. For example, if $M({\bf
  k})=(k_x\pm i k_y)^n$ ($n\ge 1$), we will get a Chern insulator with
Chern number $Z=n$ (for a 2D system). For a 2D system with TRS, the
two bands should be replaced by four bands with each color (blue and
red) representing a Kramers pair of bands, and the model Hamiltonian
becomes 4$\times$4 (block diagonal in its simplest form, such as the
BHZ Hamiltonian~\cite{Bernevig_HgTe_QW_science_2006}). If we now have
a form of $M({\bf k})=(k_x\pm i k_y)^{2n-1}$, we will expect a 2D
topological insulator with $Z_2$=1. In the presence of SOC, these
forms of coupling term $M({\bf k)}$ are in general allowed, and they
depend on the material's details. Similar discussions can be extended
to 3D systems, and in this case, we can expect either 3D
TIs~\cite{liu_model_2010} or 3D topological
semimetals~\cite{XuGang_HgCrSe_2011_PRL,
  WangZJ_Na3Bi_2012,WangZJ_Cd3As2_2013}. In real compounds, the band
inversion may happen between $s$-$p$, $p$-$p$, $p$-$d$, or $d$-$f$
states, which leads to different material classes of
TIs~\cite{ZhangHJ_2013_RRL}.

The band inversion mechanism discussed above is practically useful,
yet a crude picture for real compounds.  First, this picture comes
from the low energy effective Hamiltonian of a continuous model.  If
we put the model onto a lattice, we have to be careful about the
number of times band inversion occurs. Particularly for the Z$_2$ TIs,
if band inversion happens an even number of times or it happens around
an even number of TRIM points, we will return to a trivial insulator
phase (in the sense of $Z_2$=0, although other topological phases,
such as topological crystalline
insulator~\cite{FuLiang_Topological_Crystalline_2011PRL,
  Hsieh_EXP_TCI_2012}, may be defined).  Second, real compounds may
have many low energy bands that are strongly fixed and must be
carefully analyzed.  Third, the possible $M({\bf k})$ terms can also
be complicated and need to be treated carefully.  In any case, a final
conclusion must be carefully made based first-principles calculations
as presented in previous sections.

%
%
%
%
%
%
%
%
%
%
%
%
%
%

~\\~
\hspace*{15pt}{\bf IV.B. QAHE in Magnetic Topological Insulators}
\\


Several realistic systems, including Mn doped HgTe quantum wells
\cite{CXLiu_HgMnTe_2008}, magnetic-impurity doped Bi$_2$Se$_3$ thin
films \cite{yu_quantized_2010}, Mn doped InAs/GaSb quantum wells
\cite{LiuCX_QAHE_InAs-GaSb_2013}, magnetically doped junction Quantum
Wells \cite{ZhangHJ_QSH_JQW_2014_PRL}, etc, have been proposed to
support the QAHE. The essence of these proposals is to break the TRS
in a 2D slab of QSHI by magnetic doping. If the FM order can be
established, the introduced exchange splitting may destroy the band
inversion in half of the states (say, for one spin channel), while
keeping the band inversion in the other half of the states, leading to
the QAHE.

A HgTe quantum well with an appropriate thickness realized between two
CdTe barriers can achieve the 2D QSH state
\cite{Bernevig_HgTe_QW_science_2006,Molenkamp_HgTe_science_2007}. Based
on this structure, Liu et al. proposed that the QAHE can be obtained
by magnetic Mn-doping in HgTe/CdTe quantum wells
\cite{CXLiu_HgMnTe_2008}, if the doped Mn ions are ferromagnetically
ordered with magnetic momentum polarized along the out-of-plane
direction.  Unfortunately, (Hg$_{1-x}$Mn$_x$)Te is experimentally
paramagnetic (rather than FM), which prohibits the realization of
QAHE. Due to the paramagnetic nature, in practical measurement, a
magnetic field is required to polarize the Mn magnetic
moment. Therefore, it is hard to distinguish the Hall and the
anomalous Hall contributions. With the same idea, Zhang et
al. ~\cite{ZhangHJ_QSH_JQW_2014_PRL} proposed that a QSHI with bulk
band gap around 0.1 eV can be realized in junction quantum wells
comprising II-VI, III-V, or IV semiconductors, and proper magnetic
ordering can drive the system into the QAH state. Experimentally
again, we have to wait for the synthesis and the well-controlled
doping of a sample. In the following, we will take Bi$_2$Se$_3$ and
Bi$_2$Te$_3$ (well-known topological insulators), as paradigms to show
how the QAHE state is achieved in their thin film form with
Cr-doping. This is the only experimentally successful example so
far~\cite{chang_experimental_2013}.

\begin{figure}[th]
\begin{centering}
\includegraphics[clip,width=0.9\textwidth]{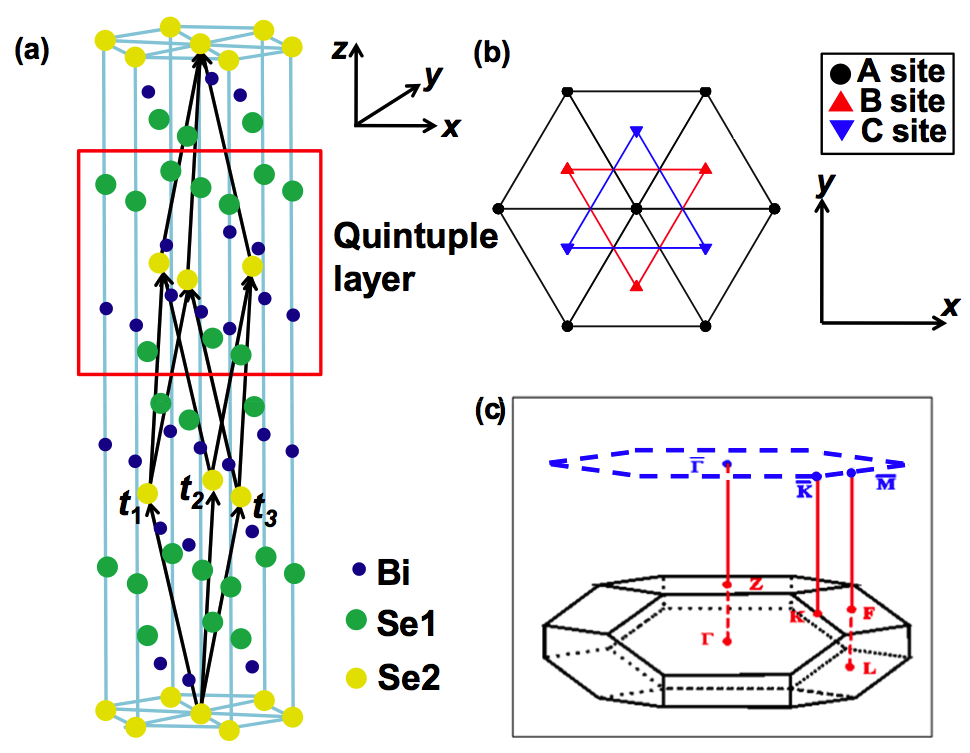}
\par\end{centering}
\caption{(Color online) Crystal structure of the Bi$_2$Se$_3$ family
  of compounds.  (a) The hexagonal supercell containing 15 atomic
  layers.  (b) Top view of a QL in the triangular lattice. Three sets
  of different sites, labeled as A, B and C sub-lattices,
  respectively, are presented.  (c) The first BZ. Four non-equivalent
  TRIM points $\Gamma$ (0, 0, 0), L($\pi$,0,0), $F$($\pi$,$\pi$,0) and
  $Z$($\pi$,$\pi$,$\pi$) are denoted in the 3D BZ. The corresponding
  surface 2D BZ is represented by the dashed hexagon, and
  $\bar{\Gamma}$, $\bar{M}$ and $\bar{K}$ are the corresponding TRIM
  special k points in the surface BZ.  Figure from
  Ref.~\onlinecite{ZhangWei_NJP_BiSe_2010}.}
\label{fig:Bi2Se3_crystal_structure}
\end{figure}

The Bi$_{2}$Se$_{3}$ family of compounds have a rhombohedral crystal
structure with space group $D^5_{3d}$ (R$\bar{3}$m).  The system has a
layered structure with five atomic layers as a basic unit, named a
quintuple layer (QL), as shown in
Fig. \ref{fig:Bi2Se3_crystal_structure}. The materials
Bi$_{2}$Se$_{3}$, Bi$_{2}$Te$_{3}$ and Sb$_{2}$Te$_{3}$ have been
theoretically predicted and experimentally observed to be topological
insulators with a large bulk band gap~\cite{ TI_exp_Yazdani_2009,
  TI_exp_XueQK_2009, TI_exp_Hasan_2009,
  chen_experimental_2009,TI_exp_SZX_2010, Hasan_Kane_RMP_2010,
  QiXL_RMP_2011, DuRuiRui_InAs_GaSb_2011, TI_exp_Yazdani_2_2011,
  ZhangHJ_Bi2Se3_2009NP}.  Recent experimental progress has shown that
well-controlled layer-by-layer MBE thin film growth can be
achieved~\cite{zhang_crossover_2010,TI_exp_XueQK_2009}, and various
magnetic transition metal elements (such as Ti, V, Cr, Fe) can be
substituted into the Bi$_{2}$Se$_{3}$ family of parent compounds with
observable ferromagnetism even above
100K~\cite{kulbachinskii_low_temperature_2001,dyck_diluted_2002,Chien_Transition_2007,niu_quantum_2011,zhou_thin_2006}.

We first discuss how to achieve a FM insulating phase in a
semiconductor doped with dilute magnetic ions.  To simplify the
discussion, we divide the whole system into two sub-systems describing
the local moments and the other band electrons, respectively. We
further assume that the magnetic exchange among local moments is
mediated by the band electrons.  If we consider only the spatially
homogeneous phase, the total free energy of the system in an external
magnetic field $H$ can be written as
\begin{equation}
F_{total}
= \frac{1}{2}\chi_{L}^{-1}M_{L}^{2}+\frac{1}{2}\chi_{e}^{-1}M_{e}^{2}
- J_{eff}M_{L}M_{e}-\left(M_{L}+M_{e}\right)H,
\end{equation}
where $\chi_{L/e}$ is the spin susceptibility of the local
moments/electrons, $M_{L/e}$ denotes the magnetization for the local
moment and electron sub-system and $J_{eff}$ is the magnetic exchange
coupling between them.  The spontaneous FM phase can be realized when
the minimization procedure of the free energy gives a non-zero
magnetization in the zero magnetic field limit ($H$=0). This leads to
the requirement of $J_{eff}^{2}-\chi_{L}^{-1}\chi_{e}^{-1}>0$, or
equivalently $\chi_{e}>1/(J_{eff}^{2}\chi_{L})$.  This result means
that in order to have a non-zero FM transition temperature ($T_{c}$),
a sizable $\chi_{e}$ is needed.  Within linear response theory, the
spin susceptibility $\chi_{e}$ for a insulator is given by the van
Vleck formula
\begin{equation}
  \chi_{e}^{zz}=\sum_{v,c,k}4\mu_{0}\mu_{B}^{2}\frac{\left\langle
      vk\right|\hat{S_{z}}\left|ck\right\rangle \left\langle
      ck\right|\hat{S_{z}}\left|vk\right\rangle }{E_{ck}-E_{vk}} ,
\end{equation}
where $\mu_{0}$ is the vacuum permeability, $\mu_{B}$ is the Bohr
magneton, $|ck\rangle$ and $|vk\rangle$ are the Bloch functions in the
conduction and valence bands respectively, $E_{ck}$ and $E_{vk}$ are
the eigen energies of $|ck\rangle$ and $|vk\rangle$, and $\hat{S_{z}}$
is the spin operator of electrons and we consider only the
$z$-direction, i.e. perpendicular to the 2D plane, for
simplicity.

\begin{figure}[ht]
\begin{centering}
\includegraphics[angle=-90,width=0.5\textwidth]{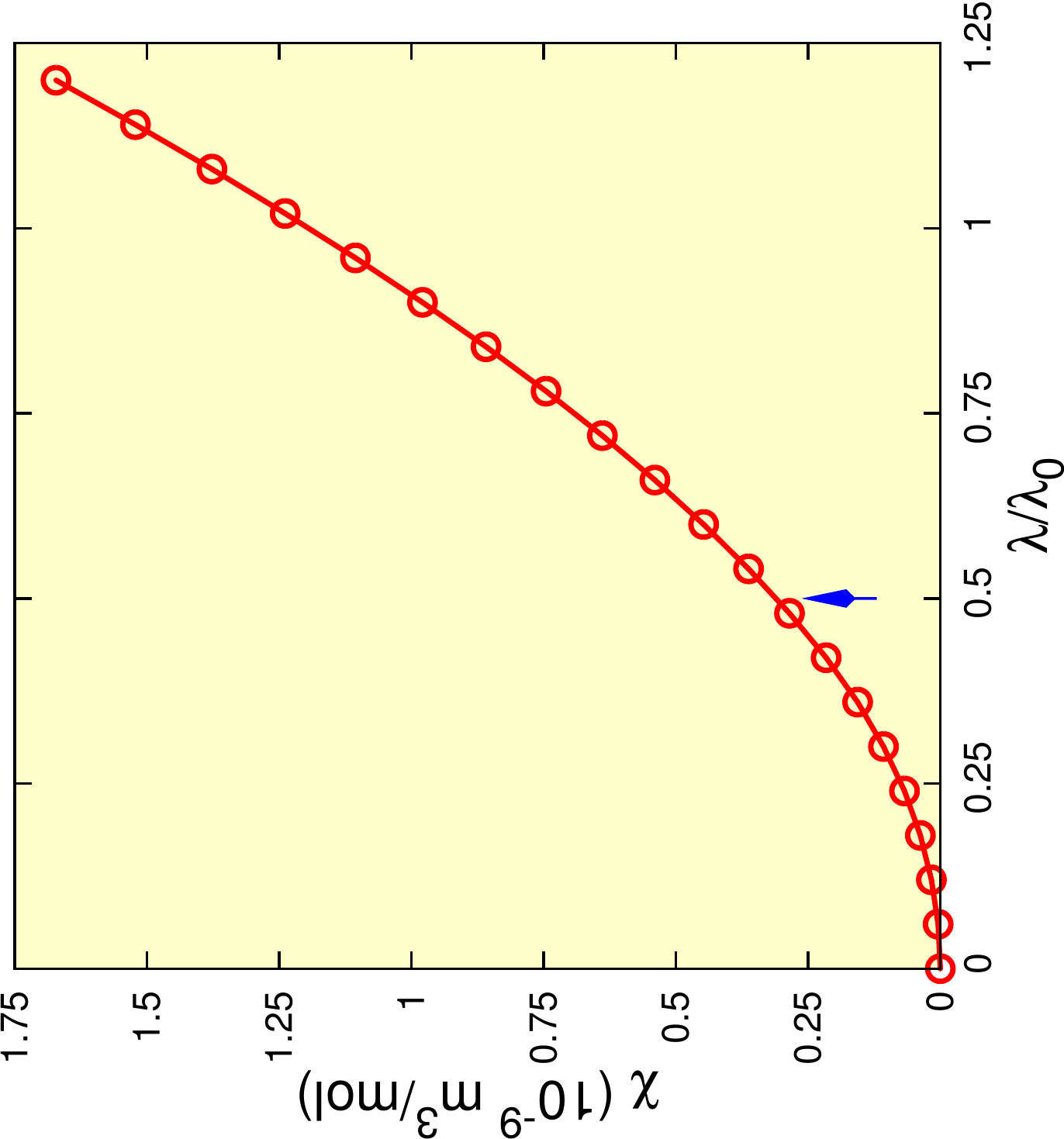}
\par\end{centering}
\caption{\label{fig:bise_spin_sus} Van-Vleck type spin
  susceptibility of Bi$_{2}$Se$_{3}$ bulk as a function of
  spin-orbit coupling (SOC) strength ($\lambda_{0}$ is the actual
  SOC strength). Figure from Ref.~\onlinecite{yu_quantized_2010}. }
\end{figure}

In most dilute magnetic semiconductors, such as
(Ga$_{1-x}$Mn$_{x}$)As, which has $s$-like conduction and $p$-like
valence bands, the matrix elements $\left\langle
  ck\right|\hat{S_{z}}\left|vk\right\rangle$ are very small.  The
electronic spin susceptibility is therefore negligible for the
insulating phase, and a finite carrier density is required to mediate
the magnetic coupling (for example, by the RKKY
mechanism~\cite{RKKY_1954}).  For the Bi$_{2}$Se$_{3}$ family of
materials, however, the semiconductor gap is opened by the SOC between
two $p$-orbital bands. The spin operator $S_{z}$ has finite matrix
elements between the valence and conduction bands, which are further
enhanced due to the band inversion~\cite{yu_quantized_2010}.  In
Fig. \ref{fig:bise_spin_sus}, we show the calculated spin
susceptibility of bulk Bi$_{2}$Se$_{3}$.  At the $\Gamma$ point, the
band inversion occurs when the relative SOC strength
$\lambda/\lambda_{0}$ exceeds 0.5, and after that the spin
susceptibility starts to increase rapidly. At the actual SOC strength,
Bi$_{2}$Se$_{3}$-family materials have a considerable spin
susceptibility $\chi_{e}^{zz}$.  Hence the local moments of the
magnetic impurities in such materials can be ferromagnetically coupled
through the van Vleck
mechanism~\cite{Vleck_Susceptibilities_1932}. Recently the interplay
between magnetism and the topological property in a material has been
discussed in more details in Ref.~\onlinecite{Tokura_Dirac-fermion-mediated_ferromagnetism_in_TI_2012,
  zhang_topology-driven_2013}.

\begin{figure}[th]
\begin{centering}
\includegraphics[angle=-90,width=0.95\textwidth]{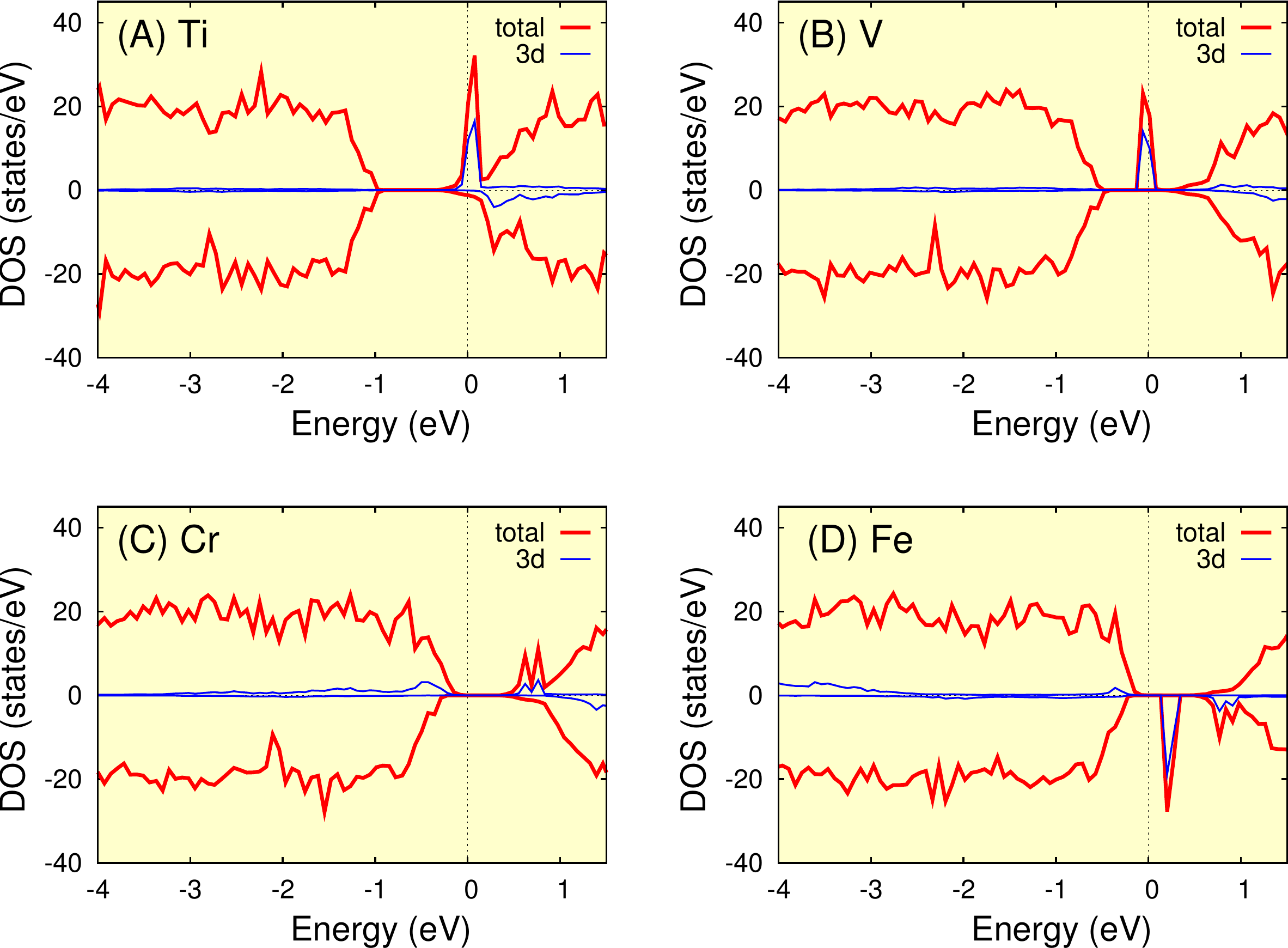}
\par\end{centering}

\caption{\label{fig:dos} Calculated densities of states (DOS) for
  Bi$_{2}$Se$_{3}$ doped with different transition metal elements. The
  Fermi level is located at energy zero, and the positive and negative
  values of DOS are used for up and down spin, respectively. The lines
  are projected partial DOS of the $3d$ states of transition metal
  ions. It is shown that Cr or Fe doping will give rise to the
  insulating magnetic state. Figure from
  Ref.~\onlinecite{yu_quantized_2010}. }
\end{figure}

\begin{figure}[bh]
\begin{centering}
\includegraphics[width=0.95\textwidth]{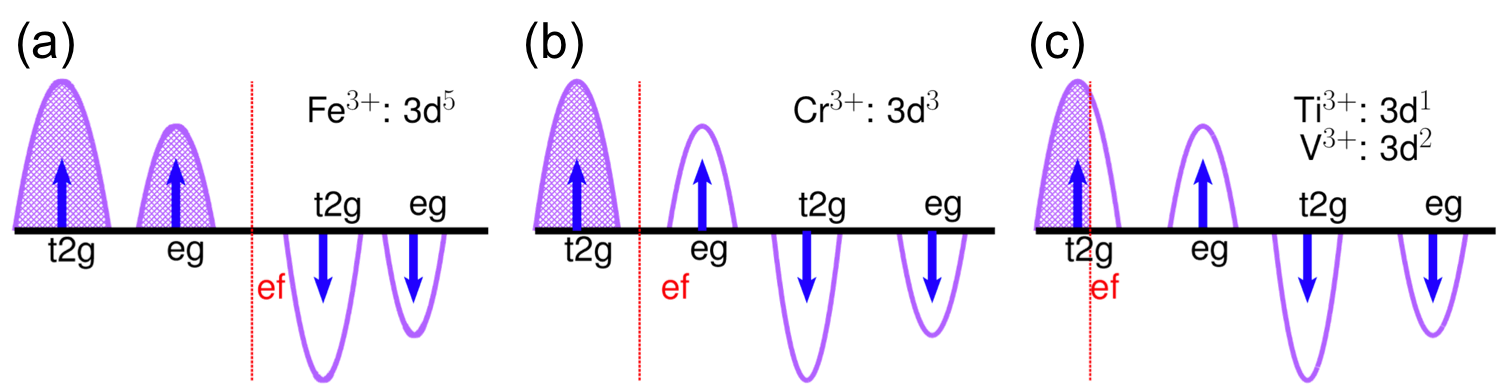}
\par\end{centering}
\caption{\label{fig:d_electrons}The configuration of the d electrons for the (a) Fe, (b) Cr, (c) Ti and V doping cases.}
\end{figure}

From first-principles calculations, we find that the insulating
magnetic ground state discussed above can indeed be obtained by a
proper choice of magnetic dopants.  Experiments
\cite{Chien_Transition_2007,kulbachinskii_low_temperature_2001,
  dyck_diluted_2002} and theoretical
calculations~\cite{yu_quantized_2010, zhang_tailoring_2012} show that
the magnetic dopants will mostly substitute for Bi ions, which has a
nominal 3+ valence state.  In order to avoid introducing free carriers
into the parent material when we dope the system, it is natural to
choose those transition metal elements that also have a stable 3+
chemical state, such as Ti, V, Cr and Fe.  Self-consistent
first-principles calculations for Bi$_{2}$Se$_{3}$ doped with Ti, V,
Cr and Fe have been performed. The calculated densities of states
(DOS) shown in Fig. \ref{fig:dos} suggest that an insulating magnetic
state is obtained only for Cr or Fe doping, while the states are
metallic for Ti or V doping cases.

The above results can be understood as illustrated in
Fig. \ref{fig:d_electrons}.  We first point out that, for all the
cases, the dopants are nearly in the 3+ valence state, and we always
obtain the high-spin state due to the large Hund's coupling of $3d$
transition metal ions. For the Fe-doped case, the Fe$^{3+}$ favors the
$d^{5\uparrow}d^{0\downarrow}$ configuration in a high-spin state,
resulting in a gap between the majority and minority spins, which
directly leads to the insulating state in Fe-doped samples as shown in
Fig. \ref{fig:d_electrons}a.  For the Cr-doped case, Cr ions
substitute in the Bi sites and feel an octahedral crystal field formed
by the six nearest neighboring Se$^{2-}$ ions.  Such a local crystal
field splits the $d$-shell into $t_{2g}$ and $e_{g}$ manifolds.  This
splitting is large enough to stabilize the
$t_{2g}^{3\uparrow}e_{g}^{0\uparrow}t_{2g}^{0\downarrow}e_{g}^{0\downarrow}$
configuration of a Cr$^{3+}$ ion, resulting in a gap between the
$t_{2g}$ and $e_{g}$ manifolds, as shown in
Fig. \ref{fig:d_electrons}(b).  For the case of Ti or V doping, the
$t_{2g}$ manifold is partially occupied, leading to the metallic
state, as shown in Fig. \ref{fig:d_electrons}(c).  We note that,
although the local density approximation in density functional theory
may underestimate the electron correlation effects, the inclusion of
electron-electron interaction $U$ (such as in the LDA+$U$ method)
should further enhance the gap (it may also reduce the $p$-$d$
hybridization and $T_{c}$~\cite{sato_magnetic_2003}).

The exchange field $\Delta$ can be estimated as
$\Delta=xJ_{eff}\langle S\rangle$ at the mean field level, where $x$
is the doping concentration and $\langle S\rangle$ is the mean field
expectation value of the local spin, $J_{eff}$ is the effective
exchange coupling between the local moments and the band electrons,
which is estimated to be around 2.7 eV for Cr-doping and 2.8 eV for
Fe-doping in Bi$_{2}$Se$_{3}$ by first-principles calculations.  With
the concentration of the magnetic dopants at 10\%, the FM Curie
temperature can be estimated in the mean field
approximation~\cite{Dietl_2001PRB} to reach the order of tens of K.

\begin{figure}[t]
\begin{centering}
\includegraphics[width=0.9\textwidth]{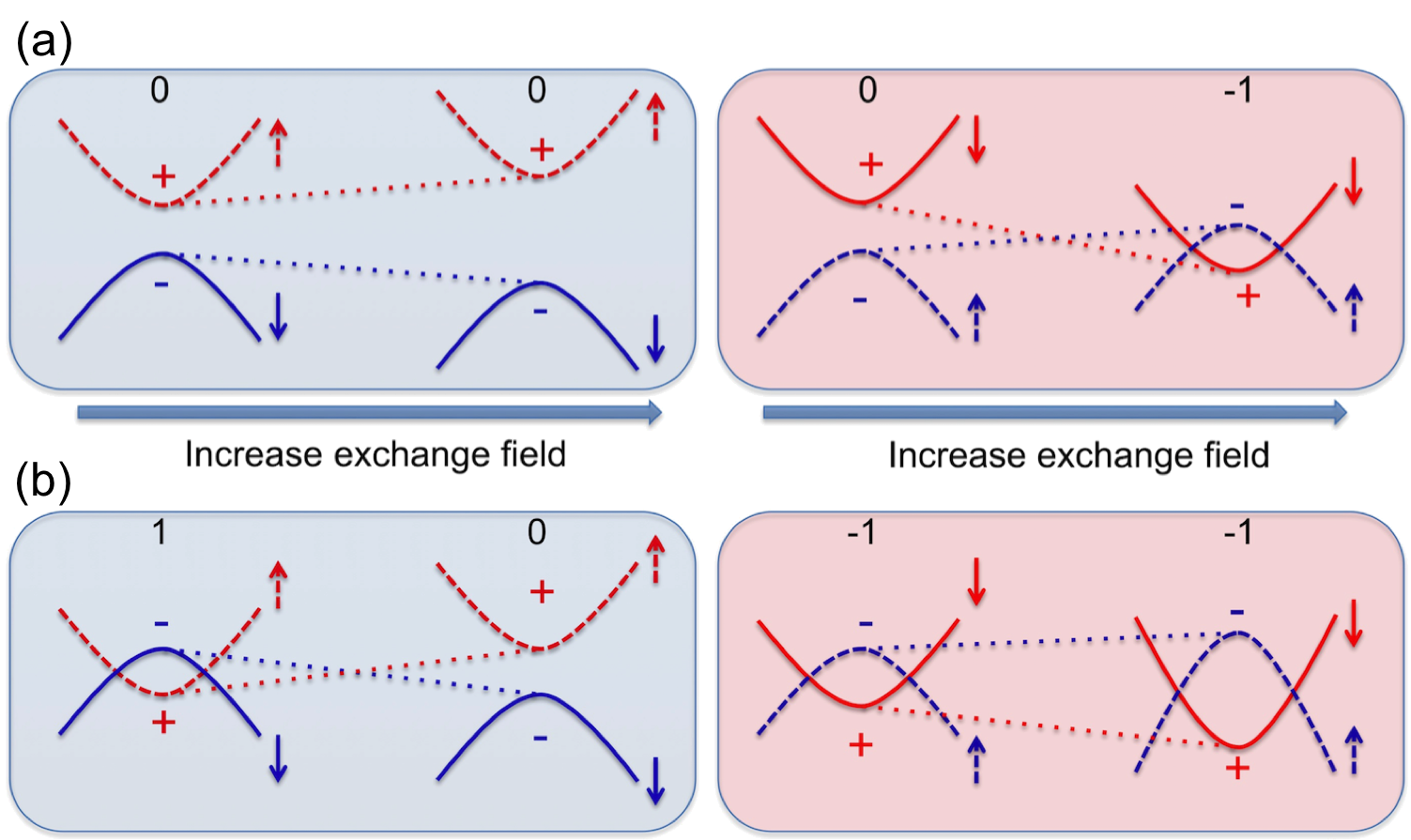}
\par\end{centering}
\caption{\label{fig:chemata} Evolution of the subbands structure upon
  increasing the exchange field. The red color denotes the subbands
  that have even parity at the $\Gamma$ point, and blue color denotes
  subbands with odd parity at the $\Gamma$ point. The dashed lines
  denote the spin up electrons; solid lines, spin down electrons. (a)
  The initial subbands are not inverted. When the exchange field is
  strong enough, a pair of inverted subbands appears (red solid line
  and blue dashed line). (b) The initial subbands are already
  inverted. The exchange field releases the band inversion in one pair
  of subbands (red dashed line and blue solid line) and increases the
  band inversion in the other pair (red solid line and blue dashed
  line).}
\end{figure}

Once the FM order is achieved in TI at a reasonable temperature, the
QAHE can in principle be realized in 2D thin films of such systems by
tuning the exchange splitting (i.e, the density of magnetic doping).
Since the bulk states are always gapped, we focus on the simplest low
energy effective Hamiltonian consisting of Dirac-type surface states
only:
\begin{eqnarray}
H & = & H_{sf}+H_{Zeeman}\nonumber \\
  & = & \left[\begin{array}{cc}
-v_{F}(k\times\mathbf{\sigma})_{z} & m_{k}^{*}\\
m_{k} & v_{F}(k\times\mathbf{\sigma})_{z}
\end{array}\right]
    +  \left[\begin{array}{cc}
gM\mathbf{\sigma}_{z} & 0\\
0 & gM\mathbf{\sigma}_{z}
\end{array}\right],
\label{eq:ham1}
\end{eqnarray}
with the basis of $\left|t\uparrow\right\rangle
,\left|t\downarrow\right\rangle ,\left|b\uparrow\right\rangle
,\left|b\downarrow\right\rangle $, where ``$t$'' (``$b$'') represents
the surface states on the top (bottom) surface, and ``$\uparrow$'',
``$\downarrow$'' represent the spin up and down states,
respectively. $v_{F}$ is the Fermi velocity, $m_{k}$ describes the
tunneling effect between the top and bottom surface states, $g$ is the
effective g-factor, $\mathbf{\sigma}$ are the Pauli matrices, and $M$
represents the exchange field along the $z$-direction, introduced by
the FM ordering.  For simplicity, spatial inversion symmetry is
assumed, which requires that $v_{F}$, $g$ and $M$ take the same values
for top and bottom surfaces.  In thick enough slab geometry
($m_{k}\approx0$), the spatially separated pairs of surface states are
well defined for the top and bottom surfaces. However, with the
reduction of the film thickness, quantum tunneling between the top and
bottom surfaces becomes more and more pronounced, giving rise to a
finite mass term $m_{k}$, which can be expanded as
$m_{k}=m_{0}+B(k_{x}^{2}+k_{y}^{2})$ up to the second order of $k$.
The Hamiltonian can be rewritten in terms of the symmetric and
anti-symmetric combination of the surface states on top and bottom
surfaces as
\begin{eqnarray}
H =  \tilde{H}_{sf}+\tilde{H}_{Zeeman}
  =  \left[\begin{array}{cc}
h_{k}+gM\sigma_{z} & 0\\
0 & h_{k}^{*}-gM\sigma_{z}
\end{array}\right],\label{eq:ham2}
\end{eqnarray}
with the following new basis: $\left|+\uparrow\right\rangle $,
$\left|-\downarrow\right\rangle $, $\left|+\downarrow\right\rangle $,
$\left|-\uparrow\right\rangle $, where
$\left|\pm\uparrow\right\rangle=$ $(\left|t\uparrow\right\rangle$
$\pm\left|b\uparrow\right\rangle)/\sqrt{2}$,
$\left|\pm\downarrow\right\rangle =$
$\left(\left|t\downarrow\right\rangle
  \pm\left|b\downarrow\right\rangle \right)/\sqrt{2}$.  Here
$h(k)=m_{k}\sigma_{z}+v_{F}\left(k_{y}\sigma_{x}-k_{x}\sigma_{y}\right)$,
which is similar to the BHZ model describing the low energy physics in
HgTe/CdTe quantum wells~\cite{Bernevig_HgTe_QW_science_2006}. When
$m_{0}B<0$, band inversion occurs, and the system will be in the QSH
phase if this is the only band inversion between two subbands with
opposite parity.  Regardless of whether this condition is satisfied or
not, a strong enough exchange field will induce the QAH effect in this
system, thanks to the presence of the $\sigma_{z}$ matrix in the
exchange field ($gM\sigma_{z}$) and the opposite signs of the Zeeman
coupling terms in the upper and lower blocks of the effective
Hamiltonian in Eq.~(\ref{eq:ham2}).  The exchange field breaks the TRS
and increases the mass term of the upper block, while reducing the
mass term for the lower block. More importantly, a sufficiently large
exchange field can change the Chern number of one of the two blocks.
There are two cases, as illustrated in Fig. \ref{fig:chemata}. In the
first case, the four-band system is originally in the topologically
trivial phase; the exchange field will induce band inversion in the
lower block and push the two sub-bands in the upper block even farther
away from each other.  Therefore, the 2D model with a negative mass in
the lower block contributes $-e^{2}/h$ for the Hall conductance (see
Fig. \ref{fig:chemata}(a)). In the other case, the system is
originally in the topologically non-trivial phase, and both blocks
have inverted band structures. A sufficiently large exchange field can
increase the band inversion in the lower block and release the band
inversion in the upper block.  Again the negative mass in the lower
block contributes $-e^{2}/h$ for the Hall conductance (see
Fig. \ref{fig:chemata}(b)).  Such a mechanism is general for the thin
films of TIs with FM ordering, it is guaranteed by the fact that the
surface states on the top and bottom surfaces have the same
$g$-factor.

\begin{figure}[h]
\begin{centering}
\includegraphics[angle=-90,width=0.95\textwidth]{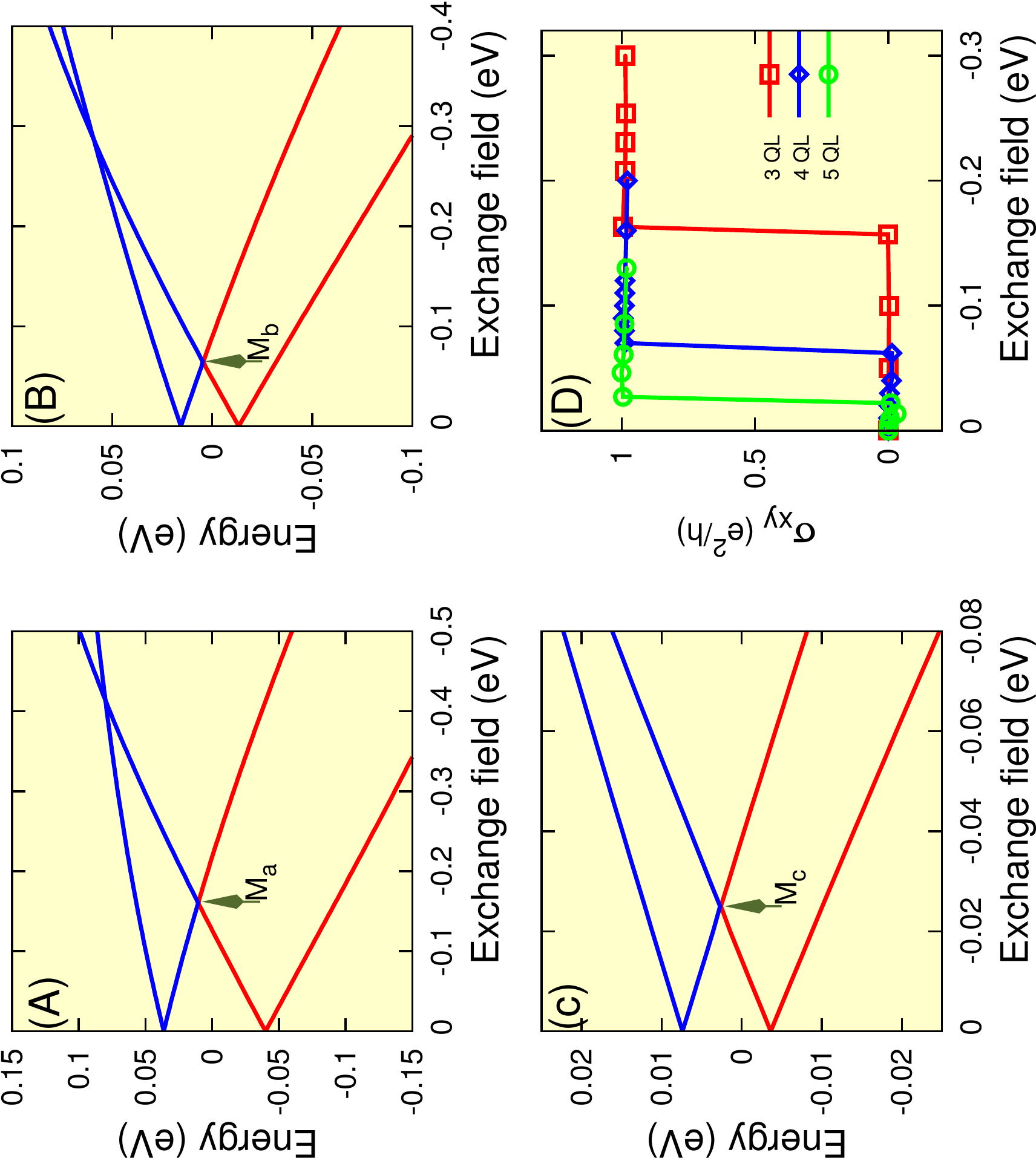}
\par\end{centering}

\caption{\label{fig:bise_m_gap} Quantized anomalous Hall (QAH)
  conductance. {(A), (B)} and (C) are the lowest sub-bands at the
  $\Gamma$ point, plotted versus the exchange field, for
  Bi$_{2}$Se$_{3}$ films with thicknesses of 3, 4 and 5 quintuple
  layers (QL), respectively. The red lines denote the top occupied
  (bottom unoccupied) states. With an increase of the exchange field,
  a level crossing {occurs (indicated by arrows)}, indicating a
  quantum phase transition to QAH state. (D) Calculated Hall
  conductance for 3, 4 and 5 QL Bi$_{2}$Se$_{3}$ films under
  ferromagnetic exchange field. As we expect, the Hall conductance is
  zero before the QAH transition, but $\sigma_{xy}=e^{2}/h$
  afterward. Figure from Ref.~\onlinecite{yu_quantized_2010}. }
\end{figure}

\begin{figure}[h]
\begin{centering}
\includegraphics[width=0.95\textwidth]{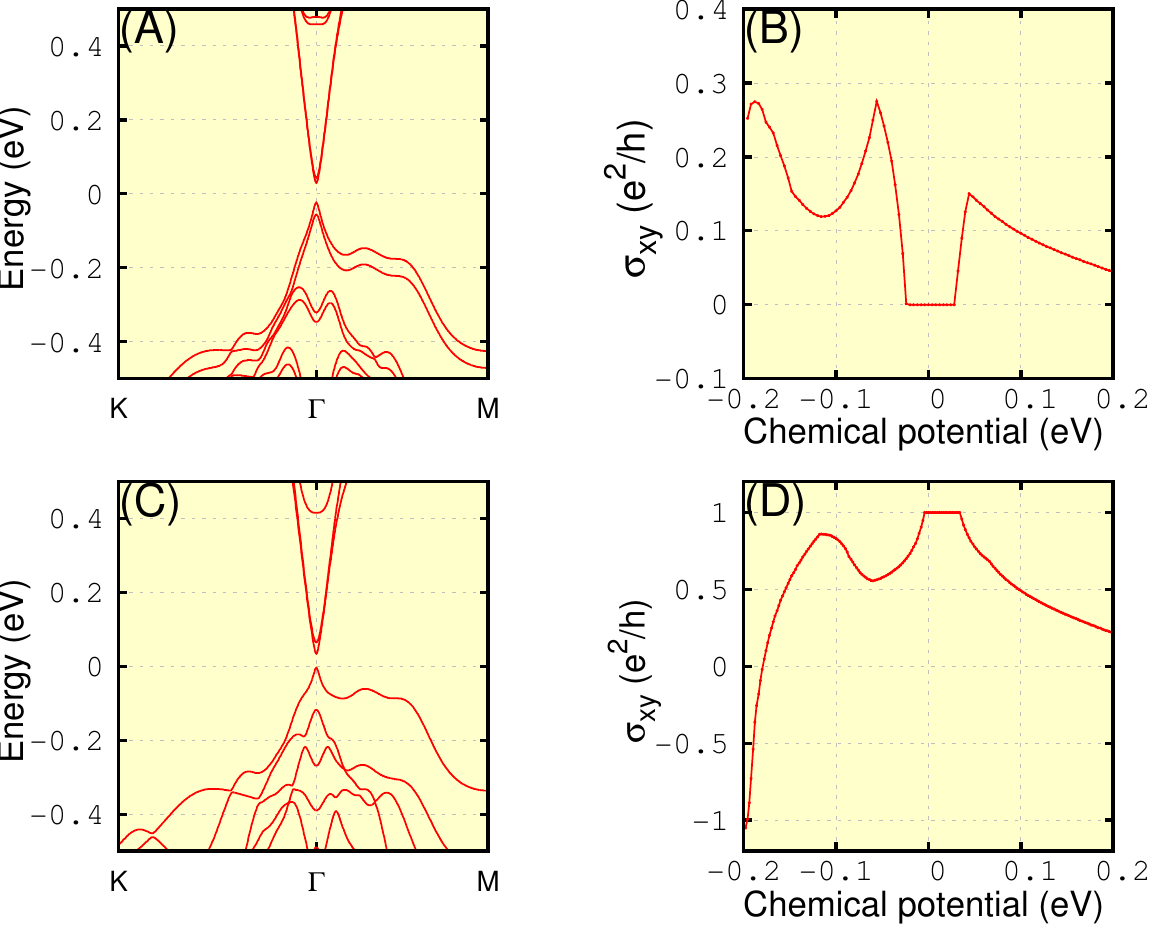}
\par\end{centering}

\caption{\label{fig:bise_hall_Pl...}The sub-bands dispersion of 2D
  Bi$_{2}$ Se$_{3}$ thin film are plotted for 3QL slab with exchange
  field (A) -0.05 eV, (C) -0.24 eV.  The corresponding Hall
  conductance are plotted in the right panels as the function of the
  chemical potential. The zero chemical potential corresponds to the
  ideal stoichiometry case. The systems in (A) is normal insulator,
  which has zero platforms in Hall conductance when the chemical
  potential locates in the energy gap. The system in (C) is QAH
  insulators, in which the Hall conductance has quantized value
  $e^{2}/h$ . Figure from Ref.~\onlinecite{yu_quantized_2010}. }
\end{figure}

In Fig. \ref{fig:bise_m_gap} A-C, we plot the lowest four sub-band
levels at the $\Gamma$ point as functions of exchange field. Level
crossings between the lowest conduction bands (blue lines) and valence
bands (red lines) are found for all three values of layer thickness,
signaling the quantum phase transition to the QAH state.  In the
insulator case, where the chemical potential is located inside the
energy gap between the conduction and valence sub-bands, the Hall
conductance is determined by the Chern number of the occupied bands
and must be an exact integer multiple of $e^{2}/h$.  The calculated
Hall conductance for Bi$_{2}$Se$_{3}$ thin films of three typical
thicknesses (Fig. \ref{fig:bise_m_gap} D) jumps from 0 to 1 at the
corresponding critical exchange field for the level crossing. The QAH
effect with higher Hall conductance plateaus in the Cr-doped
Bi$_{2}$(Se,Te)$_{3}$ was further discussed in
Ref. \onlinecite{QAHE_higher_Plateaus}, where the stronger exchange field
leads to more band crossings and a larger Chern number.

In Fig. \ref{fig:bise_hall_Pl...}, we plot the sub-band dispersion and
corresponding Hall conductance as functions of chemical potential for
3 QL Bi$_{2}$Se$_{3}$ thin films with different exchange fields.  A
quantized plateau in Hall conductance should be observed when the
chemical potential is inside the gap.  For the QAH phases
(Fig. \ref{fig:bise_hall_Pl...} (D)), {a} non-zero integer plateau is
observed when the chemical potential falls into the energy gap; while
for the non-QAH phases (Fig. \ref{fig:bise_hall_Pl...}  (B)), the
corresponding Hall conductance is zero.  In a recent experiment,
Chang, et al. \cite{chang_experimental_2013} prepared the Cr-doped
(Bi,Sb)$_{2}$Te$_{3}$ thin films with well controlled chemical
potential and long-range ferromagnetic order.  Careful control of the
Bi and Sb concentration is needed to separate the surface states from
the bulk states.  By tuning the gate voltage (and therefore, the
chemical potential), they observed the quantization of the Hall
resistance at $h/e^{2}$ at zero external magnetic filed with
temperature around 30 mK, accompanied by a considerable drop in the
longitudinal resistance.  A similar experiment was further performed
by Checkelsky, et al.~\cite{Tokura_2014NatPhys}, confirming the
existence of QAHE in this system. For readers who want to know more
experimental details, please refer to their original papers. To
conclude this part, we finally point out that the QAHE can be induced
even by in-plane magnetization, as discussed in
Ref.~\onlinecite{liu_inplane_2013}.


%
%
%
%
%
%
%
%
%
%
%
%
%


~\\~
\hspace*{15pt}{\bf IV.C. QAHE in thin film of Weyl semimetals}
\\


As discussed in sec. II.C, Weyl semimetal is a new
topological state of 3D quantum matter, unlike the 3D topological
insulators.  It can be characterized by Weyl nodes at the Fermi level
in the bulk and Fermi arcs on surfaces.  Around Weyl nodes, the
low-energy physical behavior is like that of 3D two-component Weyl
fermions~\cite{weyl_elektron_1929} and can be described by the
Hamiltonian $H=v_f \bm{\sigma}\cdot\bf{k}$, where $v_f$ is the Fermi
velocity, ${\bm\sigma}$ is the vector of the Pauli matrices, and
$\bf{k}$ is the momentum as measured from the Weyl node. A Weyl
fermion is half of a Dirac fermion, and Weyl fermions must appear in
pairs.  The Weyl Hamiltonian is robust against perturbations since it
uses all three of the Pauli matrices.  The perturbations only shift
the position of the gap-closing point in momentum space and the Weyl
nodes can be removed only when a pair of them meet in momentum
space. To obtain a Weyl semimetal, either time-reversal symmetry or
inversion symmetry must be broken, and there have been several
proposals to realize this~\cite{WanXG_WeylTI_2011, Burkov_Weyl_2011,
  XuGang_HgCrSe_2011_PRL, multilayerTRI, WS_Vanderbilt_2014, SeTe,
  TaAs_Weng, TaAs_Hassan}. In particular, Weyl semimetals are closely
related to the chiral anomaly~\cite{nielsen_adler-bell-jackiw_1983},
and the QAHE can naturally be expected in the quantum well structure
of Weyl semimetals with broken TRS. In this section, based on
first-principles calculations, we will show that HgCr$_2$Se$_4$ is a
FM Weyl semimetal with Fermi arcs on its surface, and its 2D thin film
is a promising system for realizing the QAHE.

\begin{figure}[ht]
\begin{centering}
\includegraphics[clip,width=0.6\textwidth]{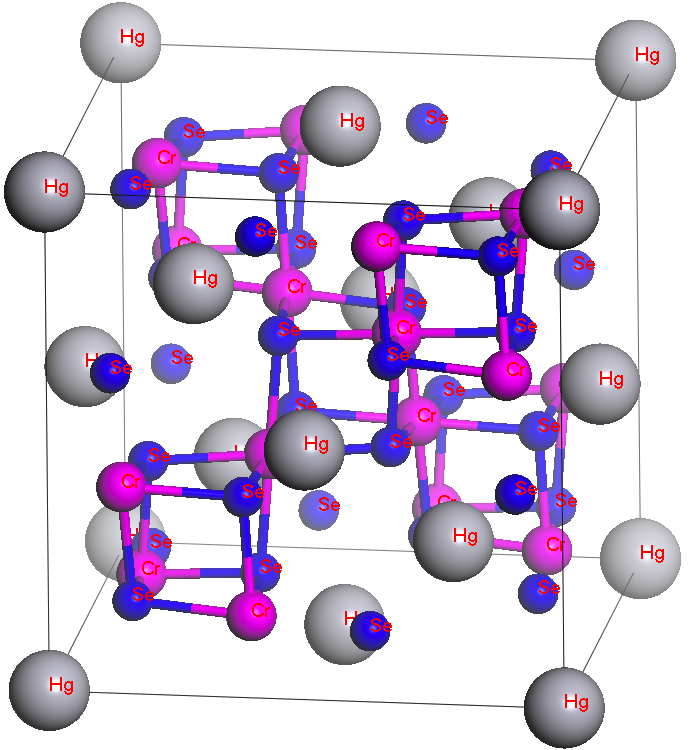}
\par\end{centering}
\caption{(Color online) Crystal structure of HgCr$_2$Se$_4$ spinels. }
\label{fig:HgCr2Se4_Crystal}
\end{figure}

\begin{figure}[hb]
\begin{centering}
\includegraphics[clip,width=0.95\textwidth]{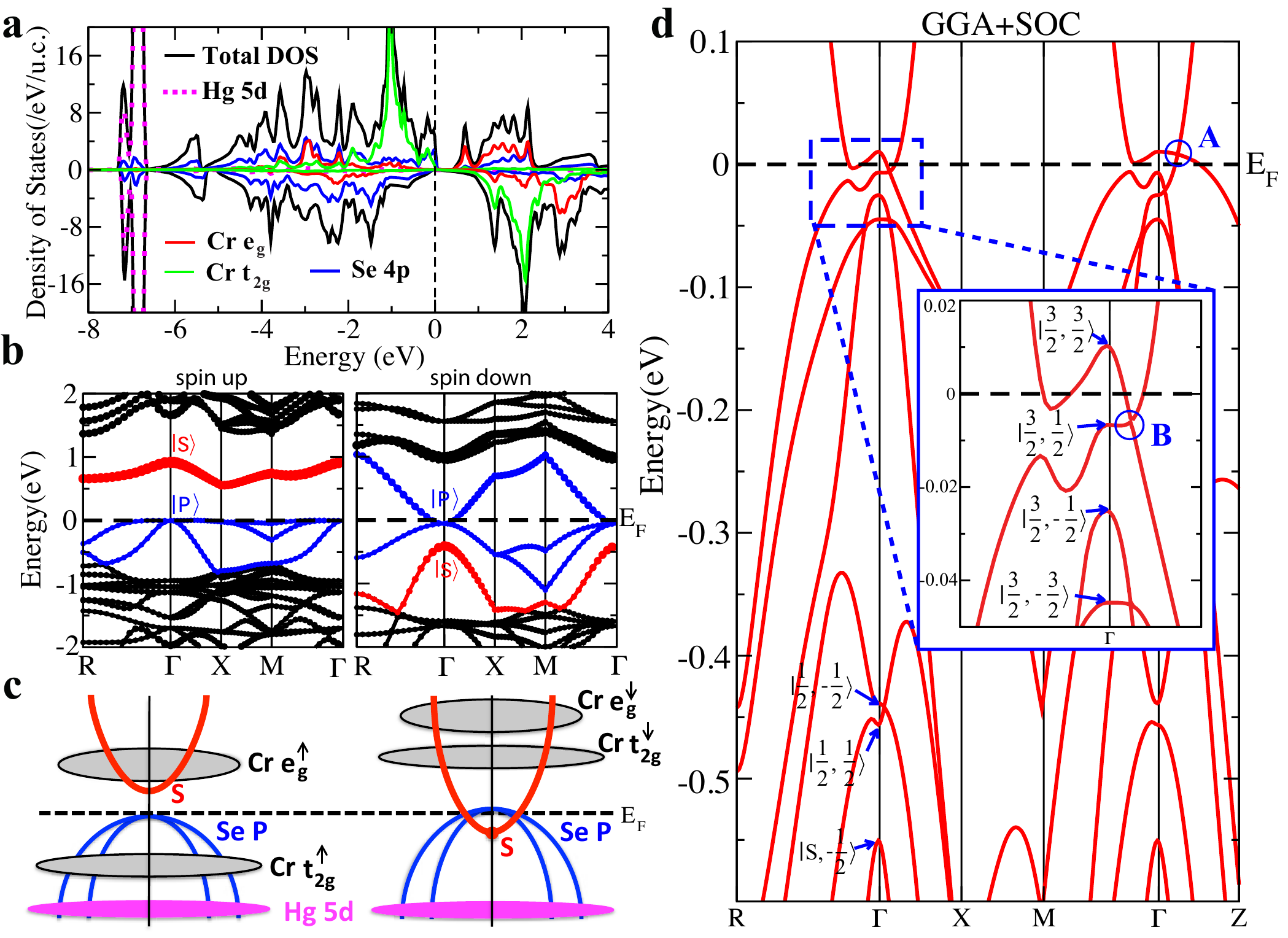}
\par\end{centering}
\caption{(Color online) Electronic structures of HgCr$_2$Se$_4$.  (a)
  The total and projected partial density of states (DOS); (b) The
  band structures without SOC (showing the up and down spin parts
  separately); (c) Schematic plot of band-inversion, where the
  $|S,\downarrow \rangle$ state is lower than the $|P\rangle$ states;
  (d) The band structure including SOC (with majority spin along the
  (001) direction). The detail around $\Gamma$ is enlarged in
  inset. Figure from Ref.~\onlinecite{XuGang_HgCrSe_2011_PRL}. }
\label{fig:HgCrSe_band_structure}
\end{figure}

HgCr$_2$Se$_4$ is a FM spinel exhibiting large coupling effects
between electronic and magnetic properties~\cite{HgCrSe-rev}.  The
spinel structure (space group Fd$\bar{3}$m) of HgCr$_2$Se$_4$ with
Cr$^{3+}$ octahedrally surrounded by Se ions, as shown in
Fig. \ref{fig:HgCr2Se4_Crystal}, yields a half-filled $t_{2g}$ shell
and a spin number of S=3/2.  First-principles calculations
\cite{XuGang_HgCrSe_2011_PRL} confirm that FM solution is considerably
(2.8 eV/f.u.) more stable than non-magnetic solution, and the
calculated moment (6.0 $\mu_B$/f.u.) is in good agreement with
experiments~\cite{HgCrSe-Tc,HgCrSe-M}.  Without SOC, the electronic
structures shown in Fig. \ref{fig:HgCrSe_band_structure}(a) and (b)
suggest that it is nearly a ``zero-gap half-metal''.  It is almost a
half-metal because a gap exists only in the up-spin channel just above
the Fermi level; it is nearly zero-gapped because of the band-touching
around the $\Gamma$ point just below Fermi level in the down-spin
channel.  Quite large exchange splitting and strong octahedral crystal
field together lead to a high spin state of Cr$^{3+}$ ions in a
$t_{2g}^{3\uparrow}e_g^{0\uparrow}t_{2g}^{0\downarrow}e_g^{0\downarrow}$
configuration with a gap between the $t_{2g}^{3\uparrow}$ and
$e_g^{0\uparrow}$ manifolds as schematically shown in
Fig.~\ref{fig:HgCrSe_band_structure}(c).  The Se-$4p$ states (located
from about -6 eV to 0 eV) are almost fully occupied and contribute to
the top part of the valence band dominantly. The hybridization with
Cr-$3d$ states makes the Se-$4p$ slightly spin-polarized with the
opposite moment (about -0.08 $\mu_B$/Se). The zero-gap behavior in the
down spin channel is the most important characteristic here
(Fig. \ref{fig:HgCrSe_band_structure}(b)), because it suggests the
inverted band structure around $\Gamma$, similar to the case in HgSe
or HgTe~\cite{HgTe-1,HgTe-2}.

Taking the four (eight if spin is considered) low energy states at the
$\Gamma$ point as basis, which are identified as $|P_x\rangle$,
$|P_y\rangle$, $|P_z\rangle$, and $|S\rangle$, the low energy physics
becomes the same as in HgSe or HgTe, and the only difference is the
presence of exchange splitting.  The band inversion (see
Fig. \ref{fig:HgCrSe_band_structure}(c); $|S,\downarrow\rangle$ is
lower than $|P\rangle$) is due to the following two factors: 1)
Hg-$5d$ states around -7.0 eV below Fermi level are very shallow and
their hybridization with Se-$4p$ states has pushed the anti-bonding
Se-$4p$ states quite high, similar to the situation in HgSe. 2) The
hybridization between unoccupied Cr-$3d^\downarrow$ and
Hg-$6s^\downarrow$ states will push the Hg-$6s^\downarrow$ state to
lower in energy. These two key factors lead to band inversion, with
the $|S,\downarrow\rangle$ lower than the
$|P_{\frac{3}{2}},\frac{3}{2}\rangle$ states by about 0.4 eV. This
value is further enhanced to be 0.55 eV in the presence of SOC.
Further LDA+$U$ calculations (with effective $U$ around
$3.0$eV~\cite{CdCrS-ldau,ACrX}) indicate that the band-inversion
remains even if the correlation effect is properly considered, unless
$U$ is unreasonably large ($>8.0$ eV). Experimental observations of
metallic behavior at low temperature for various
samples~\cite{HgCrSe-MR,HgCrSe-Tran-1,HgCrSe-Tran} strongly support
this conclusion of inverted band structure.

\begin{figure}[tbp]
\begin{centering}
\includegraphics[clip,width=0.8\textwidth]{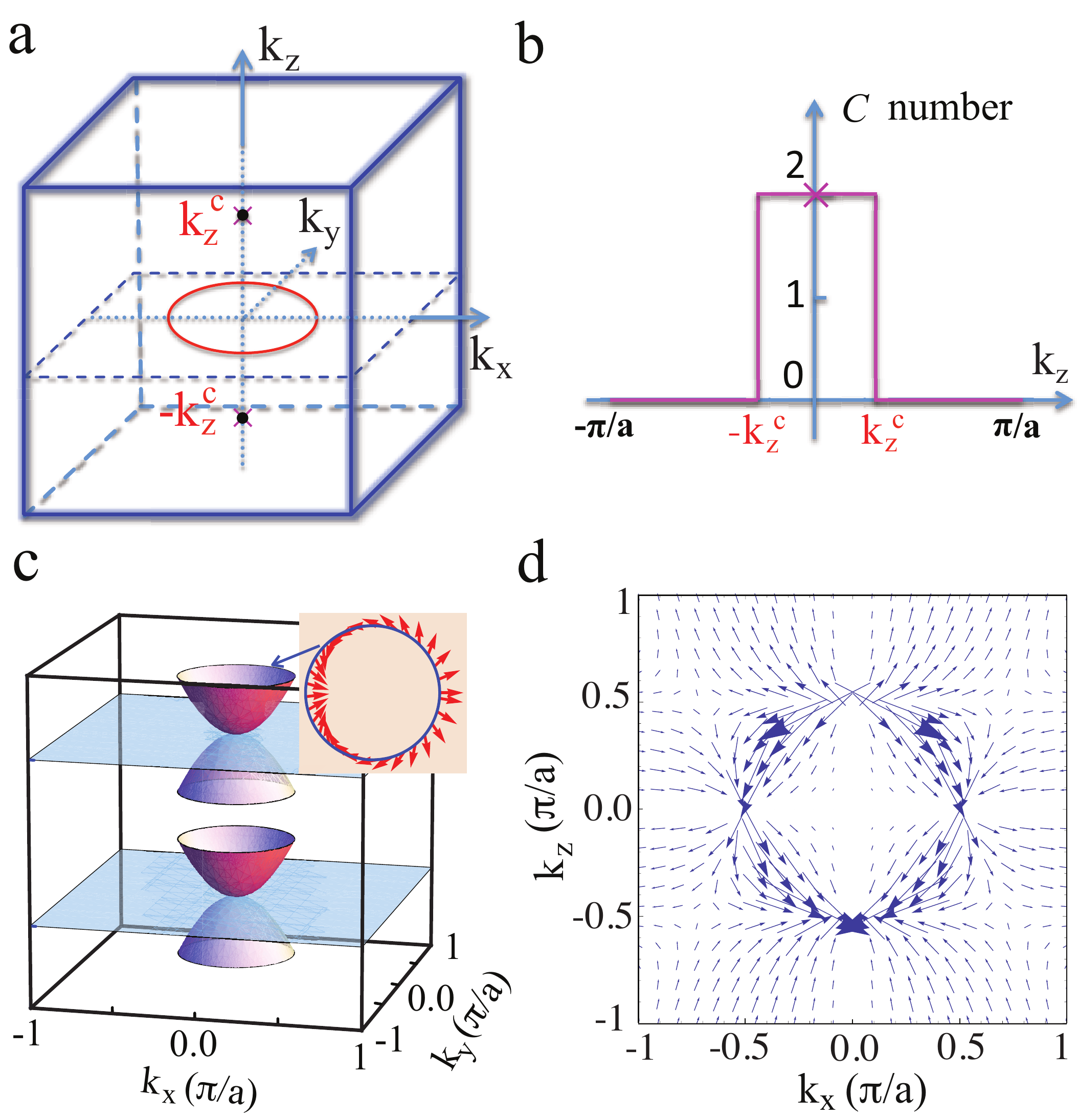}
\par\end{centering}
\caption{(Color online) Weyl nodes and gauge flux in
  HgCr$_2$Se$_4$. (a) The band-crossing points in $\vec{k}$-space; (b)
  Chern number as function of $k_z$; (c) Schematic plot of the band
  dispersion around the Weyl nodes in the $k_z$=$\pm k_z^c$ plane, and
  the inset shows the chiral spin texture. (d) Gauge flux evaluated as
  Berry curvature in the ($k_x$, $k_z$) plane. Figure from
  Ref.~\onlinecite{XuGang_HgCrSe_2011_PRL}.}
  \label{fig:HgCrSe_model_results}
\end{figure}

In the presence of SOC, the low energy eigenstates at $\Gamma$ are
given as $|\frac{3}{2},\pm\frac{3}{2}\rangle$,
$|\frac{3}{2},\pm\frac{1}{2}\rangle$,
$|\frac{1}{2},\pm\frac{1}{2}\rangle$, and $|S,\pm\frac{1}{2}\rangle$,
which can be constructed from the $|P\rangle$ and $|S\rangle$ states,
similar to the case of HgSe and HgTe again. Considering the exchange
splitting, the eight states at $\Gamma$ are separated in energy.  The
$|\frac{3}{2},\frac{3}{2}\rangle$ has the highest energy, and the
$|S,-\frac{1}{2}\rangle$ has the lowest.  Due to the band-inversion,
several band-crossings are observed in the band structure.  Among
them, however, only two kinds of band-crossings (called A and B) are
important for the states very close to the Fermi level, as shown in
Fig.~\ref{fig:HgCrSe_band_structure}(d).  As schematically shown in
Fig. \ref{fig:HgCrSe_model_results}(a), the A-type crossing gives two
points located at $k_z=\pm k_z^c$ along the $\Gamma -Z$ line.  For the
2D planes with fixed-$k_z$ ($k_z \ne 0,\pm k_z^c$), the in-plane band
structures are all gapped in the sense that a curved Fermi level is
defined. Therefore, the Chern number $C$ for each $k_z$-fixed plane
can be evaluated. It is found that the crossing A (i.e.,
$|k_z|=k_z^c$) is located at the phase boundary between the $C=2$ and
$C=0$ planes, i.e., $C=0$ for the planes with $|k_z|>k_z^c$, while
$C=2$ for the planes with $|k_z|<k_z^c$ and $k_z\ne 0$.  This change
in Chern number indicates that these two crossing points (Weyl points)
are topologically protected so long as the Weyl points remain
separated in momentum space.  The B-type crossing is a closed node
line surrounding the $\Gamma$ point in the $k_z$=0 plane. It is just
accidental and is due to the presence of crystal mirror symmetry with
respect to the $k_z$=0 plane.  The B-type crossing is not as stable as
the A-type crossing, in the sense that it can be avoided by changing
of the crystal symmetry (such as the reorientation of magnetic
momentum away from the $z$ direction).

Assuming that the magnetic momentum of the system is oriented to the
$z$ axis, the low energy physics here can be caught by the following
effective two-band model~\cite{XuGang_HgCrSe_2011_PRL,KP-method}:
\begin{equation}
H_{eff}=\left[
\begin{array}{cc}
M &  Dk_zk_-^2  \\
 Dk_zk_+^2 & -M
\end{array}
\right],
\end{equation}
in bases $|\frac{3}{2},\frac{3}{2}\rangle$ and
$|S,-\frac{1}{2}\rangle$.  Here $k_\pm=k_x\pm i k_y$, and $M=M_0-\beta
k^2$ is the mass term expanded up to the second order of $k$, with
parameters $M_0>0$ and $\beta>0$ to ensure band inversion. Since the
two bases have opposite parity, the off-diagonal element has to be an
odd function of $k$. In addition, $k_\pm^2$ has to appear to conserve
the angular moment along $z$-direction. Therefore, to the leading
order, $k_zk_\pm^2$ is the only possible form for the off-diagonal
element. The energy dispersion
$E(k)=\pm\sqrt{M^2+D^2k_z^2(k_x^2+k_y^2)^2}$ has two gapless
solutions: one is the degenerate points along $\Gamma -Z$ with
$k_z=\pm k_z^c=\pm\sqrt{M_0/\beta}$; the other is a circle around the
$\Gamma$ point in the $k_z=0$ plane determined by the equation
$k_x^2+k_y^2=M_0/\beta$. They are exactly the band-crossings obtained
from the first-principles calculations. Due to the presence of
$k_\pm^2$ in the off-diagonal element~\cite{onoda-2}, it is easy to
find that Chern number $C$ equals to 2 for the planes with
$|k_z|<k_z^c$ and $k_z\ne 0$, otherwise $C$=0.  The band dispersions
near the Weyl nodes at $k_z=\pm k_z^c$ plane are thus quadratic rather
than linear, with chiral in-plane spin texture, as shown in the inset
of Fig. \ref{fig:HgCrSe_model_results}(c).  The two Weyl nodes located
at $\pm k_z^c$ have opposite chirality and form a single pair of
magnetic monopoles. The gauge flux starting from and ending at them in
$\vec{k}$-space is shown in Fig. \ref{fig:HgCrSe_model_results}(d).
The band-crossing loop, i.e. node-line, in the $k_z=0$ plane is not
topologically unavoidable; however, its existence requires that the
gauge flux in the $k_z$=0 plane (except the loop itself) must be zero.

\begin{figure}[tbp]
\begin{centering}
\includegraphics[clip,width=0.9\textwidth]{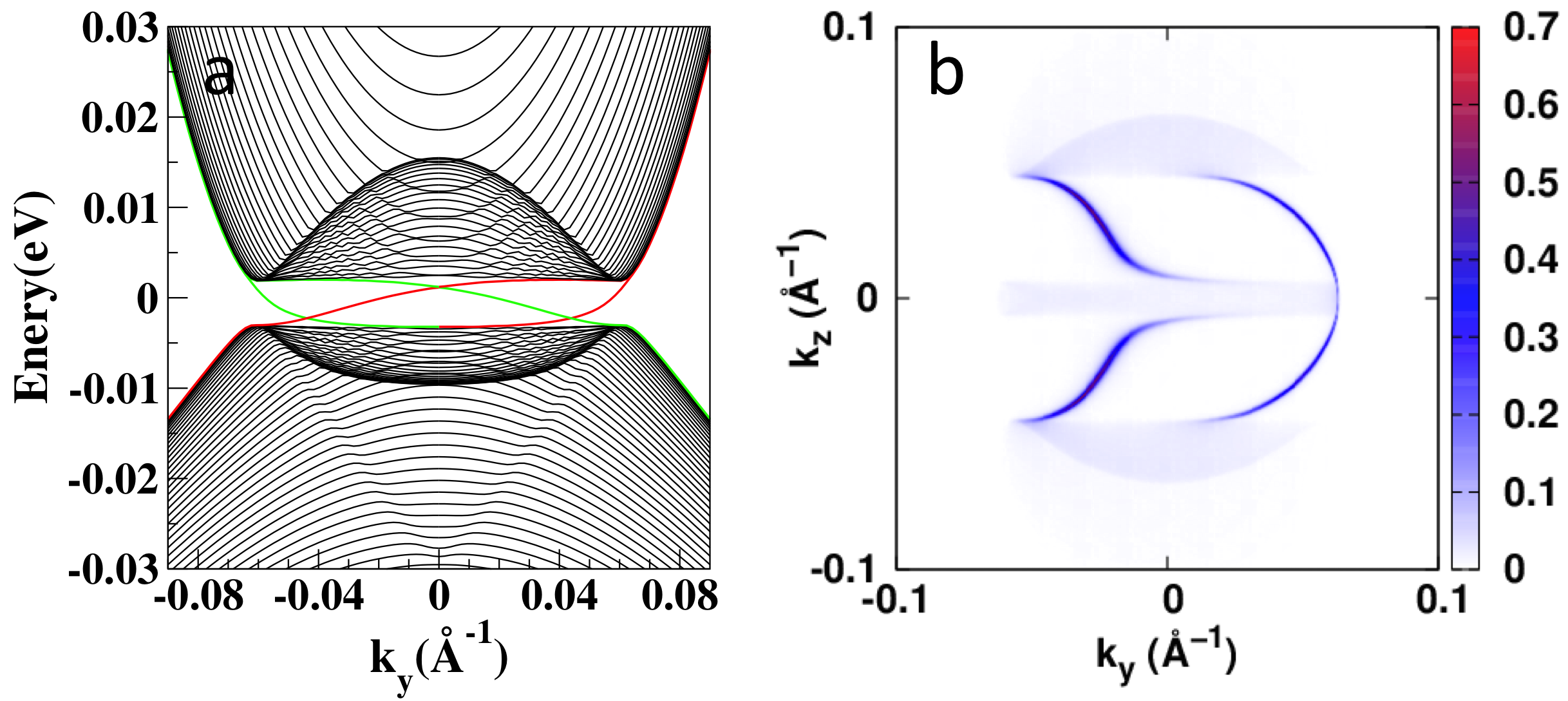}
\par\end{centering}
\caption{(Color online) Edge states and Fermi arcs of
  HgCr$_2$Se$_4$. (a) Calculated edge states for the plane with
  $k_z$=0.06$\pi$. A ribbon with two edges is used, and there are two
  edge states for each edge since $C$=2. The states located at
  different edges are indicated by different colors. (b) Calculated
  Fermi arcs for the side ($k_y$,$k_z$) surface. Figure from
  Ref.~\onlinecite{XuGang_HgCrSe_2011_PRL}.}
  \label{fig:HgCrSe_fermiarc}
\end{figure}

This Weyl semimetal (also called Chern semimetal) state realized in
HgCr$_2$Se$_4$ will lead to novel physical consequences, which can be
measured experimentally. First, each $k_z$-fixed plane with non-zero
Chern number can be regarded as a 2D Chern insulator, and there must
be chiral edge states for such plane if an edge is created. The number
of edge states is two for the case of $C$=2 (see
Fig. \ref{fig:HgCrSe_fermiarc}(a)), or zero for the case of $C$=0. If
the chemical potential is located within the gap, only the chiral edge
states can contribute to the Fermi surface, which are isolated points
for each plane with $C$=2 but nonexistent for the plane with
$C$=0. Therefore, the trajectory of such points in the ($k_x$, $k_z$)
surface or ($k_y$, $k_z$) surface form non-closed Fermi arcs, which
can be measured by ARPES.  As shown in
Fig. \ref{fig:HgCrSe_fermiarc}(b), the Fermi arcs connect the Weyl
points at $k_z$=$\pm k_z^c$, and are interrupted by the $k_z$=0 plane.
This is very different from conventional metals, in which the Fermi
surfaces must be either closed or interrupted by the Brillouin zone
boundary.  The possible Fermi arcs have also been discussed for
pyrochlore iridates~\cite{WanXG_WeylTI_2011}.

\begin{figure}[tbp]
\begin{centering}
\includegraphics[clip,width=0.99\textwidth]{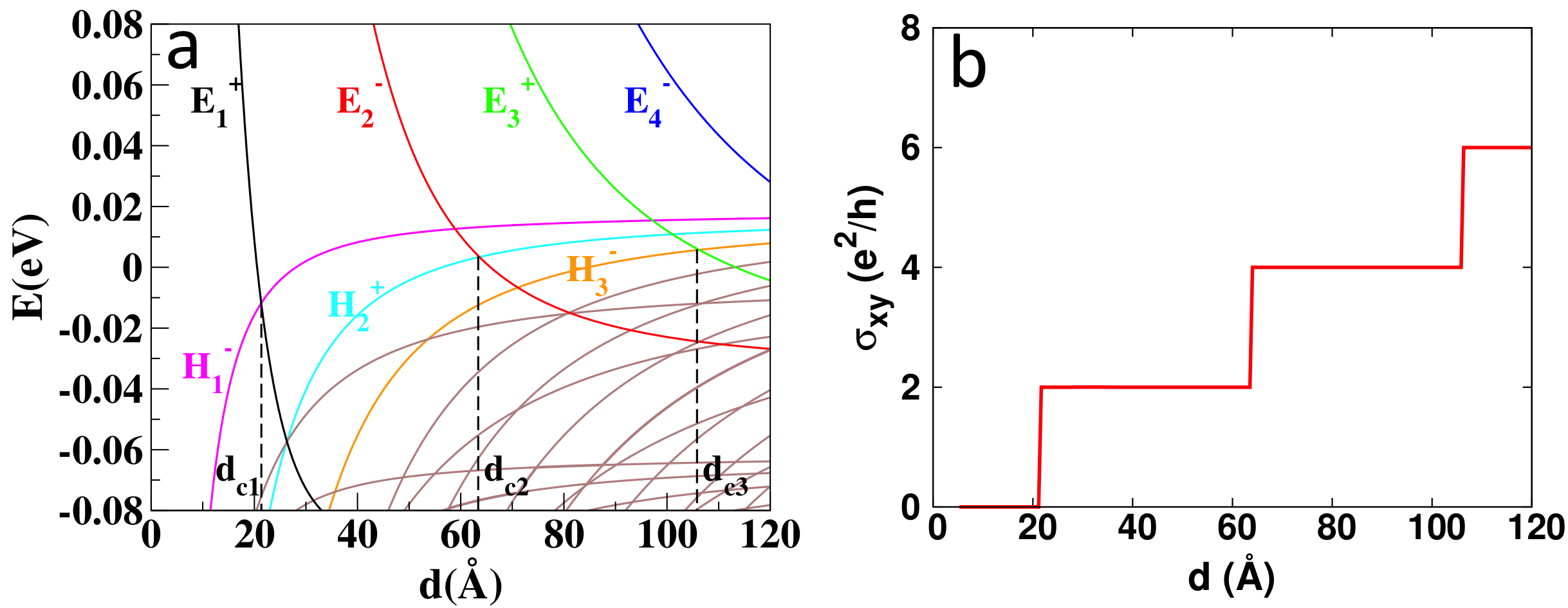}
\par\end{centering}
\caption{(Color online)  Quantized Anomalous Hall effect in
    HgCr$_2$Se$_4$ thin film.  (a) The subband energy levels at
  $\Gamma$ point as function of film thickness. (b) The Hall
  conductance as function of film thickness. Figure from Ref.~\onlinecite{XuGang_HgCrSe_2011_PRL}.}
  \label{fig:HgCrSe_Hall_plateau}
\end{figure}

The QAHE, on the other hand, is a unique physical consequence
characteristic of the Chern semimetal nature of HgCr$_2$Se$_4$ by
considering its quantum-well structure. For 2D Chern insulators, the
transverse Hall conductance should be quantized as
$\sigma_{xy}=C\frac{e^2}{h}$, where $C$ is the Chern number.
Considering the $k_z$-fixed planes, the Chern number $C$ is non-zero
for limited regions of $k_z$, and this is due to the band inversion
around $\Gamma$ as discussed above.  In the quantum well structure,
however, those low energy states around $\Gamma$ should be further
quantized into subbands (labeled as $|H_n\rangle$ and $|E_n\rangle$
for hole and electron subbands, respectively), whose energy levels
change as functions of film thickness.  As shown in
Fig. \ref{fig:HgCrSe_Hall_plateau}(a), when the film is thin enough,
the band inversion in the bulk band structure will be removed entirely
by the finite size effect.  With the increment of film thickness,
finite size effect gets weaker and the band inversion among these
subbands is restored subsequently, which leads to jumps in the Chern
number or the Hall coefficient $\sigma_{xy}$~\cite{CXLiu_HgMnTe_2008}.
As shown in Fig. \ref{fig:HgCrSe_Hall_plateau}(b), if the film is
thinner than 21$\rm \AA$, the $\sigma_{xy}$ is zero; once the film
thickness is larger than the critical thickness, we find subsequent
jumps of $\sigma_{xy}$ in units of $2e^2/h$.  In fact, the anomalous
Hall effect has been observed for the bulk samples of
HgCr$_2$Se$_4$~\cite{HgCrSe-AHE_1997}, and more recently, n-type
HgCr$_2$Se$_4$ was confirmed to be a single $s$-band half
metal~\cite{guan_single_2015}. This is in sharp contrast to pyrochlore
iridates, in which the anomalous Hall effect should be vanishing due
to the AF ordering~\cite{WanXG_WeylTI_2011}.

This proposal of realizing QAHE in thin films of FM Weyl semimetals
has some advantages compared to the previous one for the magnetic-ion
doped TIs thin films. First, it is based on the stoichiometric
material and avoids the difficulty of doping, which is in general good
for raising the mobility of the sample. Second, FM ordering has been
established in bulk HgCr$_2$Se$_4$ at temperature as high as 100
K~\cite{Tc_HgCr2Se4_1985}, which makes QAHE plausible at high
temperatures. Third, the proposed system can reach a QAHE with a high
Chern number (i.e., Chern number 2 or even higher) and higher
plateau~\cite{QAHE_higher_Plateaus}. Therefore, further experimental
efforts in realizing QAHE in thin film of HgCr$_2$Se$_4$ are much
anticipated and very promising.

%
%
%
%
%
%
%
%
%
%

%
%
%
%
%
%

~\\~
\hspace*{15pt}{\bf IV.D. QAHE on Honeycomb Lattice}
\\


Since honeycomb lattice plays a unique role in exploiting non-trivial
topology \cite{haldane_model_1988,Kane_Mele_Graphene_PRL_2005}, it is
instructive to take a look at the geometric property of its electronic
band structures, paying attention to possible configurations of Berry
curvatures. There are two sites in a unit cell of honeycomb lattice,
and with the nearest neighbor hoppings, the valance band and
conduction band touch linearly and form the Dirac cones at the six
corners of the Brillouin zone.  Taking into account the C$_3$
symmetry, one can consider the two inequivalent Dirac cones at
$K,K'=\frac{4\pi}{3d}(\pm 1,0)$ with lattice constant $d$, which are
called valleys and denoted by $\tau_z=\pm 1$. The Hamiltonian for
low-energy physics is given by
\begin{equation}
H_{\tau_z}=\left(
             \begin{array}{cc}
               0 & \tau_z k_x + ik_y \\
               \tau_z k_x - ik_y & 0 \\
             \end{array}
           \right),
\label{Hmatrix}
\end{equation}
where the momenta are measured from the valleys. This is the 2D
version of Eq. (\ref{eq:weyl}) with the basis of the 2$\times$2 matrix
referring to AB sublattices of honeycomb lattice. Therefore, the
electrons are chiral at the valleys, which makes the honeycomb lattice
special in generating non-trivial topology.

Let us introduce a staggered field $U$ on the $AB$ sublattices, and
rewrite the Hamiltonian (\ref{Hmatrix}) as
\begin{equation}
H_{\tau_z}=\tau_z k_x\sigma_x + k_y \sigma_y + U\sigma_z,
\label{Hsigma}
\end{equation}
with Pauli matrices $\sigma_{x,y,z}$. The dispersion is given by $
E=\pm\sqrt{U^2+k^2}, $ with a gap opened by the staggered potential
$U$. For this insulating state, one can define a fictitious magnetic
field $\mathbf{d}=(\tau_z k_x, k_y, U)$ in Hamiltonian (\ref{Hsigma})
on the pseudo spin $\sigma$, which exhibits a topological structure
around each valley known as \textit{meron}.  Its topological property
can be characterized by counting the winding number of the
\textit{meron} structure (i.e., the integration of Berry curvature
over a patch covering the region of the \textit{meron}, or in other
words, the Berry flux passing through the 2D BZ in the vicinity of the
each valley, divided by 2$\pi$) as~\cite{XiaoDi_RMP_2010}
\begin{equation}
c=\frac{1}{4\pi}\int dk_xdk_y\mathbf{n}\cdot \left(\partial_{k_x}\mathbf{n} \times \partial_{k_y}\mathbf{n} \right)={\rm sgn}(\tau_z U)/2,
\label{cnmeron}
\end{equation}
with $\mathbf{n}=\mathbf{d}/|\mathbf{d}|$.

\begin{figure}[t]
\centering
\includegraphics[clip,width=0.8\textwidth]{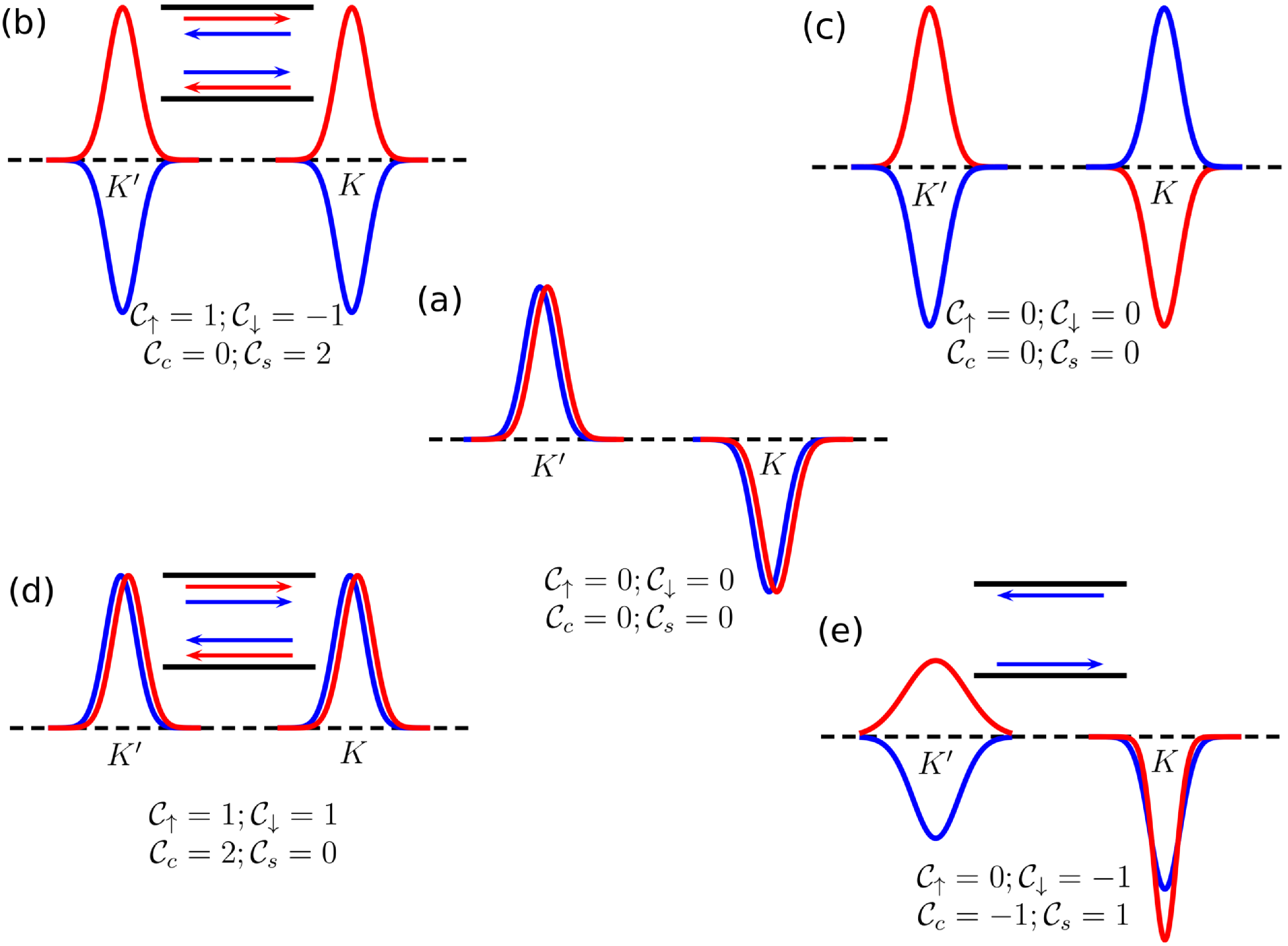}
\caption{(Colour online) Possible configurations of Berry curvatures
  on honeycomb lattice: (a) CDW, (b) QSHE, (c) SDW, (d) QAHE, (e) spin
  resolved QAHE characterized by simultaneous non-zero charge and spin
  Chern numbers {\cite{XiaoHu_NJP_2013}.} Red/blue for spin up/down
  channels. Edge currents are displayed in insets in (b), (d) and (e)
  for topologically non-trivial states. $C_{\uparrow}$ and
  $C_{\downarrow}$ are Chern numbers in spin-up and -down channels,
  and $C_{\rm c/s}=C_{\uparrow}\pm C_{\downarrow}$ are charge and spin
  Chern numbers respectively. Figure from Ref.~\onlinecite{XiaoHu_NJP_2013}.
}
\label{fig:berryphase}
\end{figure}

~\\~
\hspace*{15pt}{\it IV.D.1 Combinations of Berry curvatures on honeycomb lattice}
\\

For a staggered electric potential $H_{\rm elec}=V \sigma_z$ where $V$
is a constant, the signs of the Berry curvature around two valleys are
opposite, as seen in Eq.~(\ref{cnmeron}), which gives a topologically
trivial insulator called a charge density wave (CDW) as displayed in
Fig.~\ref{fig:berryphase}(a). On the other hand, if a staggered
magnetic flux is introduced to spinless electrons on honeycomb
lattice~\cite{haldane_model_1988}, the staggered potential term can be
written as $H_{\rm sf}=t \tau_z \sigma_z$ (around two valleys), which
depends on the valley variable in contrast to the staggered electric
field.  Since the topological charges are now given by $c={\rm
  sgn}(t)/2$, the Berry curvatures around two valleys are aligned,
which characterizes a QAHE state (see Fig.~\ref{fig:berryphase}(d)
where spin is included in addition to the Haldane model).  This
staggered field can be generated by shining circularly polarized light
on graphene, as revealed by the Floquet theorem \cite{OkaAoki}.
Taking into account the true spin degree of freedom, the SOC term of
$H_{\rm so}=\lambda s_z \tau_z \sigma_z$ can also be included as
suggested in the Kane-Mele model~\cite{Kane_Mele_Graphene_PRL_2005},
and the SOC term again takes the form of a staggered field, and
depends additionally on spin variable $s_z$. In this case, the Berry
curvatures around two valleys in the spin up and down channels are
aligned individually and oppositely for different spin channels,
resulting in the configuration in Fig.~\ref{fig:berryphase}(b), which
characterizes the QSHE state \cite{Kane_Mele_Graphene_PRL_2005}.
Antiferromagnetic (AFM) exchange field $H_{\rm afm}=M s_z \sigma_z$ is
another possible staggered field, which reverses the sign of the Berry
curvature in the spin-down channel for both valleys (compared to the
CDW state shown in Fig.~\ref{fig:berryphase}(a), assuming the same
sign for $M$ and $V$), resulting in the configuration in
Fig.~\ref{fig:berryphase}(c) associated with a topologically trivial
state called spin density wave (SDW).  Coupling of the valley degree
of freedom to the AFM order in manganese chalcogenophosphates was
recently discussed~\cite{PNASFengji}.

We observe that the configuration of Berry curvature given in
Fig.~\ref{fig:berryphase}(e) corresponds to an additional topological
state \cite{XiaoHu_NJP_2013} (see also \cite{EzawaPRB13}), by flipping
the sign of the Berry curvature around K$'$ valley in the spin up
channel from the CDW state.  In contrast to QSHE with a non-zero spin
Chern number $C_{\rm
  s}=C_{\uparrow}-C_{\downarrow}$~\cite{Haldane_2006PRL,
  XingDingYu_PRL11}, which is equivalent to the $Z_2$ index, and QAHE
with a non-zero charge Chern number $C_{\rm
  c}=C_{\uparrow}+C_{\downarrow}$, this state is characterized by
simultaneous non-zero charge and spin Chern numbers.  We may call the
state in Fig.~\ref{fig:berryphase}(e) \textit{spin-resolved} QAHE, as
opposed to other QAHE states where spin information is lost.

Note that configurations of the Berry curvature in
Figs.~\ref{fig:berryphase}(a), (b), (c) and (d) are converted to each
other by flipping the sign of two Berry curvatures either at the same
valley or with the same spin. This reflects certain symmetries in
these states: TRS in Fig.~\ref{fig:berryphase}(a), TRS and spacial IS
in Fig.~\ref{fig:berryphase}(b), IS in Fig.~\ref{fig:berryphase}(d),
and the combination of TRS and IS in Fig.~\ref{fig:berryphase}(c)
(although both symmetries are broken individually). All symmetries are
broken in the state portrayed in Fig.~\ref{fig:berryphase}(e).  One
can realize the state in Fig.~\ref{fig:berryphase}(e) by using three
fields simultaneously -- SOC, AFM exchange field and the staggered
electric potential -- which makes fine control on electronic states
possible:
\begin{equation}
H=\tau_z k_x\sigma_x + k_y\sigma_y +(\lambda \tau_z s_z +M s_z + V)\sigma_z.
\label{myhamiltonian}
\end{equation}

\begin{figure}[t]
\centering
\includegraphics[clip,width=0.8\textwidth]{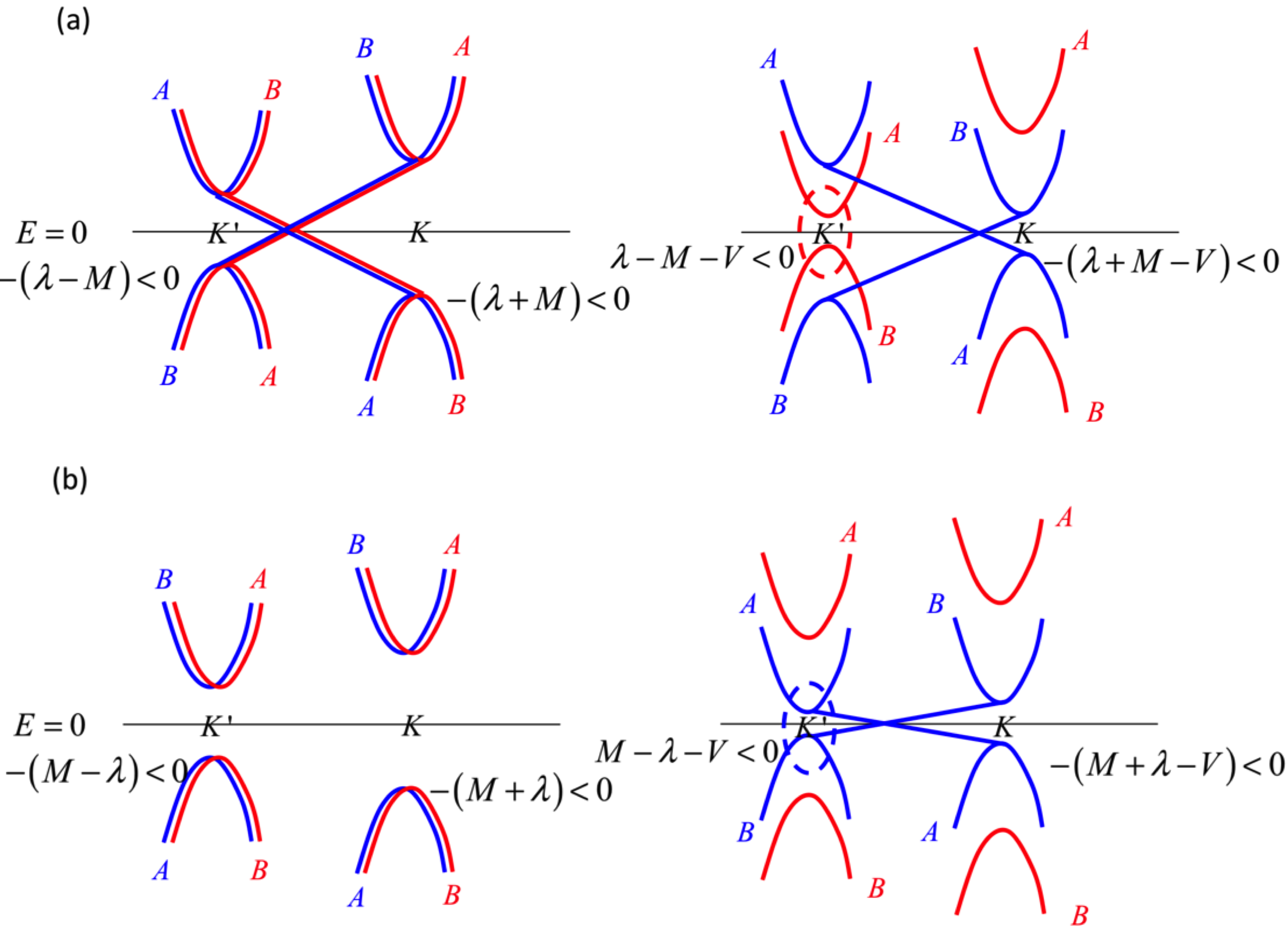}
\caption{(Color online) Band configurations for $V=0$ (left) and
  $|\lambda-M|<V<\lambda+M$ (presuming $\lambda, M>0$) for
  (a)   $M<\lambda$ and (b) $M>\lambda$. Red (blue) curves are for spin-up
  (-down) channels, and $A$ and $B$ for the sublattices. The straight
  lines are for topological edge states in ribbon systems, and the
  dashed ellipses indicate the valleys where band inversions occur
  upon increasing $V$.  }
\label{fig:quantumtransition}
\end{figure}

We reveal behaviors of the system upon tuning the electric potential
$V$ with the $M$ field and $\lambda$ fixed.  For $\lambda>M>0$, the
system takes a QSHE state at $V=0$ (see Fig.~\ref{fig:berryphase}(b)).
{The degeneracy in band dispersions is protected by the product of TRS
  and IS, a pseudo TRS.}  The gaps are $2(\lambda+M)$ and
$2(\lambda-M)$ at the two valleys, respectively.  As illustrated in
Fig.~\ref{fig:quantumtransition}(a), when $V$ increases from zero, the
gap in the spin-up channel shrinks, closes at $V=\lambda-M$ and
reopens at the K$'$ valley; this quantum phase transition makes the
spin-up channel topologically trivial, while the spin-down channel
remains topologically non-trivial. This QAHE state is possible since
the electric field $V$ breaks the IS, therefore breaking the pseudo
TRS.  When $V$ increases further, the gap in the spin-down channel
shrinks, closes at $V=\lambda+M$ and reopens at K valley (not shown
explicitly). This second quantum phase transition makes the spin-down
channel trivial as well, and the system transforms into the
topologically trivial CDW state.

For $M>\lambda>0$, the system is in a topologically trivial SDW state
at $V=0$ (see Fig.~\ref{fig:berryphase}(c)).  As seen in
Fig.~\ref{fig:quantumtransition}(b), when $V$ increases, the gap in
the spin-down channel shrinks, closes at $V=M-\lambda$ and reopens at
the K$'$ valley, which brings the spin-down channel into a topological
state. When $V$ increases further, the gap in the spin-down channel
shrinks, closes at $V=\lambda+M$ and reopens at K valley, which drives
the system to the topologically trivial CDW state.  It is clear that
the spin-resolved QAHE state is realized when the sizes (absolute
values) of the three fields satisfy the condition of being able to
form a triangle, which indicates clearly that the interplay among the
three fields is crucial. In this case the sign of the total staggered
field $U=\lambda \tau_z s_z + M s_z + V$ at one valley in one spin
channel is opposite to the other three cases, which yields the
configuration of Berry curvature in Fig.~\ref{fig:berryphase}(e).

The above discussions for the possible topological phases in honeycomb
lattice by fine tuning of parameters can be generally applied to many
different material systems, such as graphene~\cite{qiao_quantum_2010,
  nandkishore_quantum_2010, Tse_QAHE_graphene_2011PhRvB,
  MacDonald_2011_PRL, ding_engineering_2011, Qiao:2011fe,
  zhang_electrically_2012, QiaoZhenhua_PRB_2012,
  ZHQiao_QAHE_AFM_2014_PRL},
silicene~\cite{ezawa_valley-polarized_2012, EzawaNJP12, EzawaPRL13},
perovskite bilayers~\cite{DiXiao_NatComm_2011_LaAlO3_LaAuO3,
  KYYangPRB11, GAFietePRB11, GAFietePRB12, Fiete_PRB_2012,
  XiaoHu_NJP_2013, wang_interaction-induced_2014}, etc.  Valleys, as
an additional degree of freedom of electrons on honeycomb lattice,
have attracted considerable interest, for which valleytronics has been
coined as a new concept~\cite{XiaoDi_PRL_2007_valley,
  XuXiaodong_2014NatPhys}.  We will leave the discussions of
topological states in graphene and silicene to readers, and in the
following, we concentrate on the perovskite bilayer systems, which are
interesting and promising due to the presence of transition metal
elements.

%
%
%
%
%
%

~\\~
\hspace*{15pt}{\it IV.D.2 Honeycomb lattices by perovskite structures}
\\

Material realization of topological insulators (TI) has been mainly
focused on narrow band-gap semiconductors with heavy elements such as
Hg or Bi, in which the electronic properties are dominated by $s$ and
$p$ orbitals.  Recently a completely different materials class --
heterostructures of transition-metal oxides (TMOs) involving $d$
electrons -- was proposed~\cite{DiXiao_NatComm_2011_LaAlO3_LaAuO3}. The TMO heterostructures
are readily available due to achievements in the fields of oxide superlattices and oxide
electronics~\cite{mannhart_oxide_2010, zubko_interface_2011,
  hwang_emergent_2012}. The sample of layered structure of TMOs can be obtained
  with atomic precision~\cite{UedaScience98,
  Kawasaki,Gray}. This make the fine tuning of sample properties, such as 
 the lattice constant, carrier density, SOC and electron-electron correlation, very plausible. 
 Possible 2D TIs have been proposed for bilayers of perovskite-type 
 TMOs grown along the [111] direction of the crystallographic axis~\cite{DiXiao_NatComm_2011_LaAlO3_LaAuO3}.

\begin{figure}[t]
\centering
\includegraphics[clip,width=0.99\textwidth]{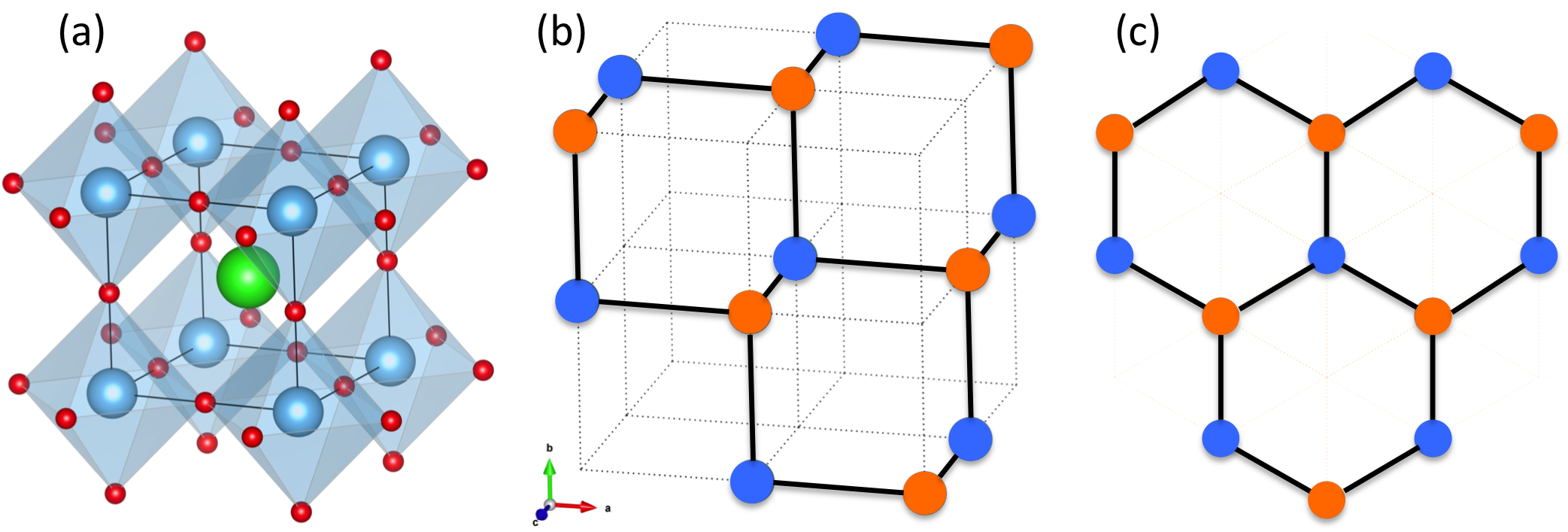}
\caption{(Color online) Perovskite structure as a buckled honeycomb
  lattice grown along the [111] crystal direction. Adapted from
  Ref.~\onlinecite{DiXiao_NatComm_2011_LaAlO3_LaAuO3}.}
\label{fig:perovskite}
\end{figure}

Perovskite compounds~\cite{Mitchellbook} are in the chemical formula of ABO$_3$, where A stands for alkaline or rare earth
metal, B stands for transition metal (TM) and O stands for oxygen. Its structure is shown in Fig.~\ref{fig:perovskite}(a) and (b).  
The ideal oxygen octahedral crystal field splits the TM $d$ orbitals into $e_g$ (with $d_{3z^2-r^2}$ and $d_{x^2-y^2}$ orbitals) 
and $t_{2g}$ (with $d_{yz}$, $d_{zx}$ and $d_{xy}$) manifolds, separated by $\sim$3eV in energy. Such crystal geometry 
usually is not good for band inversion. However, for a bilayer of the perovskite structure grown along [111] direction~\cite{DiXiao_NatComm_2011_LaAlO3_LaAuO3}, the TM ions are on a buckled honeycomb lattice as shown in
Fig.~\ref{fig:perovskite}(c). Based on tight-banding models and first-principles calculations,
D. Xiao {\it et al.} have demonstrated that both $t_{2g}$ and $e_{g}$ orbitals can
generate topologically non-trivial states for TM ions on such structure. Especially, they showed that 
LaAuO$_3$ bilayers sandwiched by two LaAlO$_3$ substrates might possess a topologically non-trivial gap around 0.15
eV~\cite{DiXiao_NatComm_2011_LaAlO3_LaAuO3}.

Various possible topological phases have been explored for the
perovskite bilayer systems~\cite{GAFietePRB11, GAFietePRB12,
  Fiete_PRB_2012}. Except the QSH phase, the QAH phase can be readily
expected by breaking the TRS, as discussed before. Thanks to the
presence of transition metal elements, different magnetic orderings
may be realized in such systems with no need of additional doping.  If
the strong electron interaction among the $d$ electrons is properly
taken into account, we can expect more interesting
physics~\cite{KYYangPRB11} for such honeycomb lattices formed by
bilayer perovskite structures. In a recent study, it was
predicted~\cite{wang_interaction-induced_2014} by the LDA+Gutzwiller
method~\cite{deng_local_2009}, that (111) LaCoO$_3$ bilayer grown on
SrTiO$_3$ substrate may support the QAHE. Here, the interplay between
SOC and Coulomb interaction stabilizes a robust FM insulating state,
which is a Chern insulator with non-zero Chern number. It was pointed
out in this study, that the SOC splitting (which is usually not strong
for 3$d$ elements) can be greatly enhanced by the strong Coulomb
interaction.  It has also been revealed that the possibility of some
intriguing phenomena, such as fractional quantum Hall effect,
associated with the nearly flat topologically non-trivial bands can be
found in $e_g$ systems \cite{DiXiao_NatComm_2011_LaAlO3_LaAuO3}.

\begin{figure}[t]
\centering
\includegraphics[clip,width=0.8\textwidth]{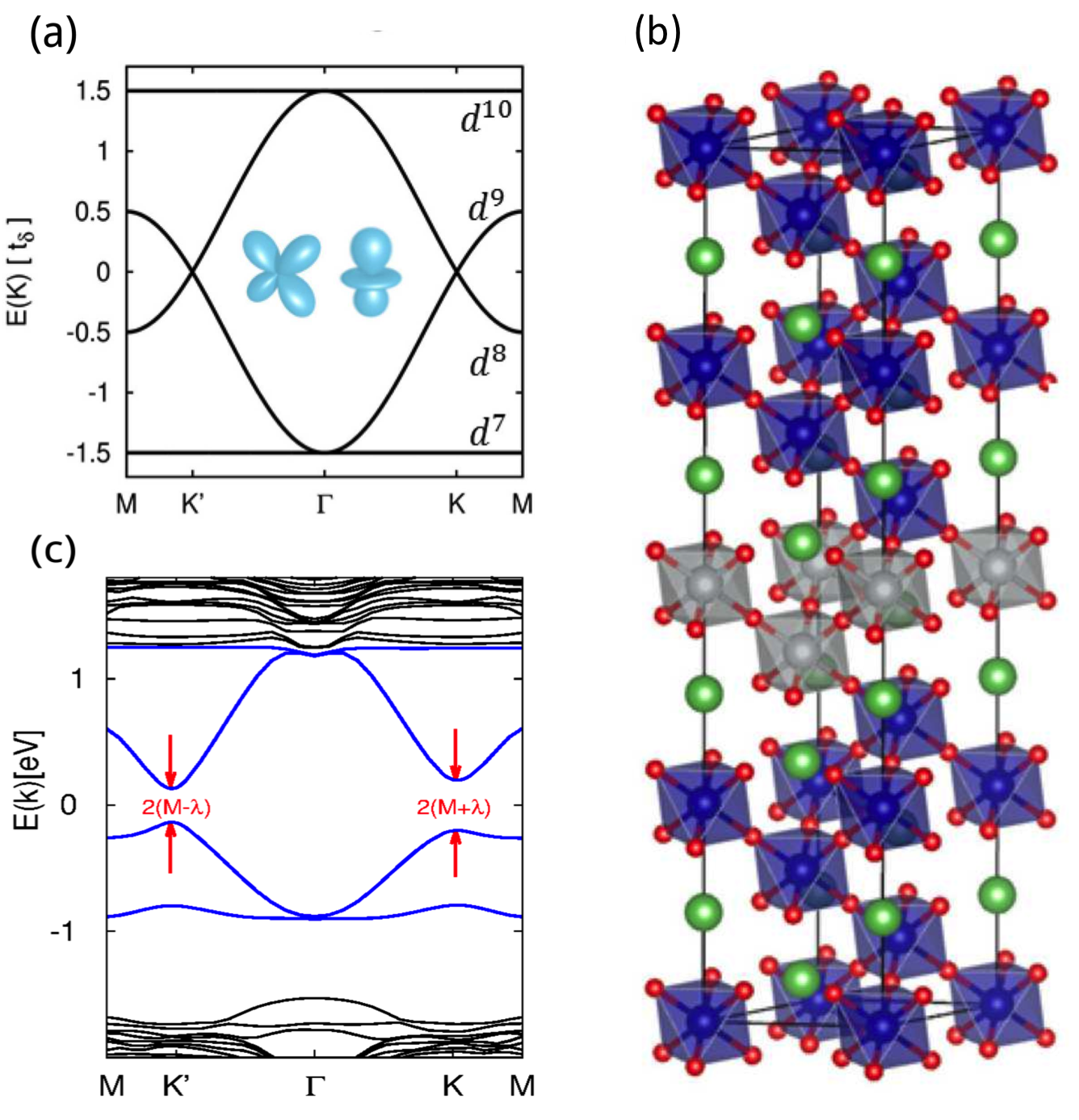}
\caption{(Color online) (a) Noninteracting tight-binding band
  structure of the e$_{\rm g}$ system, where each band is labeled with
  the corresponding electron configuration if it is occupied. Inset
  shows the two e$_{\rm g}$ orbits.  (b) Schematic of a unit cell
  (La$_2$Au$_2$O$_6$)/(La$_2$Cr$_2$O$_6$)$_{5}$.  (c) Band structures
  based on first-principles calculation. Figures from
  Ref.~\onlinecite{XiaoHu_NJP_2013}.  }
\label{fig:bandstructure}
\end{figure}

The perovskite bilayer structure can also be used to generate the
spin-resolved QAHE state.  Choosing elements adequately, one can fill
$t_{\rm 2g}$ orbitals while leaving $e_{\rm g}$ orbitals partially
occupied, and the 10Dq gap (crystal-field splitting) will prevent
$t_{\rm 2g}$ orbitals from participating in the low-energy physics.
For an atomic sheet of buckled honeycomb lattice, the two B atoms in a
unit cell contribute four $e_{\rm g}$ orbits in total. The band
structure for these four orbitals is shown in
Fig.~\ref{fig:bandstructure}(a) based on the tight-binding Hamiltonian
with the hopping integrals given by the Slater-Koster formula
\cite{XiaoHu_NJP_2013}. Two dispersive bands are contributed from
${d}^{8,9}(t_{\rm 2g}^6e_g^{2,3})$, which cross each other linearly
and form the desired Dirac cones at K and K$'$ due to the honeycomb
structure.

A candidate perovskite material which generates $d^{8}$ electrons is
LaAuO$_3$, wherein Au takes the +3 charge state.  In order to generate
an AFM exchange field, one can sandwich the atomic sheet of buckled
honeycomb lattice with $d^{8}$ electrons by two thick layers of
insulating perovskite material grown in the [111] direction.  A
candidate material for the substrates is LaCrO$_3$, a Mott insulator
with G-type AFM order, which induces opposite exchange fields on the
two sublattices in the buckled honeycomb lattice. The large Mott gap
of LaCrO$_3$ prevents undesired influences on the Dirac behavior of
$d^{8,9}$ electrons except for the exchange field.  Since La is chosen
as the common A-site cation, the lattice mismatch between substrate
and the targeting atomic sheet is minimal.  First-principles
calculations were performed in terms of VASP~\cite{VASP} code with a
supercell of (La$_2$Au$_2$O$_6$)/(La$_2$Cr$_2$O$_6$)$_{5}$ shown in
Fig.~\ref{fig:bandstructure}(b). As displayed in
Fig.~\ref{fig:bandstructure}(c), the black bands are mainly from the
substrate LaCrO$_3$ with a large gap of 3eV, characteristic of a Mott
insulator. The bands inside the Mott gap are contributed by the $e_g$
electrons of Au$^{+3}$ ions. There are two almost flat bands and two
dispersive bands, as captured by the tight-binding Hamiltonian (see
Fig.~\ref{fig:bandstructure}(a)). Gaps are opened at K and K$'$ since
the Dirac electrons now feel the AFM exchange field from LaCrO$_3$ and
SOC. The two gaps are different in magnitude and given explicitly by
2(M+$\lambda$) and 2(M-$\lambda$) respectively (see
Fig.~\ref{fig:quantumtransition}(b) for $V=0$), from which one obtains
$M=0.166$eV and $\lambda=32.91$meV for this system. SOC effect is
quite large in the present case, due to the orbital hybridization
between the two e$_{\rm g}$ orbits in the new frame along the [111]
crystalline direction.  All the states in
Fig.~\ref{fig:bandstructure}(b) are doubly degenerate due to the
product symmetry of TRS and IS.  If a gate electric field is applied
along the [111] direction, the Au ions in the buckled honeycomb
lattice should experience staggered electric potentials.  By
increasing the electric potential, the system should be driven into
the spin-resolved QAHE, as summarized in
Fig.~\ref{fig:quantumtransition}(b). The required electric field is on
the order of 0.1V/$\textup{\AA}$. As the maximal non-trivial gap is
given by $2\lambda\simeq 600$K, this topological state in the
LaCrO$_3$/LaAuO$_3$ superlattice may remain stable even around room
temperature. Interaction driven QAHE phase can also be found in SrTiO$_3$/LaCoO$_3$/SrTiO$_3$ quantum well as discussed in detail in Ref.~\onlinecite{LCOSTO2015}.

It is worth noticing that, although the edge current of the
spin-resolved QAHE state is fully spin polarized, the total system
does not show any net spin magnetization, since the numbers of
occupied states in the two spin channels of the buckled honeycomb
lattice are the same, and the substrates are AFM insulators. The spin
polarization of the edge current can be reversed by the gating
electric field, which is superior for future spintronic applications.


\section{Discussions and Future Prospects}

At this stage, the quantum Hall trio (namely, quantum Hall effect,
quantum spin Hall effect, and quantum anomalous Hall effect) has been
completed.~\cite{QuantumHallTrio} Looking back to history, it was a long term exploration
towards the final realization of QAHE. First, the understanding of the
Berry phase mechanism of intrinsic AHE was an important step, because
it established the link between the AHE and the topological properties
of electronic structure. Second, the rapid progresses in the field of
topological insulators strongly stimulated the studies on QAHE. Third,
first-principles calculations play important roles in predicting and
selecting candidate materials and systems.  This also establishes a
nice example that theoretical and computational studies can in some
cases really go ahead of and greatly help experiments. Finally, the
well-controlled growth and magnetic doping of high quality samples and
transport measurement at ultra-low temperature make the experiments
possible, thanks to the progress of modern experimental
technology. All these aspects combined together finally gave birth to
the QAHE.

On the other hand, the successful realization of QAHE is not the end
of the story. Instead, we believe this success will inspire more
extensive research in this field. Two issues become immediately
important for future studies:

(1) How can we realize a higher plateau with a Chern number larger
than 1?

(2) How can we increase the temperature range of QAHE (it is now
observed only in the tens mK range)?


From the theoretical point of view, the QAH states with higher Chern
numbers can be derived in a way similar to that described in section VI.B
if the magnetic exchange field can be made
larger~\cite{QAHE_higher_Plateaus, jiang_quantum_2012}. Larger
exchange field will, in general, generate more band inversions among
sub-bands and thus induce higher Chern numbers. While in practice this
idea is very difficult, because the system with large enough exchange
field will usually become completely metallic before the second
plateau state appears. An alternative way to realize the QAHE with a
high Chern number is to start with materials that have multiple band
inversion regimes in the BZ. An interesting proposal is the
magnetically doped topological crystalline insulators (Pb,Sn)(Te,Se),
which have band inversions around all the four $T$ points in the
BZ~\cite{BAB_QAHE_crystalineTI_2014}. Similar to the situation in
Bi$_2$Se$_3$ family compounds, here the band inversions appear between
$p$-bands with opposite parity, and the Van Vleck mechanism also
works, which can lead to a stable FM semiconductor ground state after
magnetic doping. Therefore, for thin films of this system with proper
thickness, it is possible to realize the QAHE with a Chern number
between $\pm4$~\cite{BAB_QAHE_crystalineTI_2014}.


The experimental observation of QAHE in thin films of Cr-doped
(Bi,Sb)$_2$Te$_3$ requires an extremely low temperature (approximately
30 mK), which hinders realistic applications
\cite{chang_experimental_2013}. This is presumably due to the small
band gap opened by exchange coupling and the low carrier mobility of
760 cm$^2$/Vs in the sample. Therefore, searching for realistic
systems with non-trivial band structures, strong exchange coupling and
high sample quality is essential to realize the QAH effect at a higher
temperature. Liu et al. proposed to realize QAHE in a more
conventional diluted magnetic semiconductor, the Mn-doped InAs/GaSb
type II quantum well, which has a higher carrier mobility of 6000
cm$^2$/Vs and a predicted FM transition temperature of 30
K~\cite{LiuCX_QAHE_InAs-GaSb_2013}. Other theoretical proposes for
high temperature QAHE systems are based on the stanene and germanene
materials.  These two materials are recently predicted to be QSHI with
a large insulating gap \cite{XuYong_Tin_QSH_2013}. The QSH phase
appears due to an energy inversion between the Sn(Ge)-s and
Sn(Ge)-p$_{xy}$ bands, which have opposite parities and lead to a
sizable bulk gap of 0.3 eV. In Ref. ~\onlinecite{YanBH_highT_QAHE_2014},
S. C. Wu, et al, proposed that half-passivated stanene and germanene
become Chern insulators with energy gaps of 0.34 and 0.06 eV,
respectively. The proposed FM order in this system comes from the
unpassivated sublattice that exhibits dangling bonds. Strong coupling
between the spin-polarized dangling bond states and the inherent
inverted bands opens a considerable energy gap. The estimated
mean-field Curie temperature is as high as 243 K and 509 K for Sn and
Ge lattices, respectively. In spite of those theoretical proposals, it
is still challenging to reach QAHE at higher temperature
experimentally. To achieve this goal, searching for new host materials
with simpler electronic structure, easier synthesis and less extreme
growth conditions is necessary.


The QAHE and Chern insulators can be also helpful for the possible
realization of Majorana fermions
(MF)~\cite{elliott_textitcolloquium_2015}, which are exotic
charge-neutral particles that are their own antiparticles. In recent
years, Majorana fermions have been intensively studied in the field of
condensed matter physics due to their importance to both fundamental
science and potential applications in topological quantum
computing. Theoretically, Majorana fermions have been predicted to
exist in several spin orbit coupled superconducting systems, such as
superconductor-topological insulator interfaces
\cite{LiangFu_MF_TI_2008PRL}, semiconductor-superconductor
heterostructures \cite{DasSarma_MF_2010-1, DasSarma_MF_2010-2,
  Oreg_MF_QuantumWires_2010}, nanotube-superconductor
devices\cite{DanielLoss_MF_CarbonNano_2012}, half-metal-superconductor
interface~\cite{NaCoO2_Weng}, etc. Experimentally, great efforts have
been made in this direction, and some important evidence has been
successfully observed~\cite{mourik_signatures_2012,
  nadj-perge_observation_2014}. On the other hand, in
Ref.~\onlinecite{QiXL_MF_from_QHE_2010PRB}, a different way to generate
Majorana fermions is proposed in a hybrid device that consists of a QH
or QAH insulator layer in proximity to a fully gapped superconductor
layer on top.  In this case, it can generally be expected that a
chiral topological superconductor phase with an odd number of chiral
Majorana edge modes will exist. Compared to the QH system in the
proposal, the QAH system has the advantage that it does not require
external magnetic field, which makes the realization of superconductor
proximity effect much easier. The Majorana zero mode in a model
combined with a QAHE system and an s-wave superconductor is also
discussed in Ref.~\onlinecite{YuYue_MF_QAHE_2011PRB}.

In summary of the present paper, we have reviewed the history of
achieving the QAHE. In 1880, E. H. Hall found the anomalous Hall
effect, 17 years before the discovery of the electron. In the
following 100 years, however, the developments in this field,
particularly for the understanding of AHE, were slow. There were
debates on whether intrinsic or extrinsic contribution is dominant.
The breakthrough was the discovery of the QHE in 1980, from which
people learned that there exist new quantum states characterized by
the topology of electronic structures. The concept of the Berry phase,
established in 1984 by M. V. Berry, is the solid base of various
topological quantum states. We have reviewed how the Berry phase is
related to the AHE, QHE, QAHE, QSHE, 3D TIs and 3D Weyl/Dirac
semimetals. Such a unified view has led us to a unified scheme, called
the Wilson loop method, in calculating various topological invariants,
including the Chern number, Z$_2$ index, mirror Chern number and Fermi
surface Chern number. This method is efficient in rigorously
determining the topology of real materials. However, in practical
calculations, the band inversion mechanism is much more intuitive and
efficient in predicting, selecting and designing candidate topological
materials. Finally, we have listed the necessary and complete four
conditions for realizing the QAHE. Following these conditions, we have
demonstrated several ways to achieve the QAHE.  The theoretical
prediction for the QAHE in magnetic-ion Cr and Fe doped Bi$_2$Se$_3$
and Bi$_2$Te$_3$ family TIs has been recently confirmed by
experimental observation. The key difference of this scheme from other
proposals is the FM ordering of Cr and Fe dopants induced by the van
Vleck paramagnetism, which makes this proposal successful.

\section{Acknowledgements}
H.M.W., X.D. and Z.F. are supported by the National Science Foundation of China,
the 973 program of China (No. 2011CBA00108 and 2013CB921700), and the ``Strategic Priority Research Program (B)" of the Chinese Academy of Sciences (No. XDB07020100).
R.Y. and X.H. are supported by the WPI Initiative on Materials Nanoarchitectonics, and
partially by Grant-in-Aid for Scientific Research under the Innovative Area
``Topological Quantum Phenomena" (No.25103723), MEXT, Japan.

\section{references}
\bibliography{section_references}
\end{document}